\documentclass[english, polish,11pt]{article}   
\usepackage[utf8]{inputenc}
\usepackage{lmodern}  
\usepackage[T1]{fontenc}

\usepackage{graphicx}				
\usepackage{cancel}
\usepackage{amsthm}
\usepackage{tikz}
\usepackage{pgfplots}
\usetikzlibrary{decorations.markings}
\usepackage{soul}

\usepackage{xcolor}
\definecolor{mydarkgreen}{RGB}{0,100,0}			

\usepackage{hyperref}

\usepackage{amssymb}
\usepackage{amsmath}
\usepackage{amsfonts}
\usepackage{mathbbol}
\usepackage{accents}

\newcommand{\bcc}{\color{black}}

\newcommand{\bl}{\color{black}}
\newcommand{\blue}{\color{black}}
\newcommand{\red}{\color{black}}
\newcommand{\magenta}{\color{black}}
\newcommand{\green}{\color{black}}
\newcommand{\dgreen}{\color{black}}   
\newcommand{\cyan}{\color{black}}

\newlength{\dinwidth}
\newlength{\dinmargin}

\newif\ifshowdetails
\showdetailstrue

\hypersetup{
 linkcolor=blue,citecolor=blue, urlcolor=blue}

\newcounter{todo}

\newcommand{\ppU}{A}
\newcommand{\ppV}{B}
\newcommand{\vA}{\vec{A}}
\newcommand{\vB}{\vec{B}}

\newcommand{\todobox}[1]{
  \textcolor{blue}{
    \fbox{\parbox{0.9\textwidth}{#1}}%
  }
}

\newcommand{\vpU}{\vec{\ppU}}

\newcommand{\todonotetag}{TODO\thetodo}
\newcommand{\todonote}[1]{%
\stepcounter{todo}%
{\let\thefootnote\todonotetag
\footnote{\todobox{#1}}%
}
}

\definecolor{detailsgray}{gray}{0.3}

\ifshowdetails
\newcommand{\detail}[1]{%
{\color{detailsgray}$\blacktriangleright${#1}$\blacktriangleleft$}%
}

\newcommand{\detailspar}[1]{
\par \noindent {\color{detailsgray} $\blacktriangleright$ \textit{#1} $\blacktriangleleft$ } \par
}

\newenvironment{details}{\par \noindent \begingroup\itshape\color{detailsgray} $\blacktriangleright$ \hrulefill%
\par \noindent \ignorespaces}{\par\noindent \hrulefill $\blacktriangleleft$  \endgroup \par \noindent}

\else
\newcommand{\detail}[1]{} 
\newcommand{\detailspar}[1]{} 
\usepackage{environ}
\NewEnviron{hide}{}

\fi

\newcommand{\ch}{c_{\h}}

\newcommand{\pU}{U'}

\newcommand{\pmca}{\mca'}

\newcommand{\vr}{\vec{r}}

\usepackage{tikz}
\usetikzlibrary{trees}
\usetikzlibrary{decorations.pathmorphing}
\usetikzlibrary{decorations.markings}

\newcommand{\y}{\!\!\!}
\newcommand{\mrm}{\mathrm}

\newcommand{\veps}{\epsilon}

\renewcommand{\mathbf}{\boldsymbol}

\newcommand{\mcL}{\mathcal L}

\newcommand{\mcC}{\mathcal C}

\newcommand{\mcA}{\mathcal A}

\renewcommand{\i}{\mathrm i}

\newcommand{\R}{\mrm{R}}

\newcommand{\W}{w}

\newcommand{\ti}{\tilde}

\newcommand{\Om}{\Omega}
\newcommand{\ga}{\gamma}

\newcommand{\mca}{\mathcal{A}}

\newcommand{\pa}{\partial}

\newcommand{\ov}{\overline}

\newcommand{\eps}{\varepsilon}
\newcommand{\de}{\delta}

\newcommand{\De}{\Delta}
\newcommand{\e}{\mathrm{e}}

\newcommand{\mcc}{\mathcal{C}}

\newcommand{\nin}{\noindent}
\newcommand{\si}{\sigma}
\newcommand{\ph}{\phantom}
\newcommand{\h}{\fr{1}{2}}
\newcommand{\nat}{\mathbb{N}}

\newcommand{\om}{\omega}

\newcommand{\supp}{\mathrm{supp}}
\newcommand{\fr}[2]{\frac{#1}{#2}}

\newcommand{\real}{\mathbb{R}}

\newcommand{\la}{\lambda}
\newcommand{\non}{\nonumber}

\newcommand{\lan}{\langle}
\newcommand{\ran}{\rangle}

\newcommand{\Conf}{\mrm{Conf}}

\newcommand{\mcd}{\mathcal{D}}

\def\proof{\noindent{\bf Proof. }}
\def\qed{$\Box$\medskip}

\newtheorem{theoreme}{Theorem } [section]
\newtheorem{mainresult}{Theorem}

\newtheorem{proposition}[theoreme]{Proposition}
\newtheorem{lemma}[theoreme]{Lemma}
\newtheorem{definition}[theoreme]{Definition}
\newtheorem{corollary}[theoreme]{Corollary}
\newtheorem{remark}[theoreme]{Remark}
\newtheorem{example}[theoreme]{Example}
\newtheorem{criterion}[theoreme]{Criterion}
\newtheorem{conjecture}{Conjecture}
\newtheorem{assumption}{Assumption}

\newcommand{\bea}{\begin{assumption}}
	\newcommand{\eea}{\end{assumption}}
\newcommand{\beco}{\begin{conjecture} }
	\newcommand{\eeco}{\end{conjecture} }
\newcommand{\beq}{\begin{equation}}
	\newcommand{\eeq}{\end{equation}}
\newcommand{\beqa}{\begin{eqnarray}}
	\newcommand{\eeqa}{\end{eqnarray}}
\newcommand{\ben}{\begin{arabicenumerate}}
	\newcommand{\een}{\end{arabicenumerate}}
\newcommand{\bex}{\begin{example}}
	\newcommand{\eex}{\end{example}}
\newcommand{\ber}{\begin{remark}}
	\newcommand{\eer}{\end{remark}}
\newcommand{\bec}{\begin{corollary}}
	\newcommand{\eec}{\end{corollary}}
\newcommand{\bep}{\begin{proposition}}
	\newcommand{\eep}{\end{proposition}}
\newcommand{\becr}{\begin{criterion}}
	\newcommand{\eecr}{\end{criterion}}

\newcommand{\U}{\mrm{Conf}}

\def\bel{\begin{lemma}}
	\def\eel{\end{lemma}}
\def\bet{\begin{theoreme}}
	\def\eet{\end{theoreme}}
\def\bed{\begin{definition}}
	\def\eed{\end{definition}}

\newcommand{\Z}{\mathbb{Z}}

\newcommand{\2}{\!\!&}

\newcommand{\B}{B_1}

\newcommand{\vConf}{\overrightarrow{\Conf}}

\newcommand{\G}{\mrm{G}}
  
\newcommand{\del}{s}
\newcommand{\ovR}{R^{\scriptscriptstyle -1}}
\newcommand{\Gn}{G(\Om)}
\newcommand{\NLSM}{\mrm{NLSM}}
\newcommand{\YM}{\mrm{YM}}
\newcommand{\GN}{\mrm{GN}}

\title{ The Bałaban variational problem  in the non-linear sigma model}
\author{Wojciech Dybalski\,$^a$, Alexander Stottmeister$^b$ \ and \
Yoh Tanimoto$^c$\,  \\[5mm]
\normalsize ${}^a$ Faculty of Mathematics and Computer Science, 
\normalsize Adam Mickiewicz University, \\
\normalsize ul.~Uniwersytetu Pozna\'nskiego 4, 61-614 Pozna\'n, Poland\\
\normalsize E-mail: {\tt wojciech.dybalski@amu.edu.pl} \\ [2mm]
\normalsize ${}^b$ Institut f\"ur Theoretische Physik, Leibniz Universit\"at Hannover, \\
\normalsize Appelstrasse 2, \ 30167 Hannover, Germany  \\ 
\normalsize E-mail: {\tt alexander.stottmeister@itp.uni-hannover.de} \\ [2mm] 
\normalsize ${}^c$ Dipartimento di Matematica, Universit\`a di Roma ``Tor Vergata''\\ 
\normalsize Via della Ricerca Scientifica, 1 - I--00133 Roma, Italy.\\ 
\normalsize E-mail: {\tt hoyt@mat.uniroma2.it} }

\date{}

\begin{document}

\maketitle

\begin{center}
    \emph{Dedicated to the memory of Huzihiro Araki.}
\end{center}

\vspace{0.3cm}

\begin{abstract}
The minimization of the action of a QFT with a constraint dictated
by the block averaging procedure is an important part of
Bałaban's approach to renormalization. It is particularly
interesting for QFTs with non-trivial target spaces, such as
gauge theories or non-linear sigma models on a lattice. 
We analyze this step for the $O(4)$ non-linear sigma model 
in two dimensions and demonstrate, in this case, how various ingredients 
of Bałaban's approach play together. First, using variational calculus 
on Lie groups, the equation for the critical point is derived. Then, this 
non-linear equation is solved by the Banach {\red contraction mapping} theorem. This 
step requires detailed control of lattice Green functions and their 
integral kernels via random walk expansions.  

\end{abstract}

\newpage

\newcommand{\m}{\mrm{m}}
\renewcommand{\v}{f}
\renewcommand{\u}{g}
\newcommand{\w}{h}
\newcommand{\GG}{O}
\newcommand{\mud}{\bullet}
\newcommand{\bmu}{\bar{\mu}}
\newcommand{\ovs}{\overset}
\newcommand{\ovsum}{\ovs{*}{\sum}{}}
\newcommand{\one}{\mathbb{1}}
\renewcommand{\L}{\mathcal{L}}
\renewcommand{\d}{\mrm{d}}

\tableofcontents

\newpage

\section{Introduction}\label{introduction}
\setcounter{equation}{0}

The traditional strategy towards the construction of a non-trivial QFT  in four dimensions
{\bl had} been to start from the  $P(\phi)_2$ models,
then tackle the $\phi^4$-interaction in three dimensions, and finally attack the $\phi^4_4$ model
as the seemingly simplest interacting theory in spacetime of physical dimension.  
However, since the early 80s evidence started accumulating that $\phi^4_4$ is actually trivial \cite{Ai82, AD21, Fr82}. The attention shifted towards asymptotically free theories, such as the Gross-Neveu model in two dimensions ($\GN_2$), the 
non-linear sigma model in two dimensions  ($\NLSM_2$), and the Yang-Mills 
theory in four dimensions ($\YM_4$). 
While the $\GN_2$ is by now mathematically well understood \cite{DR00,FMRS86, GK85, DY23}, 
only partial results on the latter two models have been obtained \cite{GK86, PR91, BJ85, MRS93}. 
Interestingly, the  $\NLSM_2$ bears some similarity both with the good old $P(\phi)_2$ and 
with the elusive $\YM_4$.
To indicate the similarity to the $P(\phi)_2$, recall that the action of the $O(N)$ non-linear sigma model is given by
\beqa
\mathcal{A}[\phi] = \h \int d^2x\, \pa_{\mu} \phi(x) \cdot  \pa^{\mu}\phi(x), \quad |\phi|^2=1,   \label{sigma-action}
\eeqa
i.e., it differs from the free field theory only by the constraint restricting the field $\phi: \real^2\to \real^N$
to the sphere $S^{N-1}$. The Dirac delta implements this constraint in the functional measure.
If we approximate this delta by Gaussian functions, we obtain a family of two-dimensional models
with polynomial interactions, see, e.g., \cite{BZ72, MR89}.  As for the similarity of the $\NLSM_2$ to the $\YM_4$, it is
particularly striking for $N=4$. In this case, for any $\phi\in S^3$ we have $U:=\phi_0 {\magenta 1}+\i \vec{\phi} \cdot \vec{\si}\in SU(2)$,
where $\vec{\si}$ are the Pauli matrices. In these variables, we obtain from (\ref{sigma-action}) a principal chiral model,   {\bl cf. \cite[p.19]{Po}, \cite[p. 129]{Mo}}. 
Its formal discretization gives the following action
\beqa
\mathcal{A}(U) =\sum_{b\in \Om'} \mrm{Tr}(1- \pa U(b)), \label{action-intro}
\eeqa 
where $\Om\subset \mathbb{Z}^2$ is a finite unit lattice,  the sum is over bonds $b=(b_-,b_+)$ in $\Om$ and $\pa U(b):=U(b_-)U(b_+)^*$, which is analogous to the `holonomy'  along $b$. 
The theory has a global $SU(2)\times SU(2)$ symmetry given by $\mcA(u_1Uu_2)=\mcA(U)$ for any $x$-independent $u_1,u_2\in SU(2)$.
The expression (\ref{action-intro}) resembles the Wilson action of  {\magenta the} $\YM_4$, where the sum is over plaquettes in $\mathbb{Z}^4$, $\pa U$ are the 
corresponding plaquette variables and local gauge symmetry holds. The two theories also share some important qualitative properties,
such as perturbative asymptotic freedom \cite{Po, MR89} and, more speculatively, mass transmutation \cite{Po,Fa02, Ku80, Ko98}. 
Thus, a convincing strategy toward a construction of a non-trivial QFT in four dimensions has the form 
 \beqa
 P(\phi)_2 \to \NLSM_2  \to \YM_4. \label{diagram-two}
\eeqa
This motivates our paper, which prepares the ground for non-perturbative renormalization of the non-linear sigma model.

The utility of the $\NLSM_2$ as a toy model for the $\YM_4$ was pointed out, in particular, by Bałaban in 
\cite{Ba87}. Apparently, Bałaban worked out his proof of the UV stability of the $\YM_4$ first in the case of {\magenta the} $\NLSM_2$,
but these considerations remained unpublished. {\bl As Bałaban's papers on the $\YM_4$ are not easily accessible (cf.~\cite[p.326]{MRS93}),
 we find it worthwhile to work out  in the $\NLSM_2$ one aspect of  Bałaban's method, which is the variational problem.} 
To put the variational problem
into perspective, we recall that the context of the entire construction is {\bcc the} Wilson-Kadanoff renormalization as sketched in \cite{BJ85}. That is, the unit
lattice $\Om$ is divided into boxes $B_1(y)$, which determine the coarse lattice with a spacing $L>1$. At each point
of the coarse lattice, there lives a field $\mcC(U)(y)$, which is a suitable average of the fields  $U$ inside the box $B_1(y)$.
Given this data, one computes the effective action $\mcA_1$ after one step of the renormalization group:
\beqa
\e^{-\fr{1}{g_1^2}\mcA_1(V)}:=\int dU\, \chi_{\eps}\,\de(\mcC(U)V^{-1})\,\e^{-\fr{1}{g^2}\mcA(U)}, \label{intro-effective-action}
\eeqa
where $dU$ is the product of the Haar measures on $SU(2)$ over all  lattice sites, the Dirac delta 
restricts the integration region to configurations $U$ block-averaging to $V$ \cite{Ya01}, and we ignored additive counterterms for simplicity. The characteristic function $\chi_{\eps}$ 
imposes the small field condition, which requires that the differences between
fields at neighboring points are bounded by some $\eps>0$. We note that the effective action also
enjoys the global $SU(2)\times SU(2)$ symmetry since our averaging satisfies $\mcC(u_1Uu_2)=u_1\mcC(U)u_2$ for 
$u_1,u_2\in SU(2)$, cf.~(\ref{symmetry-constraint}) below. In the regime of small $g$,  
the expression (\ref{intro-effective-action}) can be studied by the method of steepest descent: the leading contribution to the integral is due to the critical points of $U\mapsto \mcA(U)$ subject to the constraint $\mcC(U)=V$ and the
small field condition. This is {\bcc the Bałaban} variational problem. It prepares the ground for rewriting equation (\ref{intro-effective-action})
as a perturbation of a Gaussian measure, which is tractable by the usual methods of constructive QFT. 
Our solution to the variational problem is quite different from Bałaban's discussion of the corresponding
problem for the $\YM_4$ in \cite{Ba85}. We put additional emphasis on the clarity of the presentation, in particular on the 
separation of the geometric and analytic considerations.  We  {\bl intend} to demonstrate that {\bcc the Bałaban} variational
problem is an elegant topic in the variational calculus on Lie groups, cf.~Subsection~\ref{general-discussion}. 
{\bl We hope that our paper will be a useful addition
to a growing library of accessible literature on Bałaban's method, see, e.g., \cite{BJ85, Ya01, Di13, DY23, Di18, DST23}.}

{\bl To support the above comments, let us outline our solution of {\bcc the Bałaban} variational problem.} We first change the variables
from $U(x)$ to $U'(x):=U(x)V^{-1}(y_x)$, $x\in B_1(y_x)$, which describe fluctuations around the value dictated by the constraint. Then the problem
is to find the critical points of 
\beqa   
\mathcal{A'}(U') :=\sum_{b\in \Om'} \mrm{Tr}(1- \underbrace{U'(b_-)\pa V(y_b)U'(b_+)^*}_{=:W(b)} ) \quad \textrm{ with the constraint } \quad \mcC(U')=1, \label{W-variable}
\eeqa
where $\pa V(y_b):=V(y_{b_-})V(y_{b_+})^*$.  Both $U'$ and $W$ are elements of $SU(2)$, thus can be parametrized by vectors $\vec{A}$ and $\vec{W}$ multiplying the Pauli matrices, as indicated above (\ref{action-intro}).
Now, we consider the system
of equations
\beqa
\mcL_{X}\mcC(U')=0, \quad \mcL_{X}\mcA'(U')=0, \label{system-intro}
\eeqa
where $\mcL_{X}$ is the Lie derivative in the direction of the tangent vector field $X$. From the first equation in (\ref{system-intro})
we determine the tangent space of the constraint manifold, from the second one we obtain
the critical points of the action on the constraint manifold. To describe the solution, we introduce a 
transformation $R(x)\vec{v}=A_0(x)\vec{v}+\vec{A}(x)\times \vec{v}$ which is a sum of a 
rotation in the plane orthogonal to $\vec{A}$ and a rescaling determined by the length of $\vec{A}$. We also
introduce a derivation $\pa$ which maps functions on lattice sites into functions on bonds
according to $(\pa f)(b)=f(b_-)-f(b_+)$. Then  $-\pa^*\pa$ coincides with the lattice Laplacian $\De_{\Om}$
with Neumann boundary conditions. As a direct consequence of (\ref{system-intro}), the family of
vectors
\beqa
\vec{C}(x):=R^{-1}(x)^* \pa^*\vec{W}(x)    \label{intro-conservation}
\eeqa 
is constant on each block $B_1(y)$ at the critical point. 
We proceed from this relation to an equation for the critical point
via the following steps: First, we note that in the variables $\vec{A}$ the constraint has the simple form $Q(\vec{A})=0$, where $Q$ is the arithmetic
mean over boxes. Second, we decompose the variable $\vec{W}$, appearing in (\ref{W-variable}), as follows
\beqa
\vec{W}=\pa \vec{A}+\vec{r}, \label{intro-decomposition}
\eeqa
where $\pa \vec{A}$ is  the leading term and $\vec{r}_{\vec{A}}$ the remainder with respect to the parameter $\eps$ appearing in the small field condition. 
Then, using (\ref{intro-conservation}), (\ref{intro-decomposition}), we obtain  in Theorem~\ref{critical-point-thm} the following equation for the critical point:
\beqa
\vec{\ppU}=  \Gn R^* Q^* [Q\Gn R^*Q^*]^{-1}  Q \Gn \pa^* \vec{r}  -  \Gn \pa^* \vec{r}, \label{critical-equation-intro}
\eeqa 
where $\Gn:=(-\De_{\Om}+Q^*Q)^{-1}$ is a lattice Green function.  Recalling that both $R$ and $\vec{r}$ depend on $\vec{A}$,
this is a highly non-linear equation.

The problem of existence and uniqueness of solutions of equation~(\ref{critical-equation-intro}) constitutes the analytic part
of our considerations. As the equation (\ref{critical-equation-intro}) has the schematic form $\vec{\ppU}=\mrm{T}(\vec{\ppU})$, we 
apply  the Banach {\red contraction mapping} theorem. The respective metric space is given by
\beqa
\mrm{X}_{\eps}:=\big\{\, \vec{\ppU}   \,\, \,|\, \,\,   Q( \vec{A} )=0, \,\,   
\sup_{b\in \Om'}\| (\pa U)( b)- 1\| \leq \eps   \big\},
\eeqa
which is dictated by the constraint and the small field condition (stated explicitly here). We choose the $\mcL^{\infty}$-metric
on this space as suggested by the supremum over $b$ defining the small field condition. The choice of any other $\mcL^p$-metric would
lead to a mismatch with the small field condition and, thus, to estimates depending on the number of lattice points  $n^2$. We stress that we want to prove the existence and
uniqueness of solutions of equation~(\ref{critical-equation-intro}) for $\eps>0$, which is small depending on $L$ but not on $n$ as the latter should ultimately tend to infinity in the continuum limit.

Thus, to obtain that $\mrm{T}$ is a contraction, we need $\mcL^{\infty}${\magenta-}bounds on the relevant operators appearing in (\ref{critical-equation-intro}).
In particular, we have to show 
\beqa
\|G(\Om)f\|_{\infty} \leq c\|f\|_{\infty}, \quad \| (QG(\Om)Q^*)^{-1} f\|_{\infty} \leq c\|f\|_{\infty}. \label{specific-bounds}
\eeqa
It is a simple and general fact that such bounds hold for operators whose integral kernels have an exponential decay. 
It is less well known, but also true for strictly positive operators on $\mcL^2(\mathbb{Z}^d)$, that exponential
decay of the integral kernel implies the exponential decay of the integral kernel of \emph{the inverse operator}. Bałaban and Jaffe showed the latter fact in \cite{BJ85} using the method of random walk expansions. We reproduce 
their argument in Appendix~\ref{random-walk-appendix} and use it in combination with the method of images to prove (\ref{specific-bounds}).  

Another important step of the proof that $\mrm{T}$ is a contraction is to  show that $\mrm{sup}_{x\in \Om}|{\magenta \vec{A}} (x)|\leq c\eps$ {\magenta for $ \vec{A}\in \mathrm{X}_{\eps}$}.
We stress that this bound cannot follow from the small field condition alone, as the latter only controls differences of fields at neighboring points.
By exploiting in addition the constraint $Q( \vec{A} )=0$, we obtain the required bound in Theorem~\ref{configurations-theorem}. 
Interestingly, in the case of the $\YM_4$, the corresponding bound would follow quite easily using a gauge fixing condition which switches off the fields on many bonds in each box, cf. \cite[Lemma 1]{Ba85.1}. This demonstrates that the similarity of the two models has its limitations, and a rigorous analysis of the $\NLSM_2$ must not rely
 on papers on the  $\YM_4$ for technical material.

Regarding future directions, we plan to expand the expression on the r.h.s.~of (\ref{intro-effective-action}) around the obtained critical point.
After changing variables to the Lie algebra elements $\vec{A}$, it should be possible to rewrite the measure after one step as a perturbation of a Gaussian measure. 
It is an interesting question if the quadratic form defining this measure is strictly positive. As a matter of fact, the $P(\phi)_2$ models
approximating the $\NLSM_2$, which we mentioned above, have massless Goldstone bosons in their actions. We expect, however, that
in (\ref{intro-effective-action}), the resulting infrared problems will be eliminated by the  Dirac delta imposing the constraint. Then, using the cluster expansion,
we should be able to determine the behavior of the coupling constant $g\to g_1$ after the first step of the renormalization group. On the other hand, if the infrared problems
persist, we may have to introduce a gauge fixing in the functional measure, similar to the one considered in \cite{Da80}. 

In the present paper, we study only one step of the renormalization group, from a theory on a unit lattice to a theory on an $L$-lattice.
Ultimately, we would like to understand $k$ steps of the renormalization group, starting from a theory on an $L^{-k}$-lattice and going
up to a theory on a unit lattice. Using the semigroup property of the renormalization group transformations (\ref{intro-effective-action}) 
    it is easy to guess that {\bcc the  Bałaban} variational problem consists in minimizing $U\mapsto \mca(U)$ with the constraint $\mathcal{C}^k(U)=V$ in this case.
This is actually how the variational problem was originally formulated in \cite{Ba85}, disregarding additional complications related to the
large field problem. We hope to come back to this problem in the case of the $\NLSM_2$ in future work. One complication is that the linearization
of the constraint to $Q(\vec{A})=0$ is no longer automatic but requires an additional application of the Banach contraction mapping theorem. Furthermore,
one needs to establish the exponential decay {\magenta of integral kernels} for more complicated operators than those appearing in (\ref{specific-bounds}). It is likely, however, that the Bałaban-Jaffe lemma mentioned above will solve a substantial part of this problem. 
  
  Our paper is organized as follows: 
  In Subsection \ref{themainresult}, we introduce our setting and state the main result.
  In Section \ref{preparations}, we simplify the constraints by a change of variables
  and study the relations between spaces of configurations.
  In Section \ref{geometric}, we characterize the constraint manifold and write the equation for the critical point.
  In Section \ref{existence}, we show the existence and uniqueness of the solution of the critical point equation
  using the Banach contraction mapping theorem.
  We supplement the paper with  results about the exponential decay of integral kernels of certain operators in Appendix \ref{random-walk-appendix}
  and some more technical results in Appendix \ref{Lemma-SU(2)}.
  
\vspace{1cm}

{\magenta \nin \textbf{Acknowledgements:} W.D. would like to thank J\"urg Fr\"ohlich and Gian Michele Graf for useful discussions. W.D. was supported by the grant `Sonata Bis' 2019/34/E/ST1/00053  of the National Science Centre, Poland.  Y.T. is partially supported by the MUR Excellence Department Project MatMod@TOV awarded to the Department of  Mathematics, University of Rome ``Tor Vergata'' CUP E83C23000330006, by the University of Rome ``Tor Vergata'' funding OAQM
CUP E83C22001800005 and by GNAMPA–INdAM. A.S. has been funded by the MWK Lower Saxony via the Stay Inspired Program (Grant ID: 15-76251-2-Stay-9/22-16583/2022). }

\vspace{0.5cm} 
\nin\textbf{Notation}

\begin{enumerate}

\item  We introduce an odd positive integer  $L>1$ and set $I= [0,1,\ldots, n-1]$,  $n-1=L^m$, so that the parameter $m$ controls the size of the interval.

\item We denote by $\Om\subset  \mathbb{Z}^2$  the finite lattice $\Om:=I^{\times 2}= [0,1,\ldots, L^m]^{\times 2}$.

\item  We denote by  $\Om_1\subset L \mathbb{Z}^d$ the coarse lattices of the form $\Om_1=L [0,1,\ldots, L^{m-1}]^{\times 2}$.
  
\item We denote by $\Om'$ the set of oriented bonds on $\Om$.

\item We denote by $|\,\cdot\,|$ the length of a vector in $\real^\ell$ and by $\|\,\cdot\,\|$ the {\blue operator norm} on $\ell\times \ell$ matrices. (We will only
need cases $\ell=1,3$).

\item The elements of the Hilbert spaces $\L^2(\Om;\real^{\ell}), \L^2(\Om_1;\real^{\ell}), \L^2(\Om';\real^{\ell})$ are  complex-valued functions on the respective sets,  denoted $f,f', g,g'$.  The  scalar products {\magenta have} the form
\beqa
 \lan \v, \u\ran_{\Om}=\sum_{x\in \Om} \v(x) \cdot \u(x),\quad \lan \v, \u\ran_{\Om_1}=L^2\sum_{x\in \Om_1} \v(x) \cdot \u(x), \quad  
 \lan \v, \u\ran_{\Om'}=\sum_{b\in \Om'} \v(b)\cdot  \u(b). 
 \eeqa
 {\magenta We}  set $\|f\|^2_{2,\Om}=\lan f, f\ran_{\Om}$ and similarly in other cases. The dot above is the canonical scalar product in $\real^{\ell}$.
 
\item   We will write $\|f\|_{\infty {\magenta ;} \Om}:=\mrm{sup}_{x\in \Om}|f(x)|$. If there is no risk of confusion, we will
drop $\Om$.  We denote by $\mcL^{\infty}(\Om)$ the Banach space of functions on $\Om$ equipped with the norm $\|\,\cdot\,\|_{\infty {\magenta;} \Om}$.
Analogous definitions will be used for $\Om_1$. 

\item The operator norm of a map $M: \mcL^p(\Om;\real^{\ell})\to \mcL^q(\Om;\real^{\ell})$ will be denoted $\|M\|_{p,q;\Om}$. Explicitly,
\beqa
\|M\|_{p,q;\Om}=\sup_{\|f\|_{p;\Om}\leq 1} \|Mf\|_{q;\Om}.
\eeqa
An analogous definition will be used for $\Om_1$. (We will only need cases $p,q\in \{2,\infty\}$).

\item We define the boxes in the lattice $\Om$ {\blue for $y\in \Om_1$}
\beqa
\B(y):=\{\, x\in \mathbb{Z}^2  \,|\, y_{\mu}\leq x_{\mu}< y_{\mu}+L, \,\, \mu=0,1\}
\eeqa
and for any $x$ denote by $y_x$ the label $y$ of the box s.t. $x\in \B(y)$.

\item We denote by $\De_{\Om}$  the Laplacian on $\Om$ with Neumann boundary conditions.

\item $|x-x'|=\big(\sum_{\mu=0}^{1}(x_{\mu} -x'_{\mu})^2 \big)^{1/2}$, 
  $|x-x'|_{\infty}:=\sup_{\mu=0,1}|x_{\mu}-x'_{\mu}|$.

\item $\one_{\GG}$ denotes the characteristic function of a set $\GG$. 

\item By $c, c', c_1,c_2\ldots$ we denote numerical constants, independent of any parameters. Unless stated otherwise, by $C,C', C_1, C_2$
we denote constants which may depend on $L$ but independent of any other parameters (in particular independent of $n$). 
 All these constants may change from line to line.

\item {\blue We denote a scalar multiple of the identity operator on various vector spaces by the scalar. }

\item {\blue We assume summation over repeated indices e.g. $X_j\si_j:=\sum_{j=1}^3X_j \si_j$}.

\end{enumerate}

\subsection{The setting and results}\label{themainresult}

We  set $I:=[0,1,\ldots,n-1]$
and denote by $\Om=I^{\times 2}\subset \mathbb{Z}^2$ a finite lattice on which 
the model will be defined. Let $\Om'$ be the set of oriented bonds on $\Om$ denoted
by $b=(b_-,b_+)$.  Let $L>1$ be an odd integer and $\Om_1:=(L\Om)\cap \Om$ be the coarse lattice.
For every $y\in \Om_1$ we define a box in the original lattice
\beqa
B_1(y):=\{\, x\in \Om\,|\, y_{\mu}\leq x_{\mu} <y_{\mu}+L,\quad \mu=0,1\,\}, \label{box}
\eeqa
whose label $y$ is the left bottom corner.  For $x\in \Om$ we denote by $y_x$ the label of the box containing $x$.
We denote by $\Om'_1$ the set of oriented bonds on $\Om_1$.

Let  a Lie group $\G_0$ be a {\blue subgroup of} the unitary group  $U(N)$. We introduce the set of all configurations
\beqa
\Conf(\Om):= \G_0^{\times n^2} \label{conf-set}
 \eeqa
 whose elements have the form $U:=\{U(x)\}_{x\in \Om}$, $U(x)\in \G_0$. {\magenta For future reference,} we note that $\Conf(\Om)$
 corresponds to $\G$ from the general discussion in Subsection~\ref{general-discussion}.
 Now we define the action of the model as a function on $\Conf(\Om)$:
\beqa
\mca(U)= \sum_{b\in \Om'} \mrm{Re}\mrm{Tr}(1-  \pa U(b)), \quad \pa U(b):=U(b_-)U(b_+)^*. \label{action-one-x}
\eeqa
It has an important symmetry property: For any $x$-independent unitaries $u,v$ we have 
\beqa
\mca(uUv)=\mca(U). \label{symmetry}
\eeqa

We are going to {\magenta find critical points of}  this action with a constraint dictated by the block-averaging
procedure.  We follow the averaging method from   \cite{Iw85}.
As a first step we define a function 
\beqa
\mcC_0(U)(y):=\fr{1}{L^2} \sum_{x\in \B(y)} U(x) \label{C-zero}
\eeqa
which maps every configuration $U$ into a family of matrices $\{ M(y)\}_{y\in \Om_1}$ on the coarse lattice.
As these matrices need not be unitary,  we take a polar decomposition {\blue at each $y\in \Om_1$}
\beqa
\mcC_0(U)(y)=\mcC(U)(y) | \mcC_0(U)(y)|  \label{averaging-def}
\eeqa
and let the partial isometry  $\mcC$  be our averaging operation.  We set $\mcC(U)(y)=1$ whenever 
$\mcC_0(U)(y)=0$. As shown in Lemma~\ref{SU(2)-lemma}, such averaging operation is  well defined  for $\G_0=SU(2)$ and 
we have $\mcC(U)(y)\in \G_0$. 
  
 We note that  the averaging operation is consistent with the symmetry property (\ref{symmetry})
of the action. In fact, for any block-constant families of  unitaries $u,v$ 
\beqa
\mcC(uUv)=u\mcC(U)v. \label{symmetry-constraint}
\eeqa
This guarantees a consistent transformation of a constraint of the form
\beqa
\mcC(U)=V, \label{constraint}
\eeqa
where $V\in \Conf(\Om_1)$ is a given configuration on the coarse lattice. 

Finally we define  the following subset of the set of configurations (\ref{conf-set})
\beqa
\U_{\eps}(\Om)\2 = \2 \{\, U\in  \U(\Om)\,|\,   \| \pa U(b)-1\|\leq  \eps\, \,\,  \textrm{\blue for all $b\in \Om'$}  \}, \label{small-field-condition}
\eeqa
for $0<\eps\leq 1$, which encodes the small field condition.  Now we state our main theorem: 
\begin{mainresult}\label{main-theorem} Let  $\G_0=SU(2)$. Then there exist $0 < \eps, \eps_1\leq 1$  s.t.  for  $V\in \U_{\eps_{1}}(\Om_1)$ the action  $\mcA$ has a unique critical point over $\Conf_{\eps}(\Om)$ with the constraint $\mcC(U)=V$. The parameters $\eps,\eps_1$ are independent of $n$ but may depend on $L$.
\end{mainresult}
{\green We will prove this as Theorem \ref{main-theorem-intext} below.}
\begin{remark}\label{main-thm-remark} We provide several comments on Theorem~\ref{main-theorem} and the method of proof: 
\begin{enumerate}

\item It is clear from Definitions (\ref{box}), (\ref{action-one-x}), (\ref{C-zero}) that the variational
problem formulated above is in fact independent of the lattice spacing. For this reason we formulated the
problem on a unit lattice from the beginning. {\magenta This simplification is due to the fact that we look only at one step of the renormalization group.
In the full variational problem, mentioned in the Introduction, there are $k$ steps of the renormalization group, from the $L^{-k}$-lattice to the unit
lattice.  Then the dependence on the lattice spacing is encoded in the parameter $k$ and it persists after rescaling.}

\item To prove  Theorem~\ref{main-theorem} we will follow the strategy from {\magenta Sub}section~\ref{general-discussion}.  As the open set $\G_{\eps}$, on which the Lie derivatives will be computed, we take the interior of $\U_{\eps}(\Om)$.

\item {\red By setting $ \eps_1< \eps^2$ one can ensure that the critical point is not at the boundary of $\Conf_{\eps}(\Om)$.
Indeed, a critical point in $ \Conf_{\eps}(\Om)$ is also a critical point for $ \Conf_{\eps'}(\Om)$ s.t. $ \eps_1<{(\eps')}^2<\eps^2$.
Thus the unique critical point must belong to the interior of $\Conf_{\eps}(\Om)$.}

\item An essential
step of the proof of Theorem \ref{main-theorem} consists in showing that configurations from  $\U_{\eps}(\Om)$, satisfying the constraint, 
can be written as $U(x)=U'(x)V(y_x)$, where $U'(x)$ belongs to
\beqa
\U^{\eps'}(\Om) {\blue :=} \{\, U\in  \U(\Om)\,|\,    \|U(x)-1\|\leq \eps'\,  \textrm{ for all }  x\in \Om\, \} \label{U-condition}
\eeqa
for certain $\eps$, $\eps'$. We note the crucial role of the constraint in this step, which is performed in Subsection~\ref{configuration-appendix}:  the small field condition (\ref{small-field-condition})
by itself does not imply the condition from (\ref{U-condition}), as it concerns only differences of fields at neighbouring points.

\end{enumerate}

\end{remark}
In the course of our analysis we will often use a parametrization of  elements of $SU(2)$ by an axis $\hat n$ and an angle $a$ so that
\beqa
\e^{\i a(\hat n \cdot \vec{\si}) }=I \cos(a)+\i (\hat n\cdot \vec{\si})\sin(a), \label{SU2}
\eeqa
where $\vec{\si}:=(\si_1,\si_2,\si_3)$ are the Pauli matrices {\green (see e.g.\! \cite[Chapter III, Problem 5]{CDD82Analysis}).}
Consequently, we can write 
\beqa
\pU(x)= \del \ppU_0(x)I+\i \vec{\ppU}(x)\cdot \vec{\si}, \label{Pauli-matrices-decomposition}
\eeqa
where $\del:=\mrm{sgn}(\cos(a))$,  $|\vec{\ppU}(x)|\leq 1$, $A_0(x):=\sqrt{1-|\vec{\ppU}(x)|^2 }$ {\blue and $U'$ was introduced above (\ref{U-condition}).} 
It is therefore convenient to  define for $0<\eps \leq 1$ the corresponding family of configurations:
\beqa
  \vConf^{\eps}(\Om):=\{ \{ \vpU(x) \}_{x\in \Om} \,|\,  \vpU(x)\in \real^3,\, \,\,  \sup_{x\in \Om}|\vec{A}(x)|\leq \eps \,\}. \label{vector-conf}
\eeqa 
For any $\vpU\in \vConf^{\eps}(\Om)$ we have two distinct configurations $\pU$, corresponding to the two signs $\del={\blue \{\pm 1\}}$.
However, as we will see in Theorem~\ref{configurations-theorem}, only $\del=1$  plays a role in our discussion as we work close to the
unity in $\G_0$.

\section{Preparations}\label{preparations}
\setcounter{equation}{0}

\subsection{Linearization of the constraint}\label{simplification}

In this subsection we simplify the constraint.  {\green As we explained in Section \ref{introduction}}, we make a change of variables $U(x)=\pU(x)V(y_x)$, where $y_x$ is the label of the box to which $x$ belongs, i.e.,  $x\in B_1(y_x)$.  The 
action in the new variables has the form
\beqa
\pmca(\pU)= \sum_{b\in \Om} \mrm{Re}\mrm{Tr}(1-   \pU(b_-)  \pa V(y_b) \pU(b_+)^*   ), \quad \pa V( y_b):=V(y_{b_-})V(y_{b_+})^*,  
\eeqa
with the constraint
\beqa
\mcC(\pU V)=V \quad \Leftrightarrow \quad  \mcC_0(\pU)(y)= (\mcC_0(\pU)(y)^* \mcC_0(\pU)(y))^{\h}, \label{constraint-simpl}
\eeqa
where we made use of (\ref{symmetry-constraint}).
By   Lemma~\ref{SU(2)-lemma} and  the uniqueness of the polar decomposition the second relation in (\ref{constraint-simpl}) is equivalent to $\mrm{Re}\, \mcC_0(\pU) \geq 0$ and $\mrm{Im}\, \mcC_0(\pU)=0$.
The condition {\green $\mrm{Re}\, \mcC_0(\pU) \geq 0$} is satisfied
{\green automatically for $U'$ close enough to the identity (cf. Lemma \ref{configurations}).}
Thus there remains the constraint
\beqa
\mcC'(\pU)(y):=2\i \, \mrm{Im}\,  \mcC_0(\pU)(y)=0, \quad y\in \Om_{1},
\eeqa
which reads as follows
\beqa
\mcc'(\pU)(y)=\sum_{x\in \B(y)} (\pU(x)- \pU(x)^*)=0, \quad y\in \Om_1. \label{intermediate-constraint}
\eeqa
Using this, decomposition (\ref{Pauli-matrices-decomposition}) and Theorem \ref{configurations-theorem} 
we obtain 
\beqa
\sum_{x\in \B(y)} \vec{\ppU}(x)=0,\quad \del=1, \label{linearized-constraint}
\eeqa
thus we  linearized  the constraint. It is therefore convenient to define a linear averaging map $Q: \mcL^2(\Om;\real^\ell)\to \mcL^2(\Om_1;\real^\ell)$ by
\beqa
(Q f)(y):=\fr{1}{L^2} \sum_{x\in \B(y)} f(x) \label{linear-averaging}
\eeqa
to state the constraint (\ref{linearized-constraint}) as $Q(\vec{A})=0$. We note for future reference that
\beqa
(Q^* f)(x)\2=\2 f(y_x),  \quad (Q^*Q f )(x)=  \fr{1}{L^2} \sum_{x'\in \B(y_x)}f(x'),  \label{Q-properties}
\eeqa
i.e., $Q^*Q$ is the projection on block-constant functions.

\subsection{Sets of configurations}\label{configuration-appendix}

Recall the sets of configurations defined in  (\ref{small-field-condition}),  (\ref{U-condition}), (\ref{vector-conf}).
\bel \label{chain} {\blue For $U\in \U_{\eps}(\Om)$ and $x,x'\in \B(y)$ we have}
\beqa
\|U(x)U(x')^*-1\|\leq (2L) \eps.
\eeqa
\eel
\proof Let $b_1\circ \cdots \circ b_{\ell}$ be the shortest  oriented path of bonds s.t. $b_{1,-}=x$ and $b_{\ell,+}=x'$.
(We can always find such a path, possibly with $x,x'$ exchanged).  Thus we can write
\beqa
U(x)U(x')^*-1\2=\2 \pa U(b_1) \ldots \pa U( b_{\ell})-1 \non\\
\2 = \2 \pa U(b_1) \ldots \pa U( b_{\ell-1})( \pa U(b_{\ell})-1)+ \pa U( b_1) \ldots \pa U(b_{\ell-1})-1
\eeqa
where $\| \pa U(b_\ell)-1 \|\leq \eps$. After $\ell$ steps we estimate the norms and get
\beqa
\|U(x)U(x')^*-1 \|\leq \ell \eps \leq (2L) \eps.
\eeqa
This completes the proof. \qed
\bel\label{C-positivity} Let $U\in \Conf_{\eps}(\Om)$, $0<  \eps\leq 1/(4L)$. Then    $ |\mcC_0(U)(y)|$, $y\in \Om_1$, are strictly positive.
\eel
\proof We come back to definitions (\ref{C-zero}), (\ref{averaging-def})  and compute
\beqa
 \mcC_0(U)(y)^*  \, \mcC_0(U)(y) \2=\2\fr{1}{L^4} \sum_{x,x'\in \B(y)} U(x)^*U(x')\non\\
\2=\2 \fr{1}{L^2}\bigg(1 +\fr{1}{L^2}\sum_{\substack{x,x'\in \B(y) \\   x\neq x'}} U(x)^*U(x') \bigg) \label{well-definiteness}\\
\2=\2 \bigg( \fr{1}{L^2}  +\fr{1}{L^4}\sum_{\substack{x,x'\in \B(y) \\   x\neq x'}} (U(x)^*U(x')-1) +  \fr{1}{L^4} (L^4-L^2)\bigg) \\
\2=\2 \bigg( 1 +\fr{1}{L^4}\sum_{\substack{x,x'\in \B(y) \\   x\neq x'}} (U(x)^*U(x')-1) \bigg). \label{N-computation}
\eeqa  
In view of Lemma~\ref{chain}, for $(2L) \eps\leq 1/2$ the claim follows. \qed\\ 
{\magenta We recall from (\ref{constraint-simpl}) that} 
\beqa
\mcC(\pU V)=V \quad \Leftrightarrow \quad  \mcC_0(\pU)(y)= (\mcC_0(\pU)(y)^* \mcC_0(\pU)(y))^{\h}. \label{constraint-simpl-x}
\eeqa
This will be used in the following lemma.
\bel\label{configurations} Let $0<\eps \leq 1/(4L)$ and suppose that $U\in \U_{\eps}(\Om)$  satisfies the constraint (\ref{constraint}). 
Then $\| \pU(x)-{\blue 1} \|\leq  4\ch L\eps$.
\eel
\proof  Using (\ref{N-computation}), Lemma~\ref{chain} and Lemma~\ref{square-root} we have
\beqa
\|( \mcC_0(\pU) (y)^* \mcC_0(\pU) (y) )^{\h}-1\|\leq \ch (2L)\eps.
\eeqa
{\blue Only in this lemma we denote by $O(\epsilon)$ any operator satisfying $\|O(\epsilon)\|\leq \epsilon$.}
Thus  starting from (\ref{constraint-simpl-x}), 
\beqa
\mcC_0(\pU)(y)=1+O(\ch 2 L\eps), \textrm{ i.e., } \fr{1}{L^2} \sum_{x\in \B(y)} \pU(x)=1+O(\ch 2 L\eps). \label{averaging-one}
\eeqa
 Fix some $x,x'\in \B(y)$ and let $b_1\circ b_2\circ\ldots \circ b_{\ell}$ be
a  chain of bonds linking $x'$ to $x$, i.e. $b_{1,-}=x'$ and $b_{\ell,+}=x$. We have,   by Lemma~\ref{chain},
\beqa
\pU(x) +\pU(x')=(1+\pU(x')\pU(x)^*)U'(x)
=(2+O(2L\eps))\pU(x).
\eeqa
Proceeding analogously,
\beqa
\pU(x)+\pU(x')+\pU(x'')=(2+O(2L\eps)+\pU(x'')\pU(x)^*)\pU(x)=(3+2O(2L\eps))\pU(x).
\eeqa
Repeating the procedure $L^2$ times to account for all the  sites of $\B(y)$, we have
\beqa
\fr{1}{L^2}\sum_{x\in B(y)} \pU(x)=\fr{1}{L^2} \big(L^2+ (L^2-1)O(2L\eps)\big)\pU(x).
\eeqa 
Coming back to (\ref{averaging-one}), we have the following equality
\beqa
 \big(1+ O(2L\eps)\big)\pU(x)=1+O(\ch 2 L\eps)
\eeqa 
 Hence $\pU(x)-1= O(4\ch L\eps)$, which we wanted to prove. \qed
\bel\label{A-theorem} Let  $0<\eps \leq 1$. Then the following implication holds true:
\beqa
 \pU\in \Conf^{\eps}(\Om)\quad\2 \Rightarrow\2 \quad (\vec{\ppU}\in {\dgreen \vConf^{\eps}(\Om)} \textrm{ and }s=1). \label{implication-configurations} 
\eeqa
\eel
\proof We have
\beqa
\|\e^{\i a(\hat n \cdot \si) }-1\|^2 \2=\2 \|( \cos(a) -1)+\i (\hat n\cdot \vec{\si})\sin(a)\|^2\non\\
                                                 \2=\2 (\cos(a)-1)^2+\sin^2(a)\non\\
                                                \2=\2 2-2\cos(a)=2(1-\cos(a))=2\fr{\sin^2(a)}{1+\del \sqrt{1-\sin^2(a)} }.
\eeqa
Considering that $\|\e^{\i a(\hat n \cdot \si) }-1\|\leq 1$, we have $\del=1$ and
\beqa
\|\pU -1\|^2=2\fr{ |\vec{\ppU}|^2   }{1+\sqrt{1- |\vec{\ppU}|^2  } }.
\eeqa
Thus $\|\pU -1\|^2\leq \eps^2$ implies $ |\vec{\ppU}|^2\leq \eps^2$. \qed

\bet\label{configurations-theorem}  Let $0<\eps,{\blue \eps_1} \leq 1$, $U(x)=U'(x)V(y_x)$, $V\in \Conf_{\eps_1}(\Om_1)$. Then 
\beqa
 U\in \Conf_{\eps}(\Om) \textrm{ and } \mcC(U)=V \quad  \2\Rightarrow\2  \quad \pU\in \Conf^{( 4\ch  L\eps)}(\Om),  \label{first-implication-zero}\\
\pU\in \Conf^{( 4\ch  L\eps)}(\Om) \quad \2\Rightarrow\2  \quad (\vec{\ppU}\in \vConf^{(4\ch  L\eps)}(\Om) \textrm{ and }s=1). \label{first-implication} 
\eeqa
The second implication requires $4\ch  L\eps\leq 1$.  (The numerical constant $\ch$ appears in 
Lemma~\ref{square-root}).
\eet
\proof The first implication follows from Lemma~\ref{configurations}. The second from Lemma \ref{A-theorem}. \qed

\section{Geometric considerations}\label{geometric}
\subsection{Variational calculus on Lie groups}\label{general-discussion}
\setcounter{equation}{0}
\newcommand{\tin}{\ti{n}}

Consider a {\blue smooth} manifold $M$ of dimension $\tin$ with a vector field $X$ {\blue which is a linear map on $C^{\infty}(M)$}. Its flow is $\real\times M\ni (t,x)\mapsto \ga_t(x)\in M$
s.t. $\ga_0=\mrm{id}_M$ and $\ga_s\circ \ga_t=\ga_{s+t}$ for $s,t\in \real$. The vector field is {\blue characterized} by
\beqa
X(f)=\fr{d}{dt} f\circ \ga_t |_{t=0}.
\eeqa
Given a vector field $X$, the flow satisfies
\beqa
\fr{d}{dt} \ga_t(x)=X(\ga_t(x)). \label{flow}
\eeqa
Now the Lie derivative of a function $f: M\to \real$ w.r.t. $X$ is
\beqa
(\mcL_{X} f)(x)=\lim_{t\to 0} \fr{f(\ga_t(x)) -f(x)  }{t}: \quad M\to \real. 
\eeqa
Clearly if a function $f$ has a minimum at $x_0$ then $(\mcL_{X} f)(x_0)= X(f)(x_0)=0$.
Let us now  recall the standard setting for minimisation with constraints:
\bed\label{definition-manifold}  Let $0<k<\tin$. A subset $M_C\subset M$ is a $k$-dimensional  
equation-defined $C^1$-manifold if there is an open set $O\subset M$,
 functions $C_j\in C^1(O;\real)$, $j=1,\ldots \tin-k$, s.t. $M_C=\{\, x\in O\,|\, C_1(x)=\cdots=C_{\tin-k}(x)=0\}$   
and the differential forms on $M$
\beqa
dC_1(x), \ldots dC_{\tin-k}(x), \quad x\in M_C,
\eeqa  
are linearly independent. Apart from this, we define the space of normal forms
\beqa
N_{x}(M):=\mrm{Span} \{\,  d C_1(x), \ldots d C_{\tin-k}(x) \, \}.
\eeqa
\eed
\bet\label{minimizing-thm} Let $M_C$ as above and $F\in C^{1}(O; \real)$ so that $F|_{M_C}$ has a local minimum in $x_0\in M_C$, i.e.,
there is a neighbourhood $U\subset O$ of $x_0$ s.t.
\beqa
F(x)\geq F(x_0) \textrm{ for all } x\in U\cap M.
\eeqa
Then $d F(x_0)\in N_{x_0}(M_C)$ or, equivalently, $X(F)(x_0)=0$ for any $X \in T  M_C $. 
\eet
\proof  Let $\ga^{M_C}_t$ be the flow of a tangent vector $X$ understood as an element of  $T_{x_0} M_C$ and $\ga_t$ be the
flow of $X$ understood as an element of $T_{x_0} M$. 
Then 
\beqa
0=\fr{d}{dt} F(\ga^{M_C}_t(x_0) )|_{t=0}=\fr{d}{dt} F(\ga_t(x_0) )|_{t=0} =X(F)(x_0)=\lan dF(x_0), X_{x_0}\ran, \label{abstract-analysis}
\eeqa
where the first equality follows from the fact that $t\mapsto \ga^{M_C}_t(x_0) \in M_C$ near $t=0$ and $F|_{M_C}$ has a minimum there. 
The second equality follows from the definition of a tangent vector via an equivalence class of curves.  \qed

Now consider the special case of a Lie group $\G$ and a Lie algebra {\blue $\mathfrak{g}$}. 
For an element {\blue $\i X\in  \mathfrak{g}$ } we define a flow on $\G$ by
\beqa
\ga_t(U)=\e^{\i tX}U, \quad {\blue U\in \G}.
\eeqa
Considering~(\ref{abstract-analysis}), we  conclude that a necessary condition for $U_0$ to be  a minimum is
\beqa
(\mcL_{X}F)(U_0)= \fr{d}{dt} F(\e^{\i tX}U_0)|_{t=0}=0, \label{critical-point-equation}
\eeqa
for all {\blue $X\in T_{U_0} M_C$}. {\blue Such $X$ can be characterized by the conditions}
\beqa
(\mcL_{X}C_j)(U_0)=\fr{d}{dt}C_j(\e^{\i tX}U_0) |_{t=0}=0.  \label{vectors}
\eeqa
Clearly, by the same method one can look for minima in any open subset $\G_{\eps}\subset \G$ cf. {\magenta parts 2., 3. of Remark~\ref{main-thm-remark}}.
{\red The linear independence condition from Definition~\ref{definition-manifold} is easily checked in our case using the linearized formulation of the
constraint stated below~(\ref{linear-averaging}).}
For more on variational calculus on groups see \cite{Ch12}.

The plan of the remaining part of the paper is to first determine the tangent vectors $X\in M_C$ from (\ref{vectors}) in Subsection~\ref{tangent-subsection}
and then derive equation (\ref{critical-point-equation}) in Subsection~\ref{derivation}. This equation will be solved using the Banach {\blue contraction mapping} theorem in 
Section~\ref{existence}.

\subsection{Tangent space of the constraint manifold} \label{tangent-subsection}

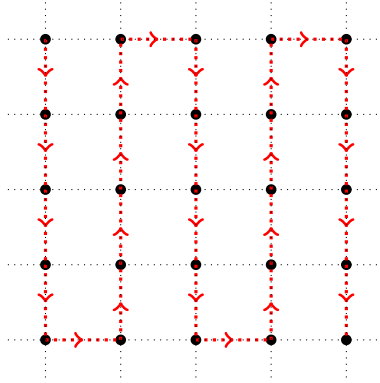
\begin{figure}[ht]\centering
\begin{tikzpicture}[scale=1]
        \begin{scope}[very thick,decoration={
                      markings,
                      mark=at position 0.5 with {\arrow{>}}}
                     ] 
         \foreach \x in {1,2,...,5}{                           
          \foreach \y in {1,2,...,5}{                       
           \node[draw,circle,inner sep=1pt,fill] at (\x,\y) {}; 
          }
         }
        \draw [dotted, very thick, postaction={decorate}, color=red] (1,5) -> (1,4);
        \draw [dotted, very thick, postaction={decorate}, color=red] (1,4) -> (1,3);
        \draw [dotted, very thick, postaction={decorate}, color=red] (1,3) -> (1,2);
        \draw [dotted, very thick, postaction={decorate}, color=red] (1,2) -> (1,1);
        \draw [dotted, very thick, postaction={decorate}, color=red] (1,1) -> (2,1);
        \draw [dotted, very thick, postaction={decorate}, color=red] (2,1) -> (2,2);
        \draw [dotted, very thick, postaction={decorate}, color=red] (2,2) -> (2,3);
        \draw [dotted, very thick, postaction={decorate}, color=red] (2,3) -> (2,4);
        \draw [dotted, very thick, postaction={decorate}, color=red] (2,4) -> (2,5);
        \draw [dotted, very thick, postaction={decorate}, color=red] (2,5) -> (3,5);
        \draw [dotted, very thick, postaction={decorate}, color=red] (3,5) -> (3,4);
        \draw [dotted, very thick, postaction={decorate}, color=red] (3,4) -> (3,3);
        \draw [dotted, very thick, postaction={decorate}, color=red] (3,3) -> (3,2);
        \draw [dotted, very thick, postaction={decorate}, color=red] (3,2) -> (3,1);
        \draw [dotted, very thick, postaction={decorate}, color=red] (3,1) -> (4,1);
        \draw [dotted, very thick, postaction={decorate}, color=red] (4,1) -> (4,2);
        \draw [dotted, very thick, postaction={decorate}, color=red] (4,2) -> (4,3);
        \draw [dotted, very thick, postaction={decorate}, color=red] (4,3) -> (4,4);
        \draw [dotted, very thick, postaction={decorate}, color=red] (4,4) -> (4,5);
        \draw [dotted, very thick, postaction={decorate}, color=red] (4,5) -> (5,5);
        \draw [dotted, very thick, postaction={decorate}, color=red] (5,5) -> (5,4);
        \draw [dotted, very thick, postaction={decorate}, color=red] (5,4) -> (5,3);
        \draw [dotted, very thick, postaction={decorate}, color=red] (5,3) -> (5,2);
        \draw [dotted, very thick, postaction={decorate}, color=red] (5,2) -> (5,1);
        \draw[step=1.0,black,thin,dotted] (0.5,0.5) grid (5.5,5.5);
        \end{scope}
\end{tikzpicture}
\caption{The spanning tree $T(y)$ of the box $B_1(y)$ is indicated in red together with its orientation. The
orientation of the bonds of the lattice is fixed by the axes of the coordinate frame.}
\label{spanning-tree}
\end{figure}

To describe the tangent space of the constraint manifold, we need some preparations. Given any $U\in \U_{\eps}(\Om)$, we 
 define a family of linear transformations on $\real^3$ by
 \beqa
 R_{\vec{A}}(x)\vec{v}= \del \ppU_0(x) \vec{v}+\vec{\ppU}(x)\times \vec{v} \label{def-R}
\eeqa
and denote their inverses, {\dgreen which exist for $A_0\neq 0$}, by $\ovR_{\vec{A}}(x)$.  Using $\vec{w}\cdot  (\vec{\ppU}(x)\times \vec{v})=-(\vec{\ppU}(x) \times \vec{w}) \cdot \vec{v}$, we easily
obtain that  
\beqa
R_{\vec{A}}^*(x)\vec{v}= \del \ppU_0(x) \vec{v}-\vec{\ppU}(x)\times \vec{v}. \label{R-T}
\eeqa
Using (\ref{SU2}) to identify $\del \ppU_0=\cos(a)$ and $\vec{\ppU}=\sin(a) \hat{n}$, we obtain
\beqa
R_{\vec{A}}\vec{v}\2=\2\cos(a) \vec{v}+ \sin(a) (\hat{n} \times \vec{v})=\cos(a) P_{\hat{n}}\vec{v}+ \cos(a)P_{\hat{n}}^{\bot}\vec{v} +   \sin(a) (\hat{n} \times P_{\hat{n}}^{\bot}\vec{v}), \\
\ovR_{\vec{A}} \vec{v}\2=\2\fr{1}{\cos(a)}  P_{\hat{n}}\vec{v}+ \cos(a)P_{\hat{n}}^{\bot}\vec{v} -  \sin(a) (\hat{n} \times P_{\hat{n}}^{\bot}\vec{v}), \label{R-inverse}
\eeqa
where $P_{\hat{n}}$ is the orthogonal  projection on $\hat{n}$. Thus  $R_{\vec{A}}(x)$ is a sum of a rotation  in the plane orthogonal to $\hat{n}$
and a  scaling transformation in the direction of $\hat{n}$. We will often write $R:=R_{\vec{A}}$ for brevity.   Now we follow the procedure from Subsection~\ref{general-discussion}:
\bep Suppose that $U\in \U_{\eps}(\Om)$ satisfies the constraint~(\ref{constraint}). Then  {\red for any} vector ${\dgreen \i}X\in \mathfrak{g}_0^{\oplus n^2}$ 
{\red the property} $\mcL_{X}\mcc'(\pU)(y)=0$ {\red is equivalent to}
\beqa
\sum_{x\in \B(y)} R(x)\vec{X}(x)=0,\quad y\in \Om_1. \label{tangent-vector-condition}
\eeqa
\eep
\proof Using (\ref{vectors}), (\ref{intermediate-constraint}), we compute
\beqa
\mcL_{X}\mcc'(\pU)(y)= \i \sum_{x\in \B(y)}  (X(x)\pU(x)+\pU(x)^*X(x) )=0. \label{tangent-equation}
\eeqa
Let us recall that for $SU(2)$ we have $X(x)=X_j(x)\si_j\in \mathfrak{su}(2)$, $\pU(x)=\ppU_0(x)+\i \ppU_j\si_j\in SU(2)$, where the coefficients $X_{j}, \ppU_0,\ppU_{j}$ are real. 
By Theorem \ref{configurations-theorem}, we could set $\del=1$ in (\ref{Pauli-matrices-decomposition}), because we are in $\Conf_{\eps}(\Om)$ and we are differentiating at a point $\pU$ of the constraint manifold.  Thus equation (\ref{tangent-equation}) gives, {\blue omitting the dependence on $x$},
\beqa
0\2=\2\sum_{x\in \B(y)} X_j (\si_j (\ppU_0+\i\ppU_{k}\si_k)   + \mrm{h.c.} )\non\\
\2=\2 \sum_{x\in \B(y)} X_j( \ppU_0\si_j+\i\ppU_{k}(\de_{j,k}+i\eps_{j,k,\ell}\si_{\ell})   + \mrm{h.c.} )\non\\
\2=\2  \sum_{x\in \B(y)} X_j( \ppU_0\si_j+\i\ppU_{j}- \ppU_{k}\eps_{j,k,\ell}\si_{\ell})   + \mrm{h.c.} )\non\\
\2=\2  \sum_{x\in \B(y)} 2( \ppU_0  X_j- \ppU_{k} X_{j'}\eps_{j',k,j}) \si_j.
\eeqa
 Thus we get that   for any $y$
\beqa
\sum_{x\in \B(y)} (\ppU_0(x) \vec{X}(x)+\vec{\ppU}(x)\times \vec{X}(x))=0
\eeqa
which concludes the proof. \qed\\
Let us now describe more explicitly families of vectors $\{{\magenta \vec{X}(x) }\}_{x\in \Om}$ satisfying (\ref{tangent-vector-condition}).
We note that for each $y$ we obtain an independent condition which depends only on  ${\magenta \vec{X}(x)},\vA(x)$ for $x\in \B(y)$.
Thus it suffices to solve (\ref{tangent-vector-condition}) in one block: We relabel the points in this block
as $x_1,\ldots, x_{L^2}$ and ask for the kernel of the matrix:
\beqa
[\mcc]=\begin{bmatrix}  R(x_1) & R(x_2) & \ldots & R(x_{L^2}) 
                                                                                   \end{bmatrix},
\eeqa
where each $R(x_j)$ symbolizes a $3\times 3$ matrix, see (\ref{def-R}).
This kernel  coincides with  the range of the following matrix, as we show in Lemma~\ref{kernel-lemma} below:
\beqa
[\mcd]= \begin{bmatrix}   \ovR(x_1) &          0         &           0        &  0   & 0     \\
                                     -\ovR(x_2) &  \ovR(x_2) &           0        &   0  & 0        \\
                                             0          &  -\ovR(x_3) &   \ddots    &  0    & 0      \\
                                             0          &          0          &     0 &  \ovR(x_{L^2-2})  & 0     \\
                                             0           &        0            &           0       &   -\ovR(x_{L^2-1})  &   \ovR(x_{L^2-1})      \\
                                             0          &         0            &          0        &   0                   &    -\ovR(x_{L^2})               \\
                                                                                         \end{bmatrix}.
\eeqa
{\dgreen Recall that for  $U\in \Conf_{\eps}(\Om)$, $\mcC(U)=V$, we have $|\vec{A}(x)|^2\leq 4c_{\h}L\eps$ by Theorem~\ref{configurations-theorem}. Thus $A_0(x)\neq 0$ and $\ovR(x)$ above exist. Now the tangent space of the constraint manifold can be described as follows: }
\bet\label{tangent-space-thm} Fix $y\in \Om_1$ and consider a spanning tree $T(y)$ of the box $B_{\magenta 1}(y)$ as in Figure~\ref{spanning-tree}. Then for every bond $c\in T(y)$, whose orientation
equals the tree's orientation, there is a three-dimensional space  of  vectors tangent to the constraint manifold
\beqa
{\magenta \vec{X}}_c=\begin{bmatrix}  0 \\ \vdots \\ 0 \\ \ovR(c_-)\vec{v}_c \\ -\ovR(c_+)\vec{v}_c  \\ 0 \\ \vdots \\ 0  \end{bmatrix}, \quad \vec{v}_c\in \real^3. 
\label{constraint-tangent-vectors}
\eeqa
If the two orientations are opposite, then $c_-$, $c_+$ should be exchanged in (\ref{constraint-tangent-vectors}).
\eet

\newcommand{\tim}{\ti{m}}

\bel\label{kernel-lemma} Suppose $R_i$, $i=1,\ldots, \tim$ are invertible transformations on $\real^3$. Let $R:=(R_1,\ldots, R_{\tim})$ be a transformation from $\real^{3\tim}$ to $\real^3$.  
Define mappings
\beqa
D_i=(0,\ldots, \ovR_i, -\ovR_{i+1}, 0,\ldots, 0)^{T}, \quad i=1, \ldots, \tim-1,
\eeqa
from $\real^3$ to $\real^{3\tim}$, where $\ovR_i$ denotes the inverse of $R_i$. Then the kernel of $R$ equals the range of $ D:=(D_1, \ldots, D_{\tim-1})$.
\eel
\proof We note that if $(\vec{v}_1,\ldots, \vec{v}_{\tim})\in \real^{3\tim}$ is in the kernel of $R$ then
\beqa
-\vec{v}_1=\ovR_1R_2\vec{v}_2+\cdots +\ovR_1R_{\tim}\vec{v}_{\tim}.
\eeqa
Thus we can parametrize this kernel with $3(\tim-1)$ parameters. Hence, the dimension of the kernel is maximally $3(\tim-1)$. To show that 
it is exactly $3(\tim-1)$, one notes that $RD=0$,  $\mrm{Ran}D_i \cap \mrm{Ran} D_j=\{0\} $ {\magenta for $i\neq j$} and the dimension of {\magenta each} $\mrm{Ran}D_i$ is $3$. 
 The last claim  follows from the linear independence of $D_i (R_ie_j)$, where $e_j$ are unit vectors in $\real^3$. \qed  

\subsection{Derivation of the critical point equation}\label{derivation}

\renewcommand{\W}{W}

Before we derive the critical point equation, we need some definitions.
\bed\label{De-0-def} We define a map $\pa: \mcL^2(\Om)\to \mcL^2(\Om')$  by
\beqa
(\pa f)(b):=f(b_-)-f(b_+).
\eeqa
The adjoint map $\pa^*:  \mcL^2(\Om')\to  \mcL^2(\Om)$ is defined by the relation
\beqa
\lan \pa f, g\ran_{\Om'}=\lan f, \pa^* g\ran_{\Om} \label{adjoint}
\eeqa
for $f\in \mcL^2(\Om)$, $g\in \mcL^2(\Om')$. It is given explicitly by
\beqa
(\pa^* f)(x)=\sum_{b\in \Om', b\ni x }\si_{b}(x) f(b), \label{lap}
\eeqa
where $\si_{b}(x)=+1/-1$ for bonds incoming/outgoing from x, according to the orientation of $\Om$. 
Finally, $\De_{\Om}:=-\pa^*\pa$ coincides with  the lattice Laplacian on $\Om$ with  
Neumann boundary conditions and thus satisfies
\beqa
\lan f,(-\De_{\Om})f\ran=\sum_{b\in \Om'} \lan \pa f(b),\pa f(b)\ran,  \label{D-N-bracketing}
\eeqa
cf. e.g. \cite[Lemma 2.5] {DST23}. {\magenta The above definitions extend naturally to vector-valued functions.}
\eed
\nin Now we fix a configuration $U\in \Conf_{\eps}(\Om)$ satisfying the constraint (\ref{constraint}) with some $V\in \Conf_{\eps_1}(\Om_1)$.
{\dgreen We recall that $\pa U(b)=U'(b_-) \pa V(y_b) U'(b_+)^*$, $\pa V(y_b):=V(y_{b_-})V(y_{b_+})^*$ and $y_{b_{\pm}}$ was defined below (\ref{box}).
In the following definition we set $\W(b):=\pa U(b)$ for brevity, decompose it into the Pauli matrices and extract the leading part of the resulting vector $\vec{W}(b)$.}
\bed\label{W-def} For any $b\in \Om'$ we define   $\W(b):=\pU(b_-) \pa V(y_b) \pU(b_+)^*= \del_{W}\W_0(b)+\i \vec{\W}(b) \cdot \vec{\si}$ as in formula~(\ref{Pauli-matrices-decomposition}). We also define a {\dgreen remainder} $\vec{r}_{\vec{A}}(b)$ by the decomposition
\beqa
 \vec{\W}(b)=:\pa\vec{A}(b)+\vec{r}_{\vec{A}} (b), \label{W-def-form}
\eeqa
We will often write $\vec{r}:=\vec{r}_{\vec{A}}$ for brevity. For an explicit formula for $\vec{r}_{\vec{A}}(b)$ see (\ref{terms}) {\magenta below}.
\eed
Let us now move towards the derivation of the critical point equation. The key step is the 
following proposition. It identifies quantities which are constant along the spanning trees of the
respective boxes, see  Figure~\ref{spanning-tree}.

\bep\label{conservation-prop}  Fix $y\in \Om_1$ and consider a spanning tree $T(y)$ of the box $B(y)$ as in Figure~\ref{spanning-tree}. Then, {\green $\vec{A}$ is a critical point } of the action~(\ref{action-one-x}) in $\Conf_{\eps}(\Om)$, with the constraint~(\ref{constraint}), {\green if and only if} we have   
\beqa
\ovR(c_-)^* \pa^* \vec{\W}(c_-) =\ovR(c_+)^* \pa^* \vec{\W}(c_+)
\eeqa
for any bond $c\in T(y)$.
\eep
\proof We proceed as explained in {\magenta Sub}section~\ref{general-discussion}.  We consider the functional:
\beqa
\pmca(\pU)= \sum_{b\in \Om'} \mrm{Re}\mrm{Tr}(1-   \pU(b_-) \pa V(y_b) \pU(b_+)^*   ), \quad \pa V(y_b):=V(y_{b_-})V(y_{b_+})^*,
\eeqa
and compute the Lie derivative
\beqa
\mcL_{X}\pmca(\pU)\2=\2 - \fr{d}{dt} \sum_{b\in \Om'} \mrm{Re}\mrm{Tr}(  \e^{\i tX(b_-)}\pU(b_-) \pa V(y_b) \pU(b_+)^*\e^{-\i t X( b_+)}   )|_{t=0}\non\\
\2=\2  \sum_{b\in \Om'} \mrm{Im}\mrm{Tr}(   X(b_-) \pU(b_-) \pa V(y_b) \pU(b_+)^* -   \pU(b_-) \pa V( y_b) \pU(b_+)^*X(b_+)  )\non\\
\2=\2  \sum_{b\in \Om'} \mrm{Im}\mrm{Tr}(   {\dgreen \pa X(b)} \pU(b_-) \pa V(y_b) \pU(b_+)^* ). \label{Lie-derivative-A}
\eeqa
Thus, referring to Definition~\ref{W-def}, we  have
\beqa
\mcL_{X}\pmca(\pU)=  {\dgreen 2}\sum_{b\in \Om'} (  {\dgreen \pa X_k(b)}) \W_k(b) 
={\dgreen 2} \sum_{x\in \Om} X_k(x) \pa^*\W_k(b), \label{integration-by-parts}
\eeqa
where the first equality follows from  
\beqa
(\vec{X} \cdot \vec{\si}) (\vec{Y} \cdot \vec{\si} )= (\vec{X}\cdot \vec{Y}) +\i (\vec{X}\times  \vec{Y}  )\cdot \vec{\si} \label{Pauli-multiplication}
\eeqa
and the fact that the Pauli matrices are traceless.  The second equality follows from~(\ref{adjoint}).   

Now we come back to the tangent vectors ${\magenta \vec{X}}_c$ of~(\ref{constraint-tangent-vectors}) and consider a bond $c$ 
on a spanning tree $T(y)$.  Formula~(\ref{integration-by-parts}) gives
\beqa
\mcL_{X_{c}}\pmca(\pU)=-{\dgreen 2}\vec{v}_c\cdot \big( \ovR(c_-)^* \pa^* \vec{\W}(c_-) -\ovR(c_+)^*  \pa^* \vec{\W}(c_+)    \big),
\eeqa
possibly with $c_-, c_+$ interchanged depending on the orientation, cf. Theorem~\ref{tangent-space-thm}. As the Lie derivatives 
vanish at a critical point and $\vec{v}_c\in \real^3$ are arbitrary, this concludes the proof. \qed

Next, we will use the conservation property from Proposition~\ref{conservation-prop}  to derive the 
critical point equation. To this end, we  recall the linear averaging operator $Q$
stated in (\ref{linear-averaging}) and define the lattice Green function
\beqa
\Gn:=(-\De_{\Om}+ Q^*Q)^{-1}. \label{Green}
\eeqa
The existence of the inverse defining $\Gn$ is a standard fact, see Lemma~\ref{inverse-lemma} below. 
In the statement of the following theorem we also use that   $Q\Gn R^*Q^*$ is an invertible map on  $\mcL^2(\Om_1;\real^3)$, which is
shown in Lemma~\ref{Q-G-R-Q-lemma}.
\bet\label{critical-point-thm} At a critical point  of the action~(\ref{action-one-x})  in $\Conf_{\eps}(\Om)$, with the constraint~(\ref{constraint}),  the following equation holds
 \beqa
\vec{\ppU}= \Gn R^*_{\vec{A}} Q^* [Q\Gn R^*_{\vec{A}}Q^*]^{-1}  Q \Gn \pa^* \vec{r}_{\vec{A}}  -  \Gn \pa^* \vec{r}_{\vec{A}}. \label{fixed-point-eq-them}
\eeqa
{\green Conversely, any solution of this equation is a critical point.}
\eet
\proof By Proposition~\ref{conservation-prop}, we can write for some block-constant family of vectors $\vec{C}(y_x)$
\beqa
 \ovR(x)^*(-\pa^*\vec{\W})(x)\2=\2\vec{C}(y_x),  \label{cons-law-proof} \\
 (\De_{\Om} \vec{\ppU})(x)-(\pa^*\vec{r})(x)\2=\2 R(x)^* \vec{C}(y_x),   \label{second-equation-critical}
\eeqa
where we used (\ref{W-def-form}).
As we checked in Section~\ref{simplification}, the constraint has the form $Q(\vec{A})=0$. Thus we can replace  $\De_{\Om}$  with $\De_{\Om}-Q^*Q$ in (\ref{second-equation-critical}), which gives, in  terms of the Green function (\ref{Green}),
\beqa
\vec{\ppU}\2=\2 - \Gn R^* \vec{C} - \Gn \pa^*\vec{r}. \label{intermediate-equation}
\eeqa
Now by applying $Q$ to both sides, we obtain 
\beqa
 Q\Gn R^* \vec{C} + Q \Gn \pa^* \vec{r}=0.
\eeqa 
Since $Q^*Q$ is the projection on block-constant vectors,  cf.  (\ref{Q-properties}), we have $\vec{C}=Q^*Q\vec{C}$.
Since the map $Q \Gn R^* Q^*$ is invertible, cf. Lemma~\ref{Q-G-R-Q-lemma}, we have
\beqa
 Q\vec{C}= -[Q \Gn R^* Q^*]^{-1}  Q \Gn \pa^* \vec{r}.
\eeqa
Substituting this to (\ref{intermediate-equation}), {\blue and using $C=Q^*QC$}, we have
\beqa
\vec{\ppU}=  \Gn R^* Q^* [Q\Gn R^*Q^*]^{-1}  Q \Gn \pa^* \vec{r}  -  \Gn \pa^* \vec{r}. \label{fixed-point-eq}
\eeqa
This concludes the proof {\green of the first part of the theorem.   The last statement is shown by reversing the steps and recalling that Proposition~\ref{conservation-prop}
is an if and only if statement.} \qed

\subsection{Structure of the expressions $\protect\overrightarrow{W}$} \label{conf-W}
\newcommand{\vC}{\vec{C}}
\newcommand{\WZ}{Z}
\newcommand{\Uuv}{\vec{\pU}(b_-)}
\newcommand{\Vvv}{ \pa\vec{V}( y_b)} 
\newcommand{\WZzv}{\vec{\pU}(b_+)}
\newcommand{\Uun}{\pU(b_-)}
\newcommand{\Vvn}{\pa V(y_b)} 
\newcommand{\WZzn}{\pU(b_+)}

In this  {\magenta  subsection} we will derive a formula of the form:
\beqa
 \vec{\W}(b)=\vec{\ppU}(b_-) - \vec{\ppU}(b_+)  +\vec{r}(b)
\eeqa
and state a formula for the {\dgreen remainder} $\vec{r}=\vec{r}(b)=\vec{r}(b_-,b_+)$. Differently than in the main part of the paper,  we will denote a vector in $\real^3$ corresponding to $U,V,Z \in SU(2)$ by $\vec{U}$,  $\vec{V}$, $\vec{Z}$, respectively.   
In the representation  (\ref{Pauli-matrices-decomposition}) we have $U= U^0+\i \vec{U}\cdot  \vec{\si}$, where $U^0:=\del_U U_0$. The multiplication table for  a product  $UV$  is:
\beqa
\2 \2(UV)^0=U^0V^0-\vec{U} \cdot \vec{V}, \label{multi-zero}\\
\2 \2\vec{(UV)}=U^0 \vec{V}+  V^0 \vec{U}  - (\vec{U}\times \vec{V}), \label{multi-one}
\eeqa
where we used (\ref{Pauli-multiplication}). 
 Let us write for brevity $\ppU_{\pm}:=\ppU(b_{\pm})$, $\ppV:={\cyan B(b):= \pa V( y_b)}$  and $\de(M):=1-M$ for any {\cyan $M\in \real$}. Then, by a straightforward
 application of (\ref{multi-zero}), (\ref{multi-one}), {\cyan postponed to Lemma~\ref{simple-computations}},  we obtain  
\beqa
\vec{r}(b)\2=\2 - \de(\ppU^0_{-}\ppV^0) \vec{\ppU}_+  + \de(\ppU^0_{+} \ppV^0) \vec{\ppU}_-   + \ppU^0_{+} \ppU^0_{-} \vec{\ppV}\non\\        
                         \2 \2- \ppU^0_{+} (\vec{\ppU}_-\times \vec{\ppV}) + \ppU^0_{-} (\vec{\ppV} \times  \vec{\ppU}_+) +\ppV^0 (\vec{\ppU}_- \times  \vec{\ppU}_+)\non\\
                         \2 \2 + \vec{\ppU}_{+} (\vec{\ppU}_-\cdot \vec{\ppV}) - \vec{\ppU}_{-} (\vec{\ppV} \cdot  \vec{\ppU}_+) +\vec{\ppV} (\vec{\ppU}_- \cdot  \vec{\ppU}_+).                         \label{terms}
\eeqa
From this representation and Theorem~\ref{configurations-theorem} it is clear that for $U\in \Conf_{\eps}(\Om)$ satisfying the constraint~(\ref{constraint}) and $V\in \Conf_{\eps_1}(\Om_1)$, for $\eps, \eps_1$ sufficiently small (uniformly in $n$)
we can lower all the $0$-superscripts, since the corresponding signs $\del$ equal $1$. Therefore, we have 
 \beqa
\de(\ppU_{0,\pm}\ppV_0)=1-\sqrt{1-(\vec{\ppU}_{0,\pm})^2} \sqrt{1-\vec{\ppV}^2}=   \fr{\vec{\ppV}^2+(\vec{\ppU}_{0,\pm})^2 -\vec{\ppV}^2
(\vec{\ppU}_{0,\pm})^2  }{1+\ppU_{0,\pm}\ppV_0}. \label{delta-form}
\eeqa

\section{Analytic considerations}\label{existence}
\setcounter{equation}{0}

In this section we prove the existence and uniqueness of solutions of the critical point equation (\ref{fixed-point-eq-them}).  
In Subsection \ref{main-line} we provide the main line of the argument based on the Banach {\red contraction mapping} theorem.
One important ingredient here are $\mcL^{\infty}$-bounds on the inverse of $Q\Gn R^*Q^*$, which are established in
Subsections \ref{strict-positivity} - \ref{L-infty}  and in Appendix \ref{random-walk-appendix}. Another ingredient are
 $\mcL^{\infty}$-bounds on the {\dgreen remainder} $\vec{r}$, which are derived in  Subsection~\ref{rest-term-bounds}.

\subsection{Existence of solutions of the critical point equation} \label{main-line}
\newcommand{\XX}{\mrm{X}}
\newcommand{\xx}{\mrm{x}}
\newcommand{\TT}{\mrm{T}}

\bet \textbf{(Banach contraction mapping theorem)} Let $\XX$ be a non-empty complete space with metric $\d$.  Suppose that $\TT: \XX\to \XX$ 
is a contraction, i.e.,
\beqa
\d(\TT\xx, \TT\xx')\leq q \d(\xx,\xx')
\eeqa
for some $q\in [0,1)$. Then there is a unique $\xx^*\in \XX$ s.t. $\TT \xx^*=\xx^*$. Given the sequence $\xx_n=\TT\xx_{n-1}$ we have $\xx^*=\mrm{lim}_{n\to \infty} \TT\xx_n$.
\eet
\nin\textbf{The space $\XX_{\eps}$.} Suppose $\eps,\eps_1$ are sufficiently small, uniformly in $n$.
Then, for any $U\in  \Conf_{\eps}(\Om)$, satisfying the constraint (\ref{constraint}) with $V\in \Conf_{\eps_1}(\Om_1)$,
we have $\vec{A}\in \vConf^{\dgreen 4\ch  L\eps }(\Om)$ by {\dgreen (\ref{configurations-theorem}) }. Furthermore, the relation
(\ref{Pauli-matrices-decomposition}) is invertible in this case, {\dgreen if we specify to} $\del=1$, cf. Subsection \ref{simplification}. Thus 
 we can define the map, denoted  by the square bracket,
\beqa
[\vec{\ppU}(x)]=\pU(x). \label{bracket-relation}
\eeqa 
We note that the inverse is simple: {\blue $\h\mrm{Tr}(U'(x)\cdot \vec{\si})=  \vec{\ppU}(x)$.} Now we define the following notion of distance from zero
\beqa
\d_0( {\blue \vec{\ppU}},  0)=\sup_{b\in \Om}\| [\vec{\ppU}](b_-) \pa V({\magenta y_b})[\vec{\ppU}](b_+)^* -  {\blue 1} \|=\sup_{b\in \Om}\| (\pa U)( b)-{\blue 1}\|,
\eeqa
which is dictated by the small field condition (\ref{small-field-condition}).  (Clearly, it may vanish for $ \vec{\ppU} \neq 0$ as it probes only differences of fields at neighbouring points). Now we define the space
\beqa
\XX_{\eps}:=\bigg\{\, \vec{\ppU}   \in  \vConf^{{\red 4c_{1/2}L^2\eps}}(\Om) \,|\,   Q( \vec{A} )=0, \,\,   \d_0( \vec{\ppU},  0)\leq \eps   \bigg\} \label{def-X_f}
\eeqa
which, {\red considering (\ref{first-implication-zero}), (\ref{first-implication})}, consists precisely of configurations satisfying the small field condition and the constraint. 
{\dgreen While it is clear that any configuration from Theorem~\ref{main-theorem} is contained in $\XX_{\eps}$, let us comment on the opposite inclusion: Any configuration 
$\vec{A}\in \vConf^{{\red 4c_{1/2}L^2\eps}}(\Om)$ gives one element  $U'=[\vec{A}]$, since we specified $s=1$ in  (\ref{bracket-relation}). 
This provides us with $U=U'V\in \Conf(\Om)$.
 Then the condition $\d_0( \vec{\ppU},  0)\leq \eps$ is only satisfied  if  $U\in \Conf_{\eps}(\Om)$. Now by reversing the steps in Subsection~\ref{simplification} we check that this configuration satisfies the constraint  $\mcC(U)=V$.
This equality of configuration spaces} is important, for the following 
reason: If we made the set of configurations smaller than in Theorem \ref{main-theorem} we could not conclude the uniqueness of solutions in
the original set. If we made it larger we could not conclude the existence of solutions in the original set.

\vspace{0.2cm}

\nin\textbf{The metric  $\d$.} We equip $X_{\eps}$ with the following metric:
\beqa
\d( \vec{\ppU}_1 ,   \vec{\ppU}_2  ):=\| \vec{\ppU}_1 -\vec{\ppU}_2\|_{\infty;\Om}=\sup_{x\in \Om }  |\vec{\ppU}_1(x) -\vec{\ppU}_2(x)|. 
\eeqa
Since  $\XX_{\eps}$ is a  closed subset of $\real^{3n^2}$, the completeness of $\XX_{\eps}$ in the metric $\d$ is clear. 

\vspace{0.2cm}

\nin\textbf{The map $\TT$}. Now the map $\TT$ is dictated by equation~(\ref{fixed-point-eq-them})
\beqa
\TT( \vec{\ppU} ):= (\Gn R^*_{ \vec{\ppU}} Q^*[Q\Gn R^*_{ \vec{\ppU}}Q^*]^{-1}  Q  -1) \Gn \pa^* \vec{r}_{\vec{\ppU}}.   \label{T-def-0}
\eeqa
We have to check that $\TT$ maps $\XX_{\eps}$ into itself and satisfies
\beqa
 \|  \TT( \vec{\ppU}_1 ) - \TT( \vec{\ppU}_2 )\|_{\infty;\Om} \leq q   \|  \vec{\ppU}_1  -  \vec{\ppU}_2\|_{\infty;\Om}
\eeqa
for some $0\leq q<1$. We start by checking that $\TT$ maps $\XX_{\eps}$ into itself for sufficiently small $\eps$, $\eps_1$.

\bep\label{preserving-space} For $\eps_1\leq \eps^2$, sufficiently small (uniformly in $n$),    $\TT$ maps $\XX_{\eps}$ into itself. 
\eep
\proof  It is manifest from  definition (\ref{T-def-0}) that $\TT$ preserves the constraint, i.e., $Q(\TT( \vec{\ppU} ))=0$.
Next, we observe that, by Lemmas~\ref{Q-G-R-Q-lemma}, \ref{r-lemma-one},
\beqa
\| \TT(\vec{\ppU} )\|_{\infty{\magenta ;}\Om} \leq   C \|\pa^* \vec{r}\|_{\infty {\magenta ;} \Om}\leq 24C (\eps^2+\eps_1).  
\label{T-invariance-bound}
\eeqa
Thus for $\eps$ sufficiently small  relation (\ref{bracket-relation}) is well defined for $\TT(\vec{A})$.  
Given this, we  have  to show that, if $\d_0(\vec{\ppU},0)\leq \eps$, then also
\beqa
\d_0(  \TT(\vec{\ppU}),  0)=\sup_{b\in \Om'}\| [\TT( \vec{\ppU}) ](b_-) \pa V( y_b)[\TT(\vec{\ppU})](b_+)^*-1\|\leq \eps.
\eeqa
We have 
\beqa
\| [\TT(\vec{\ppU})](b_-) \pa V(y_b)[\TT(\vec{\ppU}) ](b_+)^*-1\|\leq \eps_1+ \| [\TT(\vec{\ppU})](b_-) [\TT(\vec{\ppU})](b_+)^*-1\|.
\eeqa
Furthermore, by Lemma~\ref{Pauli-estimates},
\beqa
\| [\TT(\vec{\ppU})](b_-) [\TT(\vec{\ppU})](b_+)^*-1\|\!&=&\!\|[\TT(\vec{\ppU})](b_-) - [\TT(\vec{\ppU})](b_+)\|\non\\
    \2\leq\2\sqrt{6}(| \TT(\vec{\ppU})(b_-)|+|\TT(\vec{\ppU})(b_+)|) \,\,\,\, \non\\
\2\leq\2 {\dgreen C'(\eps^2+\eps_1),} \label{preserving-X}
\eeqa
{\dgreen where in the last step we used (\ref{T-invariance-bound}).  Now using that we have $\eps^2$ (and not just $\eps$) on the r.h.s. of (\ref{preserving-X}), we conclude the proof.}  \qed

\bep\label{contraction} Under the assumptions of Proposition~\ref{preserving-space} and for $\eps, \eps_1$, sufficiently small (uniformly in $n$),  the map $\TT: \XX_{\eps} \to \XX_{\eps}$ is a contraction.
\eep
\proof We rewrite (\ref{T-def-0}) as follows
\beqa
\TT( \vec{\ppU} )= (M_{ \vec{\ppU} } -1)\Gn \pa^* \vr_{\vec{\ppU}},\quad M_{ \vec{\ppU} }:=\Gn R^*_{ \vec{\ppU}}Q^* [Q\Gn R^*_{ \vec{\ppU}}Q^*]^{-1}  Q.
\eeqa
We will divide the problem into two parts
\beqa
 \TT( \vec{\ppU}_1 ) - \TT( \vec{\ppU}_2) \2 =\2 (M_{ \vec{\ppU}_1 }-1) \big(\Gn \pa^* \vr_{\vec{\ppU}_1 } - \Gn \pa^* \vr_{\vec{\ppU}_2 }) \label{shift-one} \\
 \2 \2+  (  M_{ \vec{\ppU}_1 } -M_{ \vec{\ppU}_2 } )  \Gn \pa^* \vr_{\vec{\ppU}_2 }.  \label{shift-two}
 \eeqa
 We consider first the shift~(\ref{shift-two}).  We have, by Lemma~\ref{r-lemma-one},
 \beqa
 \| \Gn \pa^* \vr_{\vec{\ppU} }\|_{\infty;\Om}\leq \|\Gn\|_{\infty, \infty; \Om}  24{\red C}(\eps^2+\eps_1). \label{eps-square}
 \eeqa
Next, we write
 \beqa
   M_{ \vec{\ppU}_2 } -M_{ \vec{\ppU}_1 } \2 = \2 \Gn R^*_{ \vec{\ppU}_2} Q^*  ([Q\Gn R^*_{ \vec{\ppU}_2 }Q^*]^{-1} -  [Q\Gn R^*_{ \vec{\ppU}_1 }Q^*]^{-1} )Q \non\\
                                                                  \2   \2+ \Gn (R^*_{ \vec{\ppU}_2} - R^*_{ \vec{\ppU}_1 })  Q^*     [Q\Gn R^*_{ \vec{\ppU}_1 }Q^*]^{-1} Q.
                                                             \eeqa
Thus, by Lemmas~\ref{R-lemma}, \ref{R-lemma-two}, we have for $D_{\vA}:= Q\Gn R^*_{\vec{\ppU}}Q^*$ 
\beqa
\|M_{ \vec{\ppU}_2 } -M_{ \vec{\ppU}_1 }\|_{\infty, \infty;\Om} \2\leq\2 2\|\Gn\|_{\infty,\infty; \Om}  2\|D_{\vA_1}^{-1} \|_{\infty,\infty; \Om_1}\, \|D_{\vA_2}^{-1} \|_{\infty,\infty;\Om_1}   \| \vec{\ppU}_1-\vec{\ppU}_2 \|_{\infty; \Om_1}\non\\
                                                                                \2 \2+\|\Gn\|_{\infty, \infty; \Om} \|D_{\vA_1}^{-1}\|_{\infty,\infty;\Om_1} 
                                                                                 \| \vec{\ppU}_1-\vec{\ppU}_2\|_{\infty;\Om}.
\eeqa 
Finally, making use of Lemmas~\ref{r-lemma-two}, \ref{R-lemma-two} we can estimate (\ref{shift-one}) as follows
\beqa
\|  (\ref{shift-one}) \|_{\infty;\Om} \2\leq\2 (1+\| M_{ \vec{\ppU}_1 }\|_{\infty,\infty;\Om }) \|\Gn\|_{\infty,\infty;\Om} \| \pa^* \vr_{\vec{\ppU}_1 } - \pa^* \vr_{\vec{\ppU}_2 }\|_{\infty;\Om} \\
\2\leq \2 (1+\| M_{ \vec{\ppU}_1 }\|_{\infty,\infty; \Om}) \|\Gn\|_{\infty,\infty; \Om} 96 {\red C} ({\dgreen \eps}+\eps_1) \| \vec{\ppU}_1-\vec{\ppU}_2\|_{\infty;\Om}.
\eeqa 
 This concludes the proof, {\dgreen considering that all the  $\| \ldots \|_{\infty,\infty; \Om}$-norms above are bounded uniformly in $n$ by  Lemmas~\ref{Q-G-R-Q-lemma}, \ref{infty-bounds}, \ref{R-lemma}.}  (We remark that Lemmas~\ref{r-lemma-two}, \ref{R-lemma}, \ref{R-lemma-two}, {\magenta which we used in the proof}, require the constraint as they use implication~(\ref{first-implication-zero})).  \qed

\newcommand{\op}{\mrm{op}}

\subsection{Strict positivity of $G^{-1}, G$ and $QG Q^*$} \label{strict-positivity}

In this subsection we start working towards the $\mcL^{\infty}$-bounds on $Q\Gn Q^*$, $(Q\Gn Q^*)^{-1}$ which we used in the proofs of
Propositions~\ref{preserving-space}, \ref{contraction}. {\magenta These bounds follow from the exponential decay of the integral kernels of these
operators. This latter property is first shown for the counterparts of these operators on $\mcL^2(\mathbb{Z}^2)$ by the method
of random walk expansions  (see Appendix~\ref{random-walk-appendix}, Proposition~\ref{first-corollary}). In Subsection~\ref{method-of-images} 
we translate this property on the finite lattice using the method of images.}

We will denote by $-\De$ the Laplacian
with free boundary conditions on $\mcL^2(\mathbb{Z}^2)$, by $Q$ the averaging operator (which is a natural extension of its finite lattice counterpart (\ref{linear-averaging})) and write
\beqa
G:=(-\De+Q^*Q)^{-1} \label{G-def}
\eeqa
for the lattice Green function on $\mcL^2(\mathbb{Z}^2)$ and note that $QGQ^*$ is an operator on  $\mcL^2(L\mathbb{Z}^2)$. To substantiate definition~(\ref{G-def}) 
and also to check one assumption of Proposition~\ref{first-corollary} below,
we study the strict positivity of the relevant operators. The following lemma is standard, cf.  \cite[Lemma~29]{Di13}, \cite[Lemma~2.10]{DST23} for similar considerations.
\bel \label{inverse-lemma}   The following {\blue points hold}: 
\begin{enumerate}
  \item Let $y\in \Om_1$ and denote by  $\De_{\B(y)}$ the Laplacian on $\B(y)$ with Neumann boundary conditions, {\dgreen cf.~\cite[Section 2]{DST23}.}
  Then,   as operators on $\L^{2}(\B(y))$,
\beqa 
-\Delta_{\B(y)}  + Q^{*}Q \geq C. \label{unit-box}
\eeqa

\item  The following inequalities hold as operators on $\mcL^{2}(\Om)$
\beqa
-\Delta_{\Omega} + Q^{*}Q \2\geq\2  C,   \label{many-boxes} \\
(-\Delta_{\Omega} + Q^{*}Q)^{-1} \2\geq\2 c. \label{Green-estimate} 
\eeqa 

\item  The following inequality holds as operators on $\mcL^{2}(\Om_1)$
\beqa
Q (-\Delta_{\Omega} + Q^{*}Q)^{-1}Q^*\geq c. \label{sandwiched-Green}
\eeqa

\item The bounds (\ref{many-boxes}), (\ref{Green-estimate}), (\ref{sandwiched-Green}) also hold for the
corresponding operators on $\mcL^2(\mathbb{Z}^2)$, resp. $\mcL^2(L\mathbb{Z}^2).$

\end{enumerate}
Here $C,c>0$ are independent  of $n$, but $C$ may depend on $L$.
\eel
\proof  As for 1., if $f \in \L^{2}(\B(y))$ is constant then $-\Delta_{\B(y)} f = 0$ and
\beqa
\langle f,  Q^{*}Q f \rangle = \|f\|^{2}_{2; \B(y)}. 
\eeqa
{\blue On the other hand, $-\Delta_{\B(y)}$ is strictly positive on the orthogonal complement of the subspace of constant functions.
Specifically, its lowest eigenvalue is
given by (see e.g.  \cite[Lemma~2.3]{DST23}) 
\beqa
(-\la^{(1)})= 4 \sin^2\bigg( \fr{\pi }{2L} \bigg). 
\eeqa
}We have $(-\la^{(1)})\geq C>0$. Therefore,
\beqa
\langle f, (- \Delta_{\B(y)}) f\rangle \geq C \|f \|^{2}_{2; \B(y)}
\eeqa
and since $Q^{*}Q $ is also positive this proves (\ref{unit-box}).

Regarding 2.,  let $f \in \L^{2}(\Omega)$  and set $f_{\B(y)}:=f|_{\B(y)}$.  We have
\beqa
\lan f, (-\Delta_{\Om} + Q^{*}Q) f \ran \2\geq\2 \sum_{y\in \Om_1} 
\big\langle f_{\B(y)}, \big(-\Delta_{\B(y)} + Q^{*}Q  \big) f_{\B(y)} \big\rangle \non\\
\2\geq\2  {\dgreen C} \sum_{y\in  \Om_1} \|f_{\B(y)}\|^{2}_{2;\Om} =  {\dgreen C} \|f\|^{2}_{2;\Om}. \label{bracketing}
\eeqa
Here in the first inequality we used   the Neumann boundary conditions and (\ref{D-N-bracketing}) to justify that we can drop the bonds linking different 
 boxes $\B(y)$.  This gives (\ref{many-boxes}). To justify (\ref{Green-estimate}), we use that on a unit lattice
 \beqa
 \|\Delta_{\Omega}\|_{2,2;\Om}\leq {\blue 4}  \label{bounded-Laplacian}
 \eeqa
 and $Q^*Q$ has norm one as a projection.

 As for 3., we note that, in general, if a {\dgreen Hermitian matrix} $M$  satisfies $M\geq c $  on $\mcL^2(\Om)$, 
then $QMQ^*\geq c  $  on $\mcL^2(\Om_1)$. In fact, suppose $\lan f, QMQ^*f\ran< c\|f\|^2_{2;\Om_1}$ 
for some $f \in \mcL^2(\Om_1)$. Then, since   $QQ^*=1$, we have   $ \|f\|^2_{2;\Om_1}=\| Q^*f\|_{2;\Om}=0$. 
Setting $M{\dgreen =} \Gn$ and using item 2. we obtain the claim. {\dgreen Alternatively, this implication can
be seen using that the map $N\mapsto QNQ^*$ on the set of Hermitian matrices is completely positive and unital.
By setting $N=M-c$, the claim follows.}

Regarding~4, we observe that the proofs of items 2. and 3. can be immediately adapted to  infinite lattices.
In particular, (\ref{bracketing}), (\ref{bounded-Laplacian}) remain valid. \qed

\subsection{Exponential decay of integral kernels of $G$, $(QGQ^*)^{-1}$} \label{exponential-decay}

From Theorem~\ref{Random-walk} and Lemma~\ref{exponential-function} we obtain immediately
the following fact:  
\bep\label{first-corollary} Let $M$ on $\mcL^2(\mathbb{Z}^2)$ be strictly positive, {\magenta  i.e., $M\geq \mathrm{m}>0$,}  and let 
\beqa
|M(x,x')|\leq C\, \e^{-C_1|x-x'|} \label{A-formula}
\eeqa
for some  constants $C,C_1>0$. Then 
\beqa
|M^{-1}(x,x')|\leq C'\, \e^{- {\magenta C_1'}|x-x'|} \label{correction-one}
\eeqa
{\magenta for some constants $C', C_1'>0$.  (These constants depend on $C,C_1$ and $\mathrm{m}$)}.
\eep
\bec\label{coarse-Jaffe-Balabal-lemma} Let $N$ on $\mcL^2(L \mathbb{Z}^2)$ be strictly positive  {\magenta  i.e., $M\geq \mathrm{m}>0$,} and let
\beqa
|N(y,y')|\leq C\, \e^{-C_1|y-y'|}, \label{B-assumption}
\eeqa
for some  constants $C,C_1>0$. Then 
\beqa
|N^{-1}(y,y')|\leq C'\, \e^{- {\magenta C'_1}|y-y'|}
\eeqa
{\magenta for some constants $C', C_1'>0$.  (These constants depend on $C,C_1$, $\mathrm{m}$ and $L$)}.
\eec
\proof Consider a unitary scaling transformation $S_{L} : \mcL^2( \mathbb{Z}^2 ) \to \mcL^2(L \mathbb{Z}^2)$ given by
\beqa
S_{L} f:=L^{-1}f_{L}, \quad f_{L}(x):=f(L^{-1} x). 
\eeqa
We define
\beqa
M:=S_{L}^*N S_{L},
\eeqa
which is  strictly positive on $\mcL^2(\mathbb{Z}^2)$. We check the assumption (\ref{A-formula}) of Proposition~\ref{first-corollary}:
\beqa
|M(x,x')| = |\lan \de_{x}, S_{L}^* N S_{L} \de_{x'}\ran| = |\lan \de^{L}_{Lx}, N \de^{L}_{Lx'}\ran|=|N(Lx,Lx')|\leq  
c\, \e^{-C_1L|x-x'|}, \label{scaling}
\eeqa
where $\de^{L}_y:=\fr{1}{L^2}\de_y$ is the  delta function on the lattice $L \mathbb{Z}^2$. Thus Proposition~\ref{first-corollary} gives
\beqa
|M^{-1}(x,x')|\leq C'\, \e^{- {\magenta C_1'}|x-x'|}.
\eeqa
Now analogous steps as in (\ref{scaling}) give
\beqa
|N^{-1}(Lx,Lx')|=|M^{-1}(x,x')| \leq C'\e^{- {\magenta C'_1} |x-x'|},
\eeqa
which concludes the proof. \qed\\
Now we apply Corollary~\ref{coarse-Jaffe-Balabal-lemma} to study the integral kernels of operators $G$ and $(QGQ^*)^{-1}$. 
\bel\label{free-exponential-decay} The following properties hold true:
\beqa
|G(x,x')| \2\leq\2 C \e^{-C_1 |x-x'|},  \label{Green-function}\\
|(QGQ^*)(y,y')| \2\leq\2 C  \e^{-C_1|y-y'|}, \\
|(QGQ^*)^{-1}(y,y')| \2\leq\2 C  \e^{-C_1|y-y'|}, \label{Green-function-inverse}
\eeqa
for some constants $C, C_1>0$.

\eel
\proof We consider the operator $M:=-\De+Q^*Q$, which is strictly positive on $\mcL^2(\mathbb{Z}^2)$ by Lemma~\ref{inverse-lemma}. and 
note that its integral kernel $M(x,x')$ vanishes unless $|x-x'|\leq 1$ or
$x'\in \B(y_x)$. Hence, since $L>1$,
\beqa
|M(x,x')|\leq c\chi(|x-x'|_{\infty}\leq L) \leq c'\e^{- L^{-1}|x-x'|_{\infty}} \leq c' \e^{- \sqrt{2} L^{-1}|x-x'| },
\eeqa
where $\chi$ is the characteristic function.
Thus Proposition~\ref{first-corollary} gives
\beqa
|G(x,x')| \leq C \e^{-    {\magenta C_1} |x-x'| }, \quad {\magenta C_1>0},
\eeqa
which proves (\ref{Green-function}). Now we consider $N:=QGQ^*$ which is strictly
positive on $\mcL^2(L\mathbb{Z}^2)$  by  Lemma~\ref{inverse-lemma}.  We check
assumption~(\ref{B-assumption}) of Corollary~\ref{coarse-Jaffe-Balabal-lemma}: 
\beqa
|N(y,y')|\2=\2|\lan Q^*\de^{L}_y, G Q^*\de^{L}_{y'}\ran|  \leq  \sum_{x,x'} \one_{\B(y)}(x) |G(x,x')| \one_{\B(y')}(x')\non\\
\2\leq\2 C \sum_{x,x'} \one_{\B(y)}(x)   \e^{- {\magenta C_1}|x-x'|  }  \one_{\B(y')}(x')\non\\
\2 \leq\2 C L^4  \e^{-  {\magenta C_1} (|y-y'|-\sqrt{2} L ) }, 
\eeqa
where we use that $Q^*\de^{L}_y$ is the characteristic function of $\B(y)$ and   $|x-x'|\geq |y-y'|-\sqrt{2}L$.
Now Corollary~\ref{coarse-Jaffe-Balabal-lemma} gives (\ref{Green-function-inverse}). \qed

\subsection{Method of images}\label{method-of-images}
\newcommand{\da}{\dagger}
\newcommand{\Img}{\mrm{Img}}

\begin{figure}[t]   
\centering
\begin{tikzpicture}[scale=0.8]
	\draw[dashed, red] (-7,-0.5) to (8,-0.5);
	\draw[dashed, red] (-0.5,-4) to (-0.5,5);
	\draw[thick, blue] (-7,0) to (8,0);
	\draw[thick, blue] (0,-4) to (0,5);
	
	\foreach \i in {1,4,7,10,13}
	{
	\foreach \j in {1,4,7}
	{
	\draw[thick, blue] (-7+\i,-4+\j) to (-5+\i,-4+\j);
	\draw[thick, blue] (-7+\i,-4+\j) to (-7+\i,-2+\j);
	}
	}
	\foreach \i in {1,4,7,10,13}
	{
	\foreach \j in {3,6,9}
	{
	\draw[thick, blue] (-7+\i,-4+\j) to (-5+\i,-4+\j);
	\draw[thick, blue] (-5+\i,-6+\j) to (-5+\i,-4+\j);
	}
	}	
	
	\filldraw (0.5,1.5) circle(1pt);
	\draw (0.5,1.5) node[below]{$x$};
	\draw[thick,blue] (-1,-0.1) to (-1,0.1);
	\draw[blue] (-1.1,-0.1) node[below]{\tiny$-1$};
	\draw[thick,blue] (-0.5,-0.1) to (-0.5,0.1);
	\draw[blue] (-0.6,-0.1) node[below]{\tiny$-0.5$};
	\draw[thick,blue] (0.5,-0.1) to (0.5,0.1);
	\draw[blue] (0.5,-0.1) node[below]{\tiny$0.5$};
	\draw[thick,blue] (1,-0.1) to (1,0.1);
	\draw[blue] (1,-0.1) node[below]{\tiny$1$};
	\draw[thick,blue] (1.5,-0.1) to (1.5,0.1);
	\draw[blue] (1.5,-0.1) node[below]{\tiny$1.5$};
	\draw[thick,blue] (2,-0.1) to (2,0.1);
	\draw[blue] (2,-0.1) node[below]{\tiny$2$};
	
	\foreach \i in {0,2,6,8,12}
	{
	\foreach \j in {0,2,6}
	{
	\filldraw (-4.5+\i,-1.5+\j) circle(1pt);
	}
	}
	\draw (-4.5+12,-1.5+6) node[below]{$z_{-\!7}$};
	\draw (-4.5+8,-1.5+6) node[below]{$z_{-\!6}$};
	\draw (-4.5+6,-1.5+6) node[below]{$z_{-\!5}$};
	\draw (-4.5+2,-1.5+6) node[below]{$z_{-\!4}$};
	\draw (-4.5,-1.5+6) node[below]{$z_{-\!3}$};
	\draw (-4.5,-1.5+2) node[below]{$z_{-\!2}$};
	\draw (-4.5+2,-1.5+2) node[below]{$z_{-\!1}$};
	\draw (-4.5+6,-1.5+2) node[below]{$z_{0}$};
	\draw (-4.5+8,-1.5+2) node[below]{$z_{1}$};
	\draw (-4.5+12,-1.5+2) node[below]{$z_{2}$};
	\draw (-4.5,-1.5) node[below]{$z_{3}$};
	\draw (-4.5+2,-1.5) node[below]{$z_{4}$};
	\draw (-4.5+6,-1.5) node[below]{$z_{5}$};
	\draw (-4.5+8,-1.5) node[below]{$z_{6}$};
	\draw (-4.5+12,-1.5) node[below]{$z_{7}$};
\end{tikzpicture}
\caption{The square containing the origin is the set $\Om$. The reflections $P_{\mu}$ are defined w.r.t. the dashed lines from the figure. The points $z_j$ of the argument $z$ are as in formula (\ref{images-test-case-x}). 
}
\label{picture}
\end{figure}
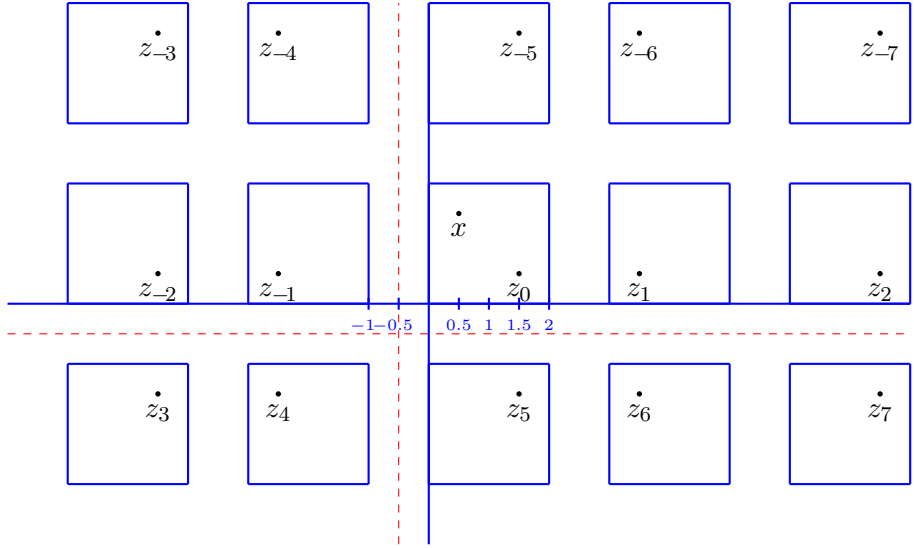

 Recall that $\Om:=I ^{\times 2}$, where $I=[0,1, 2,\ldots, n-1]$.
The boundary $\pa \Om\subset \Om$ consists of $4$ faces
\beqa
\pa\Om=(\{ 0\} \times I) \cup (I\times \{0\})\cup (\{n-1\}\times I)\cup (I\times \{n-1\}),  \label{boundary}
\eeqa
which we denote $\pa\Om_0$, $\pa\Om_1$, $\pa\Om^0$, $\pa\Om^1$, respectively. Next, we introduce the usual discrete partial derivatives
on $\mcL^2(\mathbb{Z}^2)$ and their adjoints {\dgreen consistently with Definition~\ref{De-0-def}}
\beqa
(\pa_{\mu} f)(x):=f(x+e_{\mu})-f(x), \quad (\pa^*_{\mu} f)(x):=-(f(x)-f(x-e_{\mu})),
\eeqa
where $e_{\mu}$ is the unit vector in the $\mu$-th direction.
With these definitions we can select the following subspace of functions in $\mcL^2(\mathbb{Z}^2)$:
\bed\label{D-Om-def} We say that a function $\v\in \L^2(\mathbb{Z}^d)$ satisfies Neumann boundary conditions on $\Om$, if the following relations
hold on the respective subsets of the boundary (\ref{boundary}):
\beqa
 (\pa^{*}_\mu \v)(x)\2=\2 0  \quad \textrm{ for } \quad x\in  \pa \Om_\mu,  \label{first-relation-Neumann} \\
 (\pa_\mu \v)(x)\2=\20   \quad \textrm{ for } \quad x\in \pa \Om^{\mu},  
 \eeqa
$\mu=0,1$. We denote the subspace of such functions $D_{\Om}$. 
\eed
Now we formulate an equivalent condition for $\v\in \L^2(\mathbb{Z}^{2})$ to be
 an element of $D_{\Om}$.  We let $P_{\mu}$ (resp. $\ov{P}_{\mu}$) be reflections w.r.t. axes parallel to $\pa \Om_{\mu}$ (resp. $\pa \Om^{\mu}$)
 as indicated in  Figure~\ref{picture}. (We refer to \cite[Section 3]{DST23} for formal definitions of these reflections). The following lemma is immediate: 
 \bel\label{projection-lemma-2d} Let $\v\in \L^2(\mathbb{Z}^{2})$,    $\mu=0,1$. Then $\v\in D_{\Om}$ iff it  satisfies 
 \beqa
(P_{\mu}\v)(x)\2=\2 \v(x), \quad x\in \pa\Om_{\mu}, \\
(\ov{P}_{\mu}\v)(x)\2=\2 \v(x), \quad x\in \pa\Om^{\mu}.
 \eeqa 
   \eel
 \proof  See \cite[Lemma~3.10]{DST23}.   \qed
\bel \label{Q-reflection} The following properties hold
\beqa
P_{\mu} Q^*Q P_{\mu}= Q^*Q, \quad   \ov{P}_{\mu} Q^*Q \ov{P}_{\mu} = Q^*Q. 
\eeqa
Consequently,
\beqa
G(P_{\mu}x, P_{\mu}x')=G(\ov{P}_{\mu} x, \ov{P}_{\mu} x')=G(x,x'). \label{Green-function-invariance}
\eeqa
\eel
\proof See \cite[Lemma 4.1]{DST23}. \qed \\
\nin As a useful application of Lemmas~ \ref{projection-lemma-2d}, \ref{Q-reflection}, we obtain the following:
 \bel\label{De-vs-De-Om} Suppose that $\v\in D_{\Om}$. Then, for any $\ell\in \nat$ we have
  \beqa
   ((-\De_{\Om} +Q^*Q)^{\ell} \v_{\Om})(x) = ((-\De +Q^*Q)^{\ell} \v)(x), \label{Neumann-free}
  \eeqa
where $x\in \Om$ and $\v_{\Om}:=\v |_{\Om}$.
  \eel
\proof  Property (\ref{Neumann-free}) clearly holds for $\ell=1$ by definition of $D_{\Om}$. 
 Using this, we write 
 \beqa
 \lan \de_x, (-\De_{\Om}+Q^*Q)^{\ell} \v_{\Om} \ran \2=\2 \sum_{x'\in \Om} \lan \de_x, (-\De_{\Om}+Q^*Q)^{\ell-1} \de_{x'} \ran \lan \de_{x'},  (-\De +Q^*Q )\v  \ran\non\\
\2=\2 \sum_{x'\in \Om} \lan \de_x, (-\De_{\Om}+Q^*Q )^{\ell-1} \de_{x'} \ran \lan \de_{x'},  ((-\De+Q^*Q)\v)_{\Om}  \ran \non\\
\2=\2 \lan \de_x,   (-\De_{\Om}+Q^*Q )^{\ell-1}  ((-\De+Q^*Q)\v)_{\Om}  \ran. \label{iterative-argument}
\eeqa 
 Now we obtain from Lemmas~\ref{projection-lemma-2d}, \ref{Q-reflection}  that if $f\in D_{\Om}$ then also  $(-\De+Q^*Q)\v  \in D_{\Om}$. Thus we can iterate the argument (\ref{iterative-argument}) until we obtain (\ref{Neumann-free}). \qed\\
 Using the Stone-Weierstrass theorem we can extend Lemma~\ref{De-vs-De-Om} from polynomials to continuous functions:
 \bel\label{identity-of-Laplacians-lemma}   For any $\v\in D_{\Om}$ and  $F\in C(\real)$ we have
\beqa
(F(-\De_{\Om}+Q^*Q)\v_{\Om})(x)= (F(-\De +Q^*Q) \v)(x), \quad x\in \Om, \label{arbitrary-function}
\eeqa
where $\v_{\Om}\in \L^2(\Om)$ is the restriction of $\v\in \L^2(\mathbb{Z}^2)$ to $\Om$. 
\eel
 \begin{remark} Referring to Lemma~\ref{inverse-lemma}, we obtain from (\ref{arbitrary-function})
 \beqa
 (G(\Om)\v_{\Om})(x)= ( G \v)(x), \quad  \v\in D_{\Om}, \quad x\in \Om.
  \eeqa
  \end{remark}
 \proof Since $-\De_{\Om}+Q^*Q$, $\De+Q^*Q$ are bounded operators, we can restrict $F$ to a bounded interval $[a,b]$ containing  
their spectra. Then, for any $\veps$ we can find a polynomial $F_{\veps}$ s.t.
\beqa
\sup_{\la\in [a,b]}|(F-F_{\veps})(\la)|\leq \veps.
\eeqa  
 Using this, we estimate
 \beqa
 \2 \2|(F(-\De_{\Om}+Q^*Q)\v_{\Om})(x)- (F(-\De+Q^*Q)\v)(x) |\non\\
\2   \2\ph{44444}\leq  |\lan \de_x, (F  - F_{\veps})(-\De_{\Om}+Q^*Q ) \v_{\Om} \ran|+  |\lan \de_x, (F  - F_{\veps})(-\De+Q^*Q) \v  \ran|\non\\
\2 \2 \ph{44444444444444444}+   |\lan \de_x, F_{\veps}(-\De_{\Om}+Q^*Q) \v_{\Om} \ran - \lan \de_x, F_{\eps}(-\De+Q^*Q)\v \ran|\non\\
 \2  \2\ph{44444} \leq \veps 2\|f\|_{2} + |\lan \de_x, F_{\veps}(-\De_{\Om}+Q^*Q) \v_{\Om} \ran - \lan \de_x, F_{\eps}(-\De+Q^*Q)\v \ran|. \label{F-estimate}
   \eeqa
As the last term  on the r.h.s. of (\ref{F-estimate})  vanishes by Lemma~\ref{De-vs-De-Om},  the proof is complete. \qed
   
After these preparations we can move on to the method of images. Define the set of image points  $\Img:=\{z_j\}_{j\in \nat}$ on $\mathbb{Z}^2$  by the following two 
requirements (\cite{GJ}[Section 7.4])
\begin{itemize}
\item $z\in   \Img $,
\item The set $  \Img$ is invariant under the  reflections $P_{\mu}$, $\ov{P}_{\mu}$, $\mu=0,1$, defined above.
\end{itemize}
This set is depicted in Figure~\ref{picture}. It is well known that the following relation between the Green functions with free and Neumann boundary conditions holds true:
\bel\label{method-of-images-lemma-zero} For $x,z\in \Om$ the following identity holds
\beqa
G(\Om)(x,z)=\sum_{z_j\in \Img}G(x, z_j). \label{images-test-case-x}
\eeqa
\eel
\proof  See \cite[Lemma~4.2]{DST23}. \qed\\
Our goal in this subsection is to derive an analogous formula linking  operators $(QG(\Om)Q^*)^{-1}$ and $(QGQ^*)^{-1}$,
acting on $\mcL^2(\Om_1)$ and $\mcL^2(L\mathbb{Z}^2)$, respectively. For this purpose we define the reflections on the
coarse lattice
\beqa
P_{1,\mu}:=QP_{\mu} Q^*, \quad \ov{P}_{1,\mu}:=Q\ov{P}_{\mu} Q^*. \label{coarse-reflections}
\eeqa
By acting on delta functions it is easy to see that the geometric action of these reflections is determined by
\beqa
x\in \B(P_{1,\mu}y) \quad \Leftrightarrow \quad  P_{\mu}x\in \B(y), \label{boxes-relation}
\eeqa
and analogously for $\ov{P}_{\mu}$, where we naturally extended the definition of $\B(y)$ from $\Om$ to $\mathbb{Z}^2$.
Furthermore, we have by {\dgreen the first relation in (\ref{coarse-reflections})} and $QQ^*=1$
\beqa
Q^*P_{1,\mu}= P_{\mu} Q^*.
\eeqa
This latter property combined with (\ref{Green-function-invariance}) immediately gives:
\bel\label{QGQ-invariance} The following properties hold:
\beqa
\2 \2(QGQ^*)(P_{1,\mu}y, P_{1,\mu}y')=(QGQ^*)(y,y'), \\
\2 \2(QGQ^*)^{-1}(P_{1,\mu}y, P_{1,\mu}y')=  (QGQ^*)^{-1}(y,y'),
\eeqa
and analogously for $\ov{P}_{1,\mu}$.
\eel
\nin Now we define the set of image points  $\Img_1:=\{z_j\}_{j\in \nat}$ on the coarse lattice $L\mathbb{Z}^2$,  by the following two 
requirements:
\begin{itemize}
\item $z\in   \Img_1$,
\item The set $  \Img_1$ is invariant under the  reflections $P_{1,\mu}$, $\ov{P}_{1 \mu}$, $\mu=0,1$, defined in (\ref{coarse-reflections}).
\end{itemize}
The main result of this subsection is the following:
\bel\label{method-of-images-lemma} For $x,z\in \Om$ the following identity holds
\beqa
(QG(\Om)Q^*)^{-1}(y,z)=\sum_{z_j\in \Img_1}(QGQ^*)^{-1}(y, z_j). \label{images-test-case-coarse}
\eeqa
\eel
\proof  Given  the distribution of image points $z_j\in  \Img_1$, and estimate (\ref{Green-function-inverse})  we obtain that the sum in (\ref{images-test-case-coarse})  is convergent.  It suffices to check that, in the sense of  multiplication of operators on $\mcL^2(\Om_1)$, 
\beqa
  (QG(\Om)Q^*)  \textrm{(r.h.s. of  (\ref{images-test-case-coarse})) }=1.
\eeqa
For any $\ell\in \nat$, let $I^{\ell}_1:=L [-\ell, {\blue -\ell+1, \ldots }  0,\ldots,n_1-1, {\blue \ldots,}  n_1+\ell-1]$, $n_1-1:=L^{m-1}$, and $\Om^{\ell}_1:= (I^{\ell}_1)^{\times 2}$ be a coarse finite lattice containing $\Om_1$. We will
check that for each $z\in \Om_1$ and $y\in L\mathbb{Z}^2$ the expression
\beqa
F^{\ell}_z(y):=\chi_{\Om^{\ell}_1}(y) \sum_{z_j\in \Img_1 }(QGQ^*)^{-1}(y, z_j) \label{F-y-x}
\eeqa
satisfies  $Q^*F^{\ell}_z\in   D_{\Om}\subset \L^2(\mathbb{Z}^2)$  
(cf. Definition~\ref{D-Om-def}). For this purpose we will
use the criterion from Lemma~\ref{projection-lemma-2d}: Let $\ti{P}_{\mu}$ denote the reflections $P_{\mu}$ or 
$\ov{P}_{\mu}$ and similarly for $\ti{P}_{1,\mu}$. We consider 
$x\in \pa \Om$, so $\chi_{\Om^{\ell}_1}(y_x)=\chi_{\Om^{\ell}_1}(P_{1,\mu}y_x)=1$. (We need $\chi_{\Om^{\ell}_1}$ in (\ref{F-y-x}) only to ensure that $x\mapsto F^{\ell}_z(x)$ is in 
$\L^2(L\mathbb{Z}^2)$).  Then, since $Q \de_{x}=\de^{L}_{y_x}$,
 \beqa
 (Q^*F^{\ell}_z)(\ti{P}_{\mu}x) \2=\2 \lan Q \de_{\ti{P}_{\mu} x}, \sum_{z_j\in \Img_1 } (QGQ^*)^{-1} \de^{L}_{z_j}\ran  =\sum_{z_j\in \Img_1}(QGQ^*)^{-1}(y_{\ti{P}_{\mu} x}, z_j) \non\\ \2=\2\sum_{z_j\in\Img_1}(QGQ^*)^{-1}( \ti{P}_{1,\mu} y_x, \ti{P}_{1,\mu}z_j)= \sum_{z_j\in\Img_1}(QGQ^*)^{-1}(y_x, z_j)= (Q^*F^{\ell}_z)(x), \quad\quad
 \eeqa
where in the third step we used property (\ref{boxes-relation}) and the invariance of the set $\Img_1$ under the reflections, and in the fourth step Lemma~\ref{QGQ-invariance}. 
Thus $Q^*F_z\in D_{\Om}$ and we obtain from Lemma~\ref{identity-of-Laplacians-lemma}   that, for $x\in \Om$, $F^{\ell}_{z,\Om_1}:=F^{\ell}_z|_{\Om_1}$,
\beqa
(Q \Gn Q^* F^{\ell}_{z, \Om_1})(y) \2=\2 (Q G Q^* F^{\ell}_{z})(y) \non\\
\2=\2\sum_{z_j\in \Img_1, \, \, y'\in L\mathbb{Z}^2 } \lan \de^{L}_y, Q G Q^* \de^{L}_{y'} \ran \lan \de^{L}_{y'},\chi_{\blue \Om^{\ell}_1}(QGQ^*)^{-1}  \de^{L}_{z_j} \ran\non\\
\2=\2 \de^{L}_{z}(y) - \sum_{z_j\in \Img_1,\, \, y'\in L\mathbb{Z}^2 }\lan \de^{L}_y, Q G Q^*\de^{L}_{y'} \ran \lan \de^{L}_{y'}, (1-\chi_{\Om^{\ell}_1}) (QGQ^*)^{-1}  \de^{L}_{z_j} \ran, \label{images-final-computation}
\eeqa
where in the last step we made use of the fact that $z$ is the only element of $\Img_1$ inside $\Om_1$ to obtain $\de^{L}_{z}(y) $. To conclude we 
 note that the l.h.s. of (\ref{images-final-computation}) is independent of $\ell$ due to the restriction of $F^{\ell}_z$ to $\Om_1$ and that $\lim_{\ell\to \infty}\chi_{\Om^{\ell}_1}=1$ pointwise. To enter with the limit
under the sums w.r.t. $z_j, y'$ on the r.h.s. we use the exponential decay of the kernels of  $Q G Q^*, (Q G Q^*)^{-1}$, shown in Lemma~\ref{free-exponential-decay},  and dominated convergence. \qed\\
We immediately obtain from Lemmas~\ref{method-of-images-lemma-zero}, \ref{method-of-images-lemma},  the exponential decay of kernels of the corresponding operators on a finite lattice. The following lemma is proven analogously to \cite[Theorem 4.3]{DST23}.
\bel\label{Neumann-exponential-decay} \label{decay-Neumann} The following properties hold true:
\beqa
|G(\Om)(x,x')| \2\leq\2 C \e^{-C_1 |x-x'|},  \label{Green-function-Neumann}\\
|(QG(\Om)Q^*)^{-1}(y,y')| \2\leq\2 C  \e^{-C_1|y-y'|}, \label{Green-function-inverse-Neumann}
\eeqa
for some constants $C,C_1>0$, {\blue independent of $n$.}
\eel

\begin{remark} {\dgreen By a more careful analysis one can see that {\cyan in (\ref{Green-function-Neumann})} $C\sim L^4$ and $C_1$ is independent of $L$, cf. \cite[Theorem A]{DST23}.} 
\end{remark}
\subsection{$\mcL^{\infty}$-bounds on  $\Gn$, $(Q\Gn Q^*)^{-1}$, $(Q\Gn R^*Q^*)^{-1}$}    \label{L-infty} 

In Subsections \ref{strict-positivity} -- \ref{method-of-images} we treated  $G{\cyan (\Om)}$, $(QG{\cyan (\Om)}Q^*)^{-1}$,   as operators acting on spaces $\mcL^2(\Om), \mcL^2(\Om_1)$ of
scalar-valued functions.
Let us now consider them as operators acting componentwise on functions with values in $\real^3$, as we did
in Subsection~\ref{derivation}. We recall that $\mcL^{\infty}(\Om; \real^3)$ is equipped with the norm
\beqa
\|\vec{v}\|_{\infty;\Om}=\sup_{x\in \Om} | \vec{v}(x)|,
\eeqa
where $|\,\cdot\,|$ is the Euclidean distance in $\real^3$.  Now for $M: \mcL^{\infty}(\Om; \real^3)\to \mcL^{\infty}(\Om; \real^3)$,
$N: \mcL^{\infty}(\Om_1; \real^3)\to \mcL^{\infty}(\Om_1; \real^3)$
the resulting operator norms satisfy
\beqa
\|M\|_{\infty,\infty;\Om}\2:=\2\sup_{\|\vec{f}\|_{\infty;\Om}\leq 1 } \|M \vec{f}\|_{\infty;\Om} \leq \sup_{x\in \Om} \sum_{x'\in \Om} \|M(x,x')\|, \label{norm-estimate-zero} \\
\|N\|_{\infty,\infty;\Om_1}\2:=\2\sup_{\|\vec{f}\|_{\infty;\Om_1}\leq 1 } \|N \vec{f}\|_{\infty;\Om}\leq \sup_{y\in \Om_1}L^2 \sum_{y'\in \Om_1} \|N(y,y')\|,\label{norm-estimate}
\eeqa
where $\|M(x,x')\|$ is the {\blue operator norm} {\cyan of} the $3\times 3$ matrix $M(x,x')$ {\blue w.r.t. $|\,\cdot\,|$. }
Using these relations we immediately obtain:
\bel\label{infty-bounds} The following bounds hold true
\beqa
\2 \2\|\Gn \|_{\infty,\infty;\Om}  \leq  C, \label{-G-} \\
\2 \2 \|(Q\Gn Q^*)^{-1}\|_{\infty,\infty;\Om_1} \leq C,  \label{Q-G-Q}
\eeqa
{\blue for $C$ independent of $n$.}
\eel
\proof For $M=\Gn$ we immediately obtain the bound referring to  (\ref{norm-estimate-zero}) and the decay of the kernel (\ref{Green-function-Neumann}).
For $N=(Q\Gn Q^*)^{-1}$ we use (\ref{norm-estimate}) and (\ref{Green-function-inverse-Neumann}).  
\qed\\
In the next lemma we consider an operator on $\mcL^2(\Om_1;\real^3)$ which has a non-trivial action on the target space $\real^3$.
\bel\label{Q-G-R-Q-lemma} The operator ${\dgreen D_{\vA}:= }Q\Gn R^*_{\vA}Q^*$ is invertible and  the following estimate holds true for $\vec{A}\in \vConf^{\eps}(\Om)$, $\del=1$ 
and some constant $C$ independent of $n$ {\dgreen and $\eps$}
\beqa
  \|(Q \Gn R^*_{\vA}Q^*)^{-1}\|_{\infty,\infty;\Om_1} \leq C,  \label{Q-G-R-Q}
\eeqa
 provided that $0<\eps\leq 1$ sufficiently small (uniformly in $n$). 
\eel
\proof  We  define $\de R^*_{\vec{A}} :=R^*_{\vec{A}}-1$.  It has the form 
\beqa
\de R^*_{\vec{A}}(x)\vec{v}(x)= (\ppU_0(x) -1) \vec{v}(x)-\vec{\ppU}(x)\times \vec{v}(x)
\eeqa
and satisfies, by assumption {\dgreen and the relation $A_0(x)=\sqrt{1-|\vec{A}(x)|^2}$} 
\beqa
\|\de R_{\vec{A}}(x) \|\leq  |\ppU_0(x) -1|+|\vec{A}(x)|\leq 2\eps. \label{dRA}
\eeqa

{\blue We have the following
\beqa
Q\Gn R^*_{\vA}Q^*=(Q\Gn Q^*)\big(1+(Q\Gn Q^*)^{-1}  Q\Gn \de R^*_{\vA}    Q^*\big). \label{resolvent-equation-one}
\eeqa
The first factor is invertible by part 3. of Lemma \ref{inverse-lemma}. The second 
factor is invertible provided that
\beqa
\|(Q\Gn Q^*)^{-1}  Q\Gn \de R^*_{\vA}    Q^*\|_{\infty,\infty; \Om_1}\leq \|(Q\Gn Q^*)^{-1}\|_{\infty,\infty; \Om_1}  \| Q\Gn \de R^*_{\vA}    Q^*\|_{\infty,\infty; \Om_1}  <1. \label{Neumann-condition}
\eeqa
Indeed, then the expansion of $\big(1+(Q\Gn Q^*)^{-1}  Q\Gn \de R^*_{\vA}    Q^*\big)^{-1}$ into the Neumann series is convergent. 
Given (\ref{Q-G-Q}), it suffices to control   $\| Q\Gn \de R^*_{\vA}    Q^*\|_{\infty,\infty; \Om_1}$ to show (\ref{Neumann-condition}). } In order to apply (\ref{norm-estimate}), we estimate
\beqa
\|\lan \de^{L}_{y}, Q\Gn \de R^*_{\vA} Q^* \de^{L}_{y'}\ran \| \2\leq\2  \sum_{x,x'} \one_{\B(y)}(x) |\Gn(x,x')|\,\| \de R^*_{\vA}(x')  \| \one_{\B(y')}(x')\non\\
\2\leq\2 \eps \, C  \sum_{x,x'} \one_{\B(y)}(x)   \e^{- C_1 |x-x'| }  \one_{\B(y')}(x')\non\\
\2 \leq\2 \eps\, C  L^4  \e^{- C_1 (|y-y'|-\sqrt{2} L ) }, 
\eeqa
where we used in the first step that $Q^* \de^{L}_{y'}=\one_{\B(y')}$ and in the second step we applied (\ref{dRA}). 
{\dgreen Now  (\ref{norm-estimate}), (\ref{Neumann-condition}) give
\beqa
\|(Q\Gn Q^*)^{-1}  Q\Gn \de R^*_{\vA}    Q^*\|_{\infty,\infty; \Om_1}\leq C' \eps.
\eeqa
Consequently,  for $C'\eps \leq  1/2$ we can sum up and estimate the Neumann series as follows
\beqa
\|(Q\Gn R^*_{\vA}Q^*)^{-1}\|_{\infty,\infty; \Om_1}\leq \fr{C''}{1-C'\eps}\leq 2C''.
\eeqa
Thus we have chosen $\eps$ uniformly in $n$ and established (\ref{Q-G-R-Q}) with a constant independent of $\eps$.} \qed
\bel\label{R-lemma-two}  The operator    $  D_{\vA}:= Q\Gn R^*_{\vec{\ppU}}Q^*$ satisfies the following: 
\beqa
\| (D_{\vA_1})^{-1} - (D_{\vA_2})^{-1} \|_{\infty,\infty;\Om_1} \leq 2\|\Gn \|_{\infty,\infty;\Om} \|D_{\vA_1}^{-1} \|_{\infty,\infty;\Om_1}\, \|D_{\vA_2}^{-1} \|_{\infty,\infty;\Om_1}  
 \| \vec{\ppU}_1-\vec{\ppU}_2 \|_{\infty,\infty;\Om}, \label{Q-difference}
\eeqa
where $\vec{\ppU}_1, \vec{\ppU}_2 \in \vConf^{\eps}(\Om)$  are two  configurations (with $\del_1=\del_2=1$)  and $0<\eps\leq 1$ sufficiently small, uniformly in~$n$. 
\eel
\proof  We come back to (\ref{resolvent-equation-one}), which reads in a short-hand notation
\beqa
D_{\vA}^{-1}=\big(1+D_0^{-1} \de D_{\vA}  \big)^{-1} D_0^{-1},
\eeqa 
where $\de D_{\vA}:= Q\Gn \de R^*_{\vA}Q^*$ {\cyan and $\de R^*_{\vec{A}} :=R^*_{\vec{A}}-1$}. Next, by the resolvent identity,
\beqa
D_{\vA_1}^{-1} - D_{\vA_2}^{-1} \2=\2\bigg\{ \big(1+D_0^{-1} \de D_{\vA_1}  \big)^{-1} -  \big(1+D_0^{-1} \de D_{\vA_2}  \big)^{-1}  \bigg\}D_0^{-1}\non\\
                                             \2=\2 \big(1+D_0^{-1} \de D_{\vA_1}  \big)^{-1} D_0^{-1}   \big(  \de D_{\vA_2} -  \de D_{\vA_1}    \big)\big(1+D_0^{-1} \de D_{\vA_2}  \big)^{-1} D_0^{-1}. \,\,  \,                                                         
\eeqa
Hence
\beqa
\|D_{\vA_1}^{-1} - D_{\vA_2}^{-1}\|_{\infty,\infty;\Om_1}\leq \|D_{\vA_1}^{-1} \|_{\infty,\infty;\Om_1}\, \|D_{\vA_2}^{-1} \|_{\infty,\infty;\Om_1}\, \| \de D_{\vA_2} -  \de D_{\vA_1}\|_{\infty,\infty;\Om_1}. \label{D-one}
\eeqa
 We can write
\beqa
\|  \de D_{\vA_2} -  \de D_{\vA_1}  \|_{\infty,\infty;\Om_1} \2=\2 \| Q\Gn (\de R^*_{\vA_2} -  \de R^*_{\vA_1} ) Q^*\|_{\infty,\infty;\Om_1} \non\\
                                                                   \2 \leq \2 \|\Gn \|_{\infty,\infty;\Om} \|\de R^*_{\vA_2} -  \de R^*_{\vA_1}  \|_{\infty,\infty;\Om} \non\\
                                                                   \2 = \2 \|\Gn \|_{\infty,\infty;\Om}\| R^*_{\vA_2} -  R^*_{\vA_1} \|_{\infty,\infty;\Om}\leq 2\|\Gn \|_{\infty,\infty;\Om} \|\vA_1-\vA_2\|_{\infty,\infty;\Om}, \quad\quad \label{D-two}                                                                
\eeqa
where in the last step we used (\ref{R-bound-three}) and the fact that $Q: \mcL^{\infty}(\Om)\to \mcL^{\infty}(\Om_1)$, $Q^*: \mcL^{\infty}(\Om_1)\to \mcL^{\infty}(\Om)$ have
norms bounded by one.  Now the statement follows from (\ref{D-one}), (\ref{D-two}) and Lemma~\ref{Q-G-R-Q-lemma}. \qed
\bel\label{R-lemma}  {\cyan For} $\vec{\ppU}, \vec{\ppU}_1,  \vec{\ppU}_2  \in \vConf^{\eps}(\Om)$,  $\del=\del_1=\del_2=1$,  $0<\eps\leq 1/2$, there hold the bounds
\beqa
\| R^*_{ \vec{\ppU} }\|_{\infty,\infty;\Om}  \2 \leq\2 2, \label{R-bound-one} \\ 
\|R^*_{ \vec{\ppU}_1 }- R^*_{\vec{\ppU}_2}\|_{\infty,\infty;\Om} \2 \leq \2 2  \| \vec{\ppU}_1- \vec{\ppU}_2  \|_{\infty;\Om}. \label{R-bound-three}
\eeqa
\eel
\proof Regarding (\ref{R-bound-one}), we recall (\ref{R-T}) and write 
\beqa
\| R^* _{ \vec{\ppU} }\|_{\infty,\infty;\Om}\2=\2\sup_{\|\vec{v}\|_{\infty;\Om}\leq 1} \sup_{x\in \Om}  |R^*_{ \vec{\ppU}} (x)\vec{v}(x)| \non\\
\2=\2 \sup_{\|\vec{v}\|_{\infty;\Om}\leq 1} \sup_{x\in \Om}  | \ppU_0(x) \vec{v}(x)-\vec{\ppU}(x)\times \vec{v}(x)|\leq 2,
\eeqa
where we used   the fact that $|\vec{A}(x)|\leq 1$. 

Now we move on to (\ref{R-bound-three}). We have, by (\ref{R-T}),
\beqa
(R^*_{ \vec{\ppU}_1 }(x)- R^*_{\vec{\ppU}_2}(x))\vec{v}(x) =(\ppU_{1,0}(x) -  \ppU_{2,0}(x)  )  \vec{v}(x)- (\vec{\ppU}_1(x)  -  \vec{\ppU}_2(x) )  \times \vec{v}(x).
\eeqa
Making use of  
\beqa
|\ppU_{1,0} -  \ppU_{2,0}|=\fr{|\,|\vA_2|^2-|\vA_1|^2|}{\sqrt{1-|\vA_1|^2} +\sqrt{1-|\vA_2|^2} }\leq 2\eps |\vA_1-\vA_2|,
\eeqa 
we obtain the following estimate
\beqa
|(R_{ \vec{\ppU}_1 }(x)- R_{\vec{\ppU}_2}(x))\vec{v}(x)|\leq (2\eps+1)|  \vec{v}(x)| |\vA_1(x)-\vA_2(x)|. 
\eeqa
Hence,
\beqa
\|(R_{ \vec{\ppU}_1 }- R_{\vec{\ppU}_2} )\vec{v}\|_{\infty;\Om} \leq (2\eps+1) \|\vec{v}\|_{\infty; \Om} \| \vA_1-\vA_2\|_{\infty;\Om},
\eeqa
which concludes the proof of (\ref{R-bound-three}). \qed

\subsection{$\mcL^{\infty}$-bounds for the {\dgreen remainder} $\protect\overrightarrow{r}$} \label{rest-term-bounds}

In this subsection we use the notation introduced above (\ref{terms}).
\bel\label{r-lemma-one} Suppose that $|\vec{\ppV}(b) |\leq \eps_1$, $\vec{\ppU}\in  \vConf^{\eps}(\Om)$, $\del=1$ and $0\leq \eps,\eps_1\leq 1/2$. Then 
\beqa
\|\pa^* \vec{r}\|_{\infty{\magenta ; \Om}}\leq 24(\eps^2+\eps_1). 
\eeqa
\eel
\proof Given expressions~(\ref{terms}), (\ref{delta-form})  we have, for $(x,x'):=(b_-,b_+)$, 
\beqa
|\vec{r}(x,x')| \2 \leq \2   ( |\vec{\ppU}(x)|^2 +|\vec{\ppV}({\cyan x,x'})|^2  )\,| \vec{\ppU}(x')|\non\\
               \2  \2+   ( |\vec{\ppU}(x')|^2 +|\vec{\ppV}( {\cyan  x,x' }  )|^2  )\,| \vec{\ppU}(x)| \non\\
               \2 \2 +  | \vec{\ppV}( {\cyan  x,x' }  )| \non\\
               \2  \2 + |  \vec{\ppV}( {\cyan  x,x' }  ) | \, ( |\vec{\ppU}(x)| +  |\vec{\ppU}(x')|) + |\vec{\ppU}(x)|\, | \vec{\ppU}(x')|  \non\\                            
\2 \2+3 | \vec{\ppU}(x')| \,| \vec{\ppU}(x)|\, |  \vec{\ppV}({\cyan  x,x' }) |.
\eeqa
This gives, considering that $|\vec{\ppU}(x)|\leq \eps \leq 1$,
\beqa
|\vec{r}(x,x')|\leq 2( \eps^2+\eps_1^2) \eps +\eps_1+2\eps_1\eps+ \eps^2+ 3 \eps^2 \eps_1\leq 6(\eps^2+\eps_1).
\eeqa
Now we recall from Definition \ref{De-0-def} that
\beqa
(\pa^*\vec{r})(x)=-\vec{r}(x,x+e_0)-\vec{r}(x,x+e_1)+\vec{r}(x-e_0,x)+\vec{r}(x-e_1,x),
\eeqa
possibly with some terms omitted if $x$ is close to the boundary of $\Om$. In any case,
\beqa
|(\pa^* \vec{r})(x)|\leq 24(\eps^2+\eps_1), \label{bound-on-r}
\eeqa
which was to be shown. \qed

\bel\label{r-lemma-two} Suppose that $|\vec{\ppV}(b) |\leq \eps_1$, $\vec{\ppU}_1, \vA_2\in  \vConf^{\eps}(\Om)$, $s_1=s_2=1$,  $0<\eps_1, \eps \leq 1/2$.  Then
\beqa
  \| \pa^* \vr_{\vec{\ppU}_1 }  - \pa^* \vr_{\vec{\ppU}_2 } \|_{\infty;\Om}
\leq 96(\eps+\eps_1) \| \vec{\ppU}_1-\vec{\ppU}_2\|_{\infty;\Om}.
\eeqa
\eel
\proof  We recall formula~(\ref{terms}). There we wrote for brevity $\ppU_{\pm}=\ppU(b_{\pm})$ and $\ppV={\cyan B(b)=\pa V(y_b)}$. This gives, by Appendix~\ref{conf-W},  
\beqa
\vec{r}(b)\2=\2 - \de(\ppU_{0,-}\ppV_0) \vec{\ppU}_+  + \de(\ppU_{0,+} \ppV_0) \vec{\ppU}_-   + \ppU_{0,+} \ppU_{0,-} \vec{\ppV}\non\\        
                         \2 \2- \ppU_{0,+} (\vec{\ppU}_-\times \vec{\ppV}) + \ppU_{0,-} (\vec{\ppV} \times  \vec{\ppU}_+) +\ppV_0 (\vec{\ppU}_- \times  \vec{\ppU}_+) \non\\
                   \2 \2       + \vec{\ppU}_{+} (\vec{\ppU}_-\cdot \vec{\ppV}) - \vec{\ppU}_{-} (\vec{\ppV} \cdot  \vec{\ppU}_+) +\vec{\ppV} (\vec{\ppU}_- \cdot  \vec{\ppU}_+).                                       
                         \label{terms-one}
\eeqa
We denote by $\vr_1, \ldots, \vr_9$ the respective terms in (\ref{terms-one}). We note that $\vr_1, \vr_2$ have an analogous structure, thus it suffices to estimate one of them.
Similarly, $\vr_4, \vr_5$ and  $\vr_7,\vr_8,\vr_9$ have  analogous structures.

Now we recall from Definition \ref{De-0-def} that
\beqa
(\pa^*\vec{r})(x)=-\vec{r}(x,x+e_0)-\vec{r}(x,x+e_1)+\vec{r}(x-e_0,x)+\vec{r}(x-e_1,x), \label{lap-2}
\eeqa
possibly with some terms omitted if $x$ is close to the boundary of $\Om$. We denote by $(\pa^* \vr_j)_{j'}(x)$, $j'=1,2,3,4$, 
the four terms coming from  (\ref{lap-2}) for any fixed~$j$ (or less if $x$ is close to the boundary). 

 The items below give
\beqa
\|\pa^*\vr_{\vec{A}_1} -  \pa^*\vr_{\vec{A}_2}\|_{\infty;\Om} \2\leq\2 \big(24\eps_1+ 8\eps_1+ 16\eps_1+16\eps+ 8( 3\eps^2 +  \eps_1^2) \big)
\|\vA_1-\vA_2\|_{\infty; \Om}\non\\
\2 \leq \2 24 \big(\eps_1+\eps_1+ \eps_1+\eps+  \eps^2 +\eps_1^2  \big) \|\vA_1-\vA_2\|_{\infty;\Om} \non\\
\2 \leq \2 24\times 4 \big(\eps_1+\eps) \|\vA_1-\vA_2\|_{\infty;\Om}=96 (\eps_1+\eps) \|\vA_1-\vA_2\|_{\infty;\Om}.
\eeqa
\begin{itemize}

\item Contributions $\vec{r}_1, \vec{r}_2$. We have
\beqa
(-\pa^* \vr_{\vA_1,1})_1(x)\2=\2 - \de\big(A_{1,0}(x) \ppV_0(x,x+e_0)\big) \vec{\ppU}_1(x+e_0), \\
(-\pa^* \vr_{\vA_2,1})_1(x)\2=\2 - \de\big(A_{2,0}(x) \ppV_0(x,x+e_0)\big) \vec{\ppU}_2(x+e_0).
\eeqa
We consider the difference
\beqa
\2 \2(-\pa^*\vr_{\vA_1,1})_1(x)+(\pa^*\vr_{\vA_2,1})_1(x)\non\\
\2 \2= - \de\big(A_{1,0}(x) B_0(x,x+e_0)\big) \vec{A}_1(x+e_0)-  \de\big(A_{2,0}(x) B_0(x,x+e_0)\big) \vec{A}_2(x+e_0)\non\\
\2 \2=- \de\big(A_{1,0}(x) B_0(x,x+e_0)\big) (\vec{A}_1(x+e_0) -\vec{A}_2(x+e_0))\label{rest-contribution-one} \\
\2 \2\phantom{44}  - \big( \de\big(A_{1,0}(x) B_0(x,x+e_0)\big) -  \de\big(A_{2,0}(x) V_0(x,x+e_0)\big)   \big) \vec{A}_2(x+e_0). \label{rest-contribution-two}
\eeqa
We have, by (\ref{first-delta-estimate}),
\beqa
\de(A_0B_0)\leq \eps^2+\eps_1^2.
\eeqa
This gives
\beqa
\sup_{x\in\Om}|(\ref{rest-contribution-one})(x)| \2\leq\2 (\eps^2+\eps_1^2)  \sup_{x\in\Om } |\vec{A}_1(x+e_0) -\vec{A}_2(x+e_0)| \non\\
                                                                                             \2\leq\2  (\eps^2+\eps_1^2)  \|\vec{A}_1 - \vec{A}_2 \|_{\infty;\Om}.
\eeqa
Furthermore, by (\ref{second-delta-estimate}), we have
\beqa
|\de(A_{1,0}(x)B_0(x+e_0))-\de(A_{2,0}(x)B_0(x+e_0))|\leq 2\eps | \vec{A}_1(x) -\vec{A}_2(x)|.
\eeqa
Therefore, taking into account the factor $\vec{A}_2(x+e_0)$ in  (\ref{rest-contribution-two})
\beqa
\sup_{x\in\Om}\|(\ref{rest-contribution-two})(x)| \leq 2\eps^2 \| \vec{A}_1-\vec{A}_2\|_{\infty;\Om}.
\eeqa
Thus we have
\beqa
\| (\pa^* \vr_{\vA_1,1})-(\pa^*\vr_{\vA_2,1})\|_{\infty;\Om}\leq 4( 3\eps^2 +   \eps_1^2 )  \| \vec{A}_1-\vec{A}_2\|_{\infty; \Om},
\eeqa
where the factor $4$ comes from  (\ref{lap-2}).

\item Contribution $\vr_3$. We have
\beqa
 -(\pa^*\vr_{\vA_1,3})_{1}(x) \2=\2  \de( \ppU_{1,0}(x+e_0) \ppU_{1,0}(x)) \vec{\ppV}(x,x+e_0), \\
 -(\pa^*\vr_{\vA_2,3})_{1}(x) \2=\2 \de ( \ppU_{2,0}(x+e_0) \ppU_{2,0}(x))  \vec{\ppV}(x,x+e_0).
 \eeqa
The difference gives, by (\ref{second-delta-estimate}), 
\beqa
\2 \2 |(\pa^* \vr_{\vA_1,3})_{1}(x)-(\pa^*\vr_{\vA_2,3})_{1}(x)|\non\\
\2 \2= | \de( \ppU_{1,0}(x+e_0) \ppU_{1,0}(x)) - \de( \ppU_{2,0}(x+e_0) \ppU_{2,0}(x)) | \, |\vec{\ppV}(x,x+e_0)| \non\\
\2 \2 \leq 2\eps ( |\vA_2(x+e_0)-\vA_1(x+e_0)| +|\vA_{2}(x) -  \vA_{1}(x)|  ).
\eeqa
This gives, taking into account the four terms from (\ref{lap-2})
\beqa
\|(\pa^*\vr_{\vA_1,3})-(\pa^*\vr_{\vA_2,3})\|_{\infty;\Om}\leq 16\eps  \|\vA_2-\vA_1\|_{\infty;\Om}.
\eeqa
\item Contributions $\vr_4,  \vr_5$. We have
\beqa
-(\pa^*\vr_{\vA_1,4})_{1}(x)\2=\2- A_{1,0}(x+e_0) (\vA_{1}(x)   \times \vB(x,x+e_0)), \\
-(\pa^* \vr_{\vA_2,4})_{1}(x)\2=\2- A_{2,0}(x+e_0) (\vA_{2}(x)   \times \vB(x,x+e_0)).
\eeqa
The difference has the form
\beqa
| (\pa^*\vr_{\vA_1,4})_{1}(x) - (\pa^*\vr_{\vA_2,4})_{1}(x)| \2\leq\2 |A_{1,0}(x+e_0)- A_{2,0}(x+e_0)|\, | \vA_{1}(x)   \times \vB(x,x+e_0)|\,\,\, \label{four-one} \\
                                                                                   \2 +\2   |A_{2,0}(x+e_0)|\, |\vA_{1}(x)-\vA_2(x)| \, |\vB(x,x+e_0)|.                      \label{four-two}                                                       
                                                                                       \eeqa 
Considering that {\blue $\eps\leq 1/2$, } 
\beqa
|A_{1,0}- A_{2,0}|\2 \leq \2 |\sqrt{1- \vA_{1}^2 } - \sqrt{1- \vA_{2}^2 } |\non\\
                                                    \2 \leq \2 |\vA_2 - \vA_1| |\vA_2+\vA_1| \leq  2\eps |\vA_2 - \vA_1|.
\eeqa
Thus we can  estimate
\beqa
\| (\pa^* \vr_{\vA_1,4}) - (\pa^* \vr_{\vA_2,4})\|_{\infty;\Om} \2\leq \2  4(2\eps^2\eps_1 +\eps_1) \|\vA_2 - \vA_1\|_{\infty;\Om}\non\\
\2 \leq \2 8\eps_1 \|\vA_2 - \vA_1\|_{\infty;\Om}.
\eeqa

\item Contribution $ \vr_6$. We have, {\cyan for $B_0:=B_0(x,x+e_0)$},
\beqa
 -(\pa^*\vr_{\vA_1,6})_1(x) \2 = \2   \ppV_0 (\vA_1(x) \times  \vA_1(x+e_0)), \\ 
 -(\pa^*\vr_{\vA_2,6})_1(x) \2 =\2  \ppV_0 (\vA_2(x) \times  \vA_2(x+e_0)).
 \eeqa
Hence,
\beqa
 |(\pa^*\vr_{\vA_1,6})_1(x) -   \pa^*(\vr_{\vA_2,6})_1(x)| \leq \eps_1(|\vA_1(x) - \vA_2(x)| + |\vA_1(x+e_0) - \vA_2(x+e_0)|).
\eeqa
Consequently,
\beqa
 \|(\pa^*\vr_{\vA_1,6}) -   (\pa^*\vr_{\vA_2,6})\|_{\infty;\Om} \leq 8\eps_1 \|\vA_1 - \vA_2\|_{\infty;\Om}.
\eeqa

\item Contributions $\vr_7$, $\vr_8$, $\vr_9$. We have
\beqa
 -(\pa^*\vr_{\vA_1,7})_1(x)\2=\2 \vA_1(x+e_0) (\vA_1(x)\cdot \vB(x,x+e_0) ), \\ 
 -(\pa^*\vr_{\vA_2,7})_1(x)\2=\2 \vA_2(x+e_0) (\vA_2(x)\cdot \vB(x,x+e_0) ).
  \eeqa
Hence,
\beqa
 |(\pa^*\vr_{\vA_1,7})_1(x) -   (\pa^*\vr_{\vA_2,7})_1(x)| \leq \eps_1 (|\vA_1(x+e_0)- \vA_2(x+e_0)|+ |\vA_1(x)- \vA_2(x)|  ). 
\eeqa
Consequently,
\beqa
 \|(\pa^*\vr_{\vA_1,7}) -   (\pa^*\vr_{\vA_2,7})\|_{\infty;\Om} \leq 8\eps_1 \|\vA_1 - \vA_2\|_{\infty;\Om},
 \eeqa
where the factor $4$ comes from (\ref{lap-2}).

\end{itemize}
This concludes the proof. \qed

\bel Suppose $\vA_1,\vA_2, \vC_1, \vC_2\in \real^3$ are such that $|\vA_j|\leq \eps\leq 1/2$ and $|\vC_j|\leq \ti{\eps}\leq 1/2$.
Then
\beqa
  \de(A_{j,0} C_{j',0})\2 \leq \2   \eps^2+\ti\eps^2,  \label{first-delta-estimate}\\
 |\de(A_{2,0}C_{2,0})- \de(A_{1,0}C_{1,0}) | \2 \leq \2  2\eps|\vA_2-\vA_1|+2\ti\eps |\vC_2-\vC_1|. \label{second-delta-estimate}
\eeqa
\eel
\proof Concerning~(\ref{first-delta-estimate}), we have by (\ref{delta-form})
\beqa
\de(A_{j,0} C_{j',0})=\fr{\vec{C}_{j'}^2+\vec{A}_{j}^2 -\vec{C}_{j'}^2\vec{A}_{j}^2  }{1+A_{j,0}C_{j',0}}\leq  \eps^2+\ti{\eps}^2,
\eeqa
where we estimated $-\vec{C}_{j'}^2\vec{A}_{j}^2 \leq 0$. Regarding~(\ref{second-delta-estimate})
we have
\beqa
 \de(A_{2,0}C_{2,0}) -\de(A_{1,0}C_{1,0})= \sqrt{1-\vA^2_1}\sqrt{1-\vC^2_1} - \sqrt{1-\vA^2_2}\sqrt{1-\vC^2_2}.
\eeqa
We write $\vA_{2}=:\vA_{1}+\de \vA$, $\vC_{2}=:\vC_{1}+\de \vC$ and obtain
\beqa
\de(A_{2,0}C_{2,0}) -\de(A_{1,0}C_{1,0}) =   \sqrt{1-\vA^2_1}\sqrt{1-\vC^2_1} - \sqrt{1-(\vA_1+\de \vA )^2}\sqrt{1-(\vC_1+\de \vC)^2}.
\eeqa
We define a function of {\cyan $t\in [0,1]$}
\beqa
F(t):=\sqrt{1-\vA^2_1}\sqrt{1-\vC^2_1} - \sqrt{1-(\vA_1+t\de \vA )^2}\sqrt{1-(\vC_1+t\de \vC)^2},
\eeqa
which satisfies $F(0)=0$. Thus we can write $F(1)=\int_0^1dt\, \pa_t F(t)$ which gives
\beqa
F(1)= \int_0^1 dt \bigg( \fr{  \de \vA \cdot (\vA_1+t\de \vA )  }{\sqrt{1-(\vA_1+t\de \vA )^2 } } \sqrt{1-(\vC_1+t\de \vC)^2} +\{A \leftrightarrow C\}   \bigg). 
\eeqa
Hence
\beqa
|F(1)|\leq 2( |\de \vA| \eps +|\de \vC| \ti{\eps}),
\eeqa
which concludes the proof.  \qed
\subsection{The main result}

We summarize the considerations of this paper:
\bet\label{main-theorem-intext} Let  $\G_0=SU(2)$. Then there exist $0 < \eps, \eps_1\leq 1$  s.t.  for  $V\in \U_{\eps_{1}}(\Om_1)$ the action  $\mcA$ has a unique critical point over $\Conf_{\eps}(\Om)$ with the constraint $\mcC(U)=V$. The parameters $\eps,\eps_1$ are independent of $n$ but may depend on $L$.
\eet
\proof Follows from  Theorem~\ref{critical-point-thm}
and Propositions~\ref{preserving-space}, \ref{contraction}. \qed

\appendix
\section{Random walk expansions} \label{random-walk-appendix}
\setcounter{equation}{0}

In this appendix we reproduce  Theorem 4.1 and its proof from Section IV of \cite{BJ85} with some modifications. 
We work on an infinite lattice $\mathbb{Z}^d$ and keep the dimension $d$ arbitrary.   Also,
in this appendix we simply write $\|\,\cdot\,\|_p$ for the norm in $\mcL^p(\mathbb{Z}^d)$, $1\leq p\leq \infty$, and
 $\|\,\cdot\,\|_{\op}:=\|\,\cdot\,\|_{2,2;\mathbb{Z}^d}$ for the $C^*$-norm on bounded operators on $\mcL^2(\mathbb{Z}^d)$.

\subsection{Short range localizing functions}
\newcommand{\n}{\mrm{n}}
\newcommand{\C}{\mrm{C}}
\newcommand{\A}{\mrm{A}}

\bed\label{short-range-localizing} A function\footnote{We choose $\real^d$ and not $\mathbb{Z}^d$ as a domain, because we want to
consider $b(\veps x)$ in (\ref{eps-def}).} $a:\real^d\to \real$ is short range localizing if 
\begin{enumerate}

\item[(i)] $0<a(x)$.

\item[(ii)] $a(x) \leq c_{\de}(1+|x|)^{-2(d+\de)}$ for some $\de>0, c_{\de}\geq 0$.

\item[(iii)] For $\de$ as above, define 
\beqa
b(x):=(1+|x|)^{d+\de} a(x).  \label{b-def}
\eeqa
We assume that  for some constant $K<\infty$, which may depend on $d$ and $\de$,
\beqa
\fr{b(x+y)}{b(x)} \leq K  \label{K-bounds}
\eeqa
for all $x$ and for $|y|\leq 2d^{1/2}$.

\item[(iv)] For $\delta$ as above there are constants $c$ and $\veps>0$ s.t. for all $\n$ 
\beqa
(\underbrace{b*b*\cdots *b}_{\n})(x)\leq c^{\n} b(\veps x), \label{eps-def}
\eeqa
{\blue where $(b*b)(x):=\sum_{y\in \mathbb{Z}^d} b(x-y)b(y)$, $x\in \real^d$.}

\item[(v)]  The function $b$ is eventually decreasing. That is, there exists $M_0\in \nat$ sufficiently large, possibly depending {\cyan on} $\de$, s.t. $b(x')\leq b(x)$ for all $x,x'\in \real^d$ satisfying $M_0\leq |x|\leq |x'|$.

\end{enumerate}

\eed
\nin In comparison with \cite{BJ85} we removed a lower bound by $K^{-1}$ in  (\ref{K-bounds}) and added item (v). The latter property
has the following  consequence {\dgreen which will be used in (\ref{no-covolutions}) below:}
\bel\label{b-scaling} Let $b$ be as in Definition~\ref{short-range-localizing}  above. Then, for $\ti{M}\in \nat$ sufficiently large we have 
\beqa
b(\ti{M}x / {\blue 3})\leq b(x), \quad x\in \mathbb{Z}^d.
\eeqa
\eel
\proof For $|x|\geq {\blue 3}M_0$ from property (v) we clearly have that $b(\ti{M}x /{\blue 3})\leq b(x)$, for any $\ti{M}\in \nat$. Now the set $S:=\{\, x\in \mathbb{Z}^d  \,|\,|x|\leq {\blue 3}M_0\,\}$
is finite and $b$ tends to zero as $|x|\to \infty$ by property (ii). Thus we can  find a sufficiently large $\ti{M}$ s.t. $b(\ti{M}x/ {\blue 3})\leq b(x)$
for all $x\in S$, $x\neq 0$. As for $x=0$ the statement is trivial,  this concludes the argument. \qed
\bel\label{exponential-function}  Fix $c_1>0$. The function $a(x):=\e^{-c_1 |x|}$
is short range localizing in the sense of Definition~\ref{short-range-localizing}.
A possible choice of parameters is $\de=c_1$, $M_0=2/c_1$, $\veps=1/4$.
\eel
\proof Property (i) is obvious. Regarding property (ii), we note that for any $\de>0$ there is such $c_{\de}$ that
\beqa
(1+|x|)^{2(d+\de)}\e^{-c_1 |x|} \leq c_{\de}.
\eeqa
To check (iii) we fix $\de>0$ and define $b(x):=(1+|x|)^{d+\de} a(x)$. We note that
\beqa
\e^{-c_1|x+y|} \2\leq\2 \e^{-c_1|x|} \e^{c_12d^{1/2}},  \\
(1+|x+y|)^{\blue d+\de} \2\leq \2 (1+|x|+2d)^{d+\de}, 
\eeqa
which gives the  bound in (\ref{K-bounds}). 
Now we move on to property~(iv). First, we note that
\beqa
b(x)=(1+|x|)^{d+\de} a(x)=\big( (1+|x|)^{d+\de} \e^{-\h c_1 |x|}\big) \e^{-\h c_1 |x|} \leq c_{\de}^{1/2}  a((1/2) x)=:c_{\de}^{1/2} a_{1/2}(x).
\eeqa 
Consequently, we have
\beqa
(\underbrace{b*b*\cdots *b}_{\n})(x) \2\leq\2 c_{\de}^{\n/2} (\underbrace{a_{1/2}*\ldots * a_{1/2}}_{\n} )(x) \non\\
\2=\2 c_{\de}^{\n/2} \sum_{x_1,\ldots,x_{\n-1} } \e^{-\h c_1|x-x_1|} \e^{-\h c_1|x_1-x_2|} \ldots \e^{-\h c_1|x_{\n-2}-x_{\n-1}| } \e^{ -\h c_1|x_{\n-1}| }.
\eeqa
{\blue We rewrite each exponential function  above as $ \e^{-\h c_1|x_i-x_j|}= \e^{-\fr{1}{4} c_1|x_i-x_j|}  \e^{-\fr{1}{4} c_1|x_i-x_j|}$ and
treat the two factors separately.}
We observe that, by the inverse triangle inequality $|x-x_1|+|x_1-x_2|\geq |x|-|x_1|+|x_1|-|x_2|$ and so on. Hence
\beqa
 \e^{-\fr{1}{4} c_1|x-x_1|} \e^{-\fr{1}{4} c_1|x_1-x_2|} \ldots \e^{-\fr{1}{4} c_1|x_{\n-2}-x_{\n-1}| } \e^{ -\fr{1}{4} c_1|x_{\n-1}| } \leq \e^{-\fr{1}{4} c_1|x|}.
 \eeqa
On the other hand $\|f_1*f_2\|_{\infty}\leq \|f_1\|_{\infty} \|f_2\|_1$ and $\|g_1*g_2\|_1\leq \|g_1\|_1 \|g_2\|_1$. Hence
\beqa
\|\underbrace{a_{1/4}*\ldots *a_{1/4}}_{\n}\|_{\infty} \leq \|a_{1/4}\|_{\infty} \|a_{1/4}\|_1^{\n-1}. 
\eeqa
Since $\|a_{1/4}\|_{\infty}=1$, this gives
\beqa
(\underbrace{b*b*\cdots *b}_{\n})(x) \2\leq\2 (c_{\de}^{1/2} (1+ \|a_{1/4}\|_1))^{\n}  \e^{-\fr{1}{4} c_1|x|}\leq (c{}'_{\de})^{\n}\big(1+\fr{1}{4} |x|\big)^{d+\de}   \e^{-\fr{1}{4} c_1|x|}
\eeqa
and yields the proof of (iv) {\blue with $\eps=1/4$}.   Regarding (v), we consider the following function and its derivative 
\beqa
f(w):=(1+w)^{d+\de} \e^{-c_1w}, \quad f'(w)=(d+\de)(1+w)^{d+\de-1} \e^{-c_1w}+(1+w)^{d+\de}(-c_1) \e^{-c_1w}.
\eeqa
Thus $f$ is decreasing, provided that
\beqa
(d+\de) (1+w)^{-1} -c_1\leq 0 \quad \Rightarrow \quad  w\geq \fr{d+\de}{c_1} -1=: M'_0.
\eeqa
We can choose as $M_0$ the smallest natural number larger than $M_0'$. This concludes the argument. \qed

\subsection{Exponential decay of integral kernels of operators and their inverses}

\begin{figure}[t]\centering
\begin{tikzpicture}[scale=0.2]
        \begin{scope}
         \foreach \x in {-13,-12,...,13}{                           
          \foreach \y in {-13,-12,...,13}{                       
           \node[draw,circle,inner sep=0.1pt,fill] at (\x,\y) {}; 
          }
         }
         \draw [very thick] (-10,-10) -- (0,-10) -- (0,0) -- (-10,0) -- cycle;
         \foreach \x in {-2,-1,...,2}{                           
          \foreach \y in {-2,-1,...,2}{                       
         \node[draw,circle,inner sep=1.5pt,fill, color=blue] at (5*\x,5*\y) {}; 
          }
         }

        \end{scope}
\end{tikzpicture}
\caption{The centers of $2\ti{M}$-boxes in $\mathbb{Z}^2$ as blue (thick) dots, with one of the $2\ti{M}$-boxes indicated.}
\label{fig:boxesandpoints}
\end{figure}
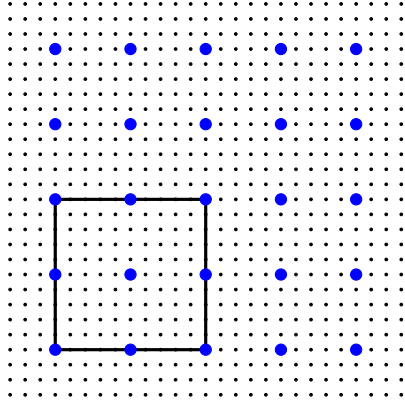

We start with some definitions:
\bed\label{boxes-def} Let $\{ \Box^j\}_{j\in \mathbb{Z}^d }$ be a cover of $\real^d$ by  {\blue closed } $2\ti{M}$-cubes where $\Box^j$ 
has a  center at   $j\ti{M}$, $j\in \mathbb{Z}^d$,  and $\ti{M}\in \nat$.   Thus cubes $\Box^i$, $\Box^j$
overlap if $|i-j|\leq 2 d^{1/2}$.  A given cube $\Box^i$ overlaps with $5^d$ neighbouring cubes, including itself. {\blue See {\magenta Figure} \ref{fig:boxesandpoints}.}

We set $\Box_j:= {\blue \mathbb{Z}^d} \cap \Box^j$.
We also let $\Box_j$ be the projection of $\mcL^2(\mathbb{Z}^d)$
onto $\mcL^2(\Box_j)$. The projections are orthogonal for $|i-j|>2d^{1/2}$. 
\eed
\bed Let  $h_j^2$, $j\in \mathbb{Z}^d$, be a smooth partition of unity on  $\real^d$ constructed
as follows: Let $h\in C_0^{\infty}(\real)$ be positive and  s.t.
\beqa
h(t) := \left\{ \begin{array}{ll}
1 & \textrm{for}\quad |t|\leq 1/3, \\
0 & \textrm{for}\quad  |t|\geq 2/3,
\end{array} \right. \label{smooth-part}
\eeqa
and 
\beqa
\sum_{j\in \mathbb{Z}} h^2(t-j)=1, \quad \sup|h'(t)|\leq 10.
\eeqa
Now we define the  functions on $\mathbb{Z}^d$
\beqa
h_j(x):=\prod_{k=0}^{d-1} h\bigg( \fr{x_k}{\ti M} -j_k \bigg)
\eeqa 
which satisfy
\beqa
h_j\Box_j=h_j, \quad \sum_{j\in \mathbb{Z}^d} h^2_j(x)=1. \label{two-h-relations}
\eeqa
\eed

\bet\label{Random-walk}  Let $\A$ on $\mcL^2(\mathbb{Z}^d)$ be strictly positive  and let 
\beqa
|\A(x,x')|\leq a(x-x'),
\eeqa
where $a$ is a short range localizing function (see Definition~\ref{short-range-localizing}).  Now fix $\de>0$
s.t.  $a(x) \leq c_{\de}(1+|x|)^{-2(d+\de)}$ for some finite $c_{\de}$ and define $b(x)=(1+|x|)^{d+\de} a(x)$.
Then, for $\veps$ as in (\ref{eps-def}), and $\ti{M}$ as in {\blue  Definition \ref{boxes-def}}, sufficiently large, we have
\beqa
|\A^{-1}(x,x')|\leq  c\, b( \veps (x-x')/{\ti{M}} ).
\eeqa
\eet
\proof  By assumption, there exists $\m^2>0$  s.t.  
$\m^2 \leq \A$. Consequently, 
\beqa
\m^2\Box_j \leq \Box_j \A \Box_j,
\eeqa
hence $\Box_j A\Box_j$ is invertible on $\mcL^2(\Box_j)$. We can define
\beqa
\C_j:=(\Box_j \A \Box_j )^{-1} |_{\mcL^2(\Box_j)}, \quad \C:=\sum_{j} h_j \C_j h_j,
\eeqa
where {\blue  $h_j$ is a multiplication operator.  We treat $\C$ as   an approximate inverse of $\A$ and define the corresponding rest term $\R$
by the relation}
\beqa
\A\C=:1-\R.
\eeqa
Now we study $\A^{-1}(x,x')$ using the series (\ref{series}) from 
Proposition~\ref{aux-proposition}. We can parametrize the terms in the expansion for $\A^{-1}(x,x')$
arising from (\ref{series}) by  paths $\om=\{\om_0,\om_1, \ldots, \om_{2\n} \}$, where $\om_j\in \mathbb{Z}^d$
indexes a cube $\Box_{\om_j}$ centered at $\ti{M}\om_j$. Here $x\in \Box_{\om_0}$, $x'\in \Box_{\om_{2\n}}$
and, using that  $\R=\sum_{i,j} \R_{ij} \C_j h_j$, (see (\ref{def-of-R}) below), 
\beqa
A^{-1}\2=\2\sum_{\n=0}^{\infty} \C \R^{\n}\non\\ 
 \2=\2 \sum_{\n=0}^{\infty}\big( \sum_{i_0} \sum_{i_1,j_1} \ldots \sum_{i_{\n},j_{\n}}\big) (h_{i_0}\C_{i_0}  h_{i_0})   (\R_{i_1 j_1} \C_{j_1} h_{j_1}) (\R_{i_2 j_2} \C_{j_2} h_{j_2})
 \ldots (\R_{i_{\n} j_{\n}} \C_{j_{\n}} h_{j_{\n}})\non\\
\2=\2  \sum_{\n=0}^{\infty}\big( \sum_{\om_0,\om_1, \ldots, \om_{2\n-1},\om_{2\n} }\big) (h_{\om_0}\C_{\om_0}  h_{\om_0} )   (\R_{\om_1 \om_2} \C_{\om_2} h_{\om_2}) 
(\R_{\om_3 \om_4} \C_{\om_4}  h_{\om_4})\ldots (\R_{\om_{2\n-1} \om_{2\n}} \C_{\om_{2\n}} h_{\om_{2\n}}).\quad\quad \label{expansion-R}
\eeqa
The $\n$-th summand vanishes, unless for all $i=0,1,\ldots, \n-1$
\beqa
|\om_{2i} -\om_{2i+1} |\leq 2d^{1/2}. \label{overlap-cond}
\eeqa
Indeed, since $\R_{\om_{2i+1} ,\om_{2i+2}}: \mcL^2(\Box_{2i+2}) \to \mcL^2(\Box_{2i+1})$, we have that $h_{\om_{2i}} \Box_{\om_{2i+1}}=0$ if (\ref{overlap-cond}) fails,
as then the cubes $\Box_{\om_{2i}}$ and $\Box_{\om_{2i+1}}$  are disjoint.

We have, using that $\|\C_j\|_{\op}\leq \m^{-2}$  and $\|h_j\|_{\op}=1$,
\beqa
|\A^{-1}(x,x')|\leq \sum_{ \{\om \}_{x,x'} } \m^{-(2\n+2)}   \|\R_{\om_1,\om_2}\|_{\op}\, \| \R_{\om_3,\om_4}\|_{\op} \ldots \| \R_{\om_{2\n-1},\om_{2\n}}\|_{\op}+|\C(x,x')|,
\label{A-1-norms}
\eeqa
where {\blue the last term is the $\n=0$ case},  $\{\om \}_{x,x'}$ are all paths satisfying the conditions (\ref{overlap-cond}) and 
\beqa
 x\in \Box_{\om_0}, \quad x'\in \Box_{\om_{2\n}}. \label{boundary-terms} 
\eeqa
We remark that for a fixed $x$ there are less than $5^d$ boxes which may contain it. We also remark, that for $x\in \Box_{\om_0}$ we have
\beqa
\big|\fr{x}{\ti{M}}-\om_0\big|_{\infty}\leq 1\,\quad \Rightarrow \quad \big|\fr{x}{\ti{M}}-\om_0\big| \leq \sqrt{d}.
\eeqa
Let us write explicitly  the $\n=4$ term of (\ref{A-1-norms}).  Using that, by Lemma~\ref{R-i-j-lemma} below, $\|\R_{i,j}\|_{\op} \leq O(\ti{M}^{-\de}) b_{\ti{M}/3}(i-j)$,
we get
\beqa
|\A^{-1}(x,x')^{(4)}| \2\leq\2 \sum_{ \{\om_0,\ldots, \om_8\}_{x,x'} } 
O(\ti{M}^{-\n \de/2}) \m^{-(2\n+2)} b_{\ti{M}/3}(\om_1-\om_2)  b_{\ti{M}/3}(\om_3-\om_4)  \times \non\\
  \2 \2\phantom{444444444444444444444444444444444444} \times b_{\ti{M}/3}(\om_5-\om_6) b_{\ti{M}/3}(\om_7-\om_8)\non\\
\2\leq\2  \sum_{ \{\om_0,\ldots, \om_8\}_{x,x'} } 
O(\ti{M}^{-\n \de/2}) \m^{-(2\n+2)} {\blue b(\om_1-\om_2)  b(\om_3-\om_4)  b(\om_5-\om_6) b(\om_7-\om_8)},\quad  \label{no-covolutions}
\eeqa
where in the second step we exploited Definition~\ref{short-range-localizing} (v) and Lemma~\ref{b-scaling}.  In the next step
we would like to express (\ref{no-covolutions}) as a convolution of functions $b$ to be able to apply property (\ref{eps-def}):
\beqa
(\ref{no-covolutions}) 
\2=\2  \sum_{ \{\om_0,\ldots, \om_8\}_{x,x'} } 
O(\ti{M}^{-\n \de/2}) \m^{-(2\n+2)} b(x/\ti{M}-\om_2)  b(\om_2-\om_4)  b(\om_4-\om_6) b(\om_6-x'/\ti{M})\times \non\\
\2 \2 \phantom{444444444}\times\bigg\{ \fr{b(\om_0-\om_2)}{b(x/\ti{M}-\om_2)}    \bigg\} \bigg\{ \fr{{\blue b(\om_1-\om_2)}}{b(\om_0-\om_2) } \bigg\} \bigg\{ \fr{\blue  b(\om_3-\om_4)}{b(\om_2-\om_4) } \bigg\}
  \times \non\\
\2 \2\phantom{44444444444444444444444}  \times \bigg\{ \fr{\blue  b(\om_5-\om_6)}{b(\om_4-\om_6) } \bigg\}\bigg\{ \fr{\blue b(\om_7-\om_8)}{b(\om_6-\om_8) } \bigg\}  \bigg\{ \fr{b(\om_6-\om_8)}{b(\om_6-x'/\ti{M}) } \bigg\}. \label{curly-brackets}
\eeqa
Now we can use Definition~\ref{short-range-localizing} (iii) to estimate the expressions in curly bracket. We have, for example,
\beqa
 \bigg\{ \fr{b(\om_7-\om_8)}{b(\om_6-\om_8) } \bigg\}=\bigg\{ \fr{b((\om_7-\om_6)+\om_6 -\om_8)}{b(\om_6-\om_8) } \bigg\} \leq K,
\eeqa
since $|\om_6-\om_7|\leq 2d^{1/2}$ by (\ref{overlap-cond}), and similarly for the remaining brackets. After estimating the curly brackets in (\ref{curly-brackets}), we can sum over $\om_i$, $i$-odd, and over the boundary terms $\om_0$, $\om_{2\n}$. From (\ref{expansion-R}) and the property that $\R_{i,j}: \mcL^2(\Box_j) \to \mcL^2(\Box_i)$
we see that, for example, the sum over $\om_1$ runs at most over $5^d$ boxes which overlap with $\om_0$.   We can argue analogously for the boundary terms, using (\ref{boundary-terms}).  Thus we have
\beqa
|\A^{-1}(x,x')^{(4)}| \2\leq\2 \sum_{ \{\om_2,\om_4,\om_6\} } O(\ti{M}^{-\n \de/2}) \m^{-(2\n+2)} (K 5^{d})^{\n+2} b(x/\ti{M}-\om_2)  b(\om_2-\om_4)\times \non\\
\2 \2 \phantom{4444444444444444444444444444444444} \times b(\om_4-\om_6) b(\om_6-x'/\ti{M}).
\eeqa
 Let us relax the conditions (\ref{overlap-cond}) in $\{\om_2,\om_4,\om_6\}$. Now that $\om_6$ runs over $\mathbb{Z}^d$, we can change variables $\om_6'=\om_6-x'/\ti M$,  which gives
 \beqa
 |\A^{-1}(x,x')^{(4)}| \2\leq\2 \sum_{  \{\om_2,\om_4,\om_6'\}   } O(\ti{M}^{-\n \de/2}) \m^{-(2\n+2)} (K 5^{d})^{\n+2} b(x/\ti{M}-\om_2)  b(\om_2-\om_4)\times \non\\ 
  \2 \2\phantom{44444444444444444444444444444444} \times b(\om_4-x'/\ti{M}- \om_6') b(\om_6') \quad\quad\non\\
\2=\2  O(\ti{M}^{-\n \de/2}) \m^{-(2\n+2)}(K 5^{d})^{\n+2}(b*b*b*b)((x-x')/ \ti{M} )\non\\
\2\leq \2  O(\ti{M}^{-\n \de/2}) \m^{-(2n+2)}  (K 5^{d})^{\n+2}   c^{\n} b( \veps (x-x') / \ti{M} ),
 \eeqa
 where we made analogous changes of variables in $\om_4, \om_2$ and in the last step we applied Definition~\ref{short-range-localizing}~(iv).
  
Guided by the above example, one immediately obtains in the general case
\beqa
|A^{-1}(x,x')|\leq |\C(x,x')|+b( \veps (x-x')/\ti{M}) \sum_{\n=1}^{\infty} O(\ti{M}^{-\n\de/2})\m^{-(2\n+2)} (K 5^{d})^{\n+2}  c^{\n}.
\eeqa
Noting that  the sum is finite for $\ti M$ sufficiently large and that $\C(x,x')=0$ unless $x,x'$ belong to the same box, we conclude the proof. \qed

\bep\label{aux-proposition} For $\ti{M}$ sufficiently large, we have $\|\R\|_{\op} \leq O(\ti{M}^{-\de/2})<1$.
Therefore,
\beqa
\A^{-1}=\C ({\cyan 1}-\R)^{-1}=\sum_{\n=0}^{\infty} \C \R^{\n} \label{series}
\eeqa
is a  convergent series in the operator norm $\|\,\cdot\,\|_{\op}$.
\eep
\proof Let us expand the product $\A\C$ as 
\beqa
\A\C\2=\2\sum_{j} \A h_j \C_j h_j\non\\
      \2=\2 \sum_j \Box_j \A h_j\C_j h_j+\sum_j (1-\Box_j) \A h_j \C_j h_j\non\\
      \2=\2 \sum_j h_j\Box_j \A \Box_j \C_j h_j +\sum_j \Box_j [\A, h_j] \Box_j \C_j h_j +\sum_j (1-\Box_j) \A h_j \C_j h_j, \label{AC-exp}
      \eeqa
where in the last step we used $h_j\Box_j=\Box_j h_j=h_j$, cf. (\ref{two-h-relations}).  
By $\C_j:=(\Box_j \A \Box_j )^{-1} |_{\mcL^2(\Box_j)}$ the first term in (\ref{AC-exp}) sums up to one. 
Since $\A\C={\cyan 1}-\R$, the last two terms in (\ref{AC-exp})
coincide with $\R$. Thus we have
\beqa
\R
\2=\2 \sum_j \Box_j [h_j, \A] \Box_j \C_j h_j +\sum_{j,i} (\Box_j-1) h_i^2 \A h_j \C_j h_j.
\eeqa
We note that $(\Box_j-1) h_i$ vanishes for $i=j$. Thus we can write
\beqa
\R=\sum_{i,j} \R_{i,j}\C_j h_j,  \label{def-of-R}
\eeqa
where 
\beqa
\R_{i,j}:=\de_{i,j}\Box_j [h_j, \A] \Box_j+  (1-\de_{i,j}) (\Box_j-1) h_i^2 \A h_j.  \label{R-i-j}
\eeqa
Now by Lemmas~\ref{aux-lemma-two}, \ref{R-i-j-lemma} we obtain
\beqa
\|\R\|_{\op} \leq O(\ti{M}^{-\de/2}) 2^d \m^{-2} \sup_i\sum_j b(\ti{M}(i-j)/3)\leq O(\ti{M}^{-\de/2}),
\eeqa
where in the last step we used  $b(x):=(1+|x|)^{d+\de}a(x) \leq (1+|x|)^{-(d+\de)}$ to show that the sum is finite
uniformly in $\ti{M}$. \qed

\bel\label{aux-lemma-one} Let $T(x,x')$  be the kernel of $T$ on $\mcL^2(\mathbb{Z}^d)$. Then
\beqa
\|T\|_{\op} \leq \bigg(\sup_x \sum_{x'} |T(x,x')| \bigg)^{1/2}  \bigg(\sup_{x'} \sum_x |T(x,x')| \bigg)^{1/2}.
\eeqa
\eel
\proof The bound is standard and elementary. See e.g. \cite[Lemma B.6.1]{DG}. \qed

\begin{figure}[t]\centering
\begin{tikzpicture}[scale=0.75]
        \begin{scope}
        \draw [thick, ->] (-6,2) --(6,2);
        \draw [thick, ->] (-6,0) --(6,0);
        \draw [dotted, very thick, ->] (-4,0) -- (-4,2);
        \draw [dotted, very thick, ->] (-1,0) -- (-1,2);
        \draw [dotted, very thick, ->] (2,0) -- (2,2);
        \draw [dotted, very thick, ->] (-3,-3) -- (-3,2);
        \draw [dotted, very thick, ->] (0,-3) -- (0,2);
        \draw [dotted, very thick, ->] (3,-3) -- (3,2);
        \draw [dotted, very thick, ->] (-2,-6) -- (-2,2);
        \draw [dotted, very thick, ->] (1,-6) -- (1,2);
        \draw [dotted, very thick, ->] (4,-6) -- (4,2);
        \draw [very thick] (-5,-1) -- (-3,-1) -- (-3,1) -- (-5,1) -- cycle;
        \draw [very thick] (-2,-1) -- (0,-1) -- (0,1) -- (-2,1) -- cycle;
        \draw [very thick] (1,-1) -- (3,-1) -- (3,1) -- (1,1) -- cycle;
        \draw [thick, ->] (-6,-3) --(6,-3);
        \draw [very thick] (-4,-4) -- (-2,-4) -- (-2,-2) -- (-4,-2) -- cycle;
        \draw [very thick] (-1,-4) -- (1,-4) -- (1,-2) -- (-1,-2) -- cycle;
        \draw [very thick] (2,-4) -- (4,-4) -- (4,-2) -- (2,-2) -- cycle;
        \draw [thick, ->] (-6,-6) --(6,-6);
        \draw [very thick] (-3,-7) -- (-1,-7) -- (-1,-5) -- (-3,-5) -- cycle;
        \draw [very thick] (0,-7) -- (2,-7) -- (2,-5) -- (0,-5) -- cycle;
        \draw [very thick] (3,-7) -- (5,-7) -- (5,-5) -- (3,-5) -- cycle;
        \end{scope}
\end{tikzpicture}
\caption{Three families of disjoint boxes in one dimension. Their union is the whole family of the $2\ti{M}$-boxes
centered at $\ti{M}\Z$ described in Definition \ref{boxes-def}.}
\label{fig:3families}
\end{figure}
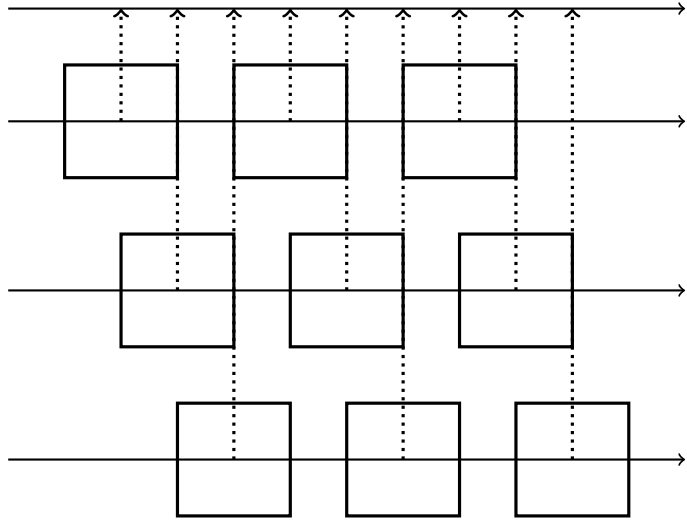

\bel\label{aux-lemma-two} Let $\R$ be as above. Then
\beqa
\|\R\|_{\op} \leq 2^d \m^{-2} \bigg( \sup_i \sum_j \|\R_{i,j}\|_{\op}\bigg)^{1/2} \bigg( \sup_j \sum_i \|\R_{i,j}\|_{\op}    \bigg)^{1/2}.
\eeqa
\eel
\proof First, we note that
\beqa
\sum_i \lan f, \Box_i f\ran{\blue \leq 3^d}\|f\|_{2}^2.
\eeqa
{\blue In fact, as these are $2\ti{M}$ boxes, in one dimension we could divide the sum into {\blue three} families
of shifted  projections s.t. in each family the projections are mutually orthogonal. In arbitrary dimension we can do it in every direction. 
See {\magenta Figure} \ref{fig:3families}.}

Next, since $\m^2  \leq \A$, we have $\m^2\Box_j\leq \Box_j \A\Box_j$, hence $\m^2 \C_j^{1/2} \Box_j \C_j^{1/2} \leq   \Box_j$.
Thus $\|\C_j\|_{\op}\leq \m^{-2}$. We can write, using $\R=\sum_{i,j} \R_{ij} \C_j h_j$, 
\beqa
|\lan f, \R g\ran|\2\leq\2 \sum_{i,j} \|\Box_i f\|_{2} \, \|\R_{ij} \C_j h_j\|_{\op}\, \|\Box_j g\|_{2} \non\\
                       \2\leq\2 \m^{-2}\sum_{i,j} (\|\Box_i f\|_{2}\, \|\R_{ij}\|_{\op}^{1/2}) \,  (   \|\R_{ij}\|_{\op}^{1/2}\|\Box_j g\|_{2})  \non\\
                       \2\leq \2 2^d \m^{-2} \bigg( \sup_i \sum_j \|\R_{i,j}\|_{\op}\bigg)^{1/2} \bigg( \sup_j \sum_i \|\R_{i,j}\|_{\op}    \bigg)^{1/2} \|f\|_2\|g\|_2,
\eeqa
 where we applied the Cauchy-Schwarz inequality in the last step. \qed
\bel\label{R-i-j-lemma} Under the assumptions of Theorem~\ref{Random-walk} and for $\ti{M}$ sufficiently large
\beqa
\|\R_{i,j}\|_{\op}\leq O(\ti{M}^{-\de/2}) b(\ti{M}(i-j)/3)
\eeqa
with $\de$, $b$ of (\ref{b-def}). {\blue (Here $O(\epsilon)$ denotes a number satisfying $|O(\epsilon)| \leq c\epsilon $). }
\eel
\begin{remark} {\dgreen This lemma only requires properties (i), (ii) {\cyan and (v)} from Definition~\ref{short-range-localizing}.}
\end{remark}
\proof We start with the $i=j$ case:
\beqa
(\R_{i,i})(x,x')=(\Box_i [h_i, A] \Box_i)(x,x')=\A(x,x')(h_i(x)-h_i(x')) \Box_i(x)\Box_i(x'). \label{R-i-i}
\eeqa
As $h$ has a bounded derivative, we obtain 
\beqa
|h_i(x)-h_i(x')|\leq O(\ti{M}^{-\de/2})|x-x'|^{\de/2}. \label{h-difference}
\eeqa
Here we used the definition $ h_j(x):=\prod_{k=0}^{d-1} h\big( \fr{x_k}{\ti M} -j_k \big)$, the identity
\beqa
\bigg|\fr{h_i(x)-h_i(x')}{|x-x'|^{\de/2}} \bigg|=\bigg|\fr{h_i(x)-h_i(x')}{|x-x'|} \bigg|^{\de/2} \big| h_i(x)-h_i(x')\big|^{1-\de/2}
\eeqa
and the boundedness of $h$.

Now we recall that, by assumption, $|\A(x,x')|\leq a(x-x')$ and   $a(x-x')=(1+|x-x'|)^{-(d+\de)} b(x-x')$. Hence, by (\ref{R-i-i}), (\ref{h-difference}),
\beqa
|\R_{i,i}(x,x')|\2\leq\2 O(\ti{M}^{-\de}) |x-x'|^{\de/2} |\A(x,x')|\leq O(\ti{M}^{-\de/2}) |x-x'|^{\de/2} a(x-x') \non\\
\2\leq\2   O({\blue \ti{M}}^{-\de/2}) (1+|x-x'|)^{-(d+\de/2)}b(x-x'), \label{i-i-case} 
\eeqa
{\dgreen where we expressed $a$ by $b$ in order to compensate the factor $|x-x'|^{\de/2}$.} 
Therefore, by Lemma~\ref{aux-lemma-one} and the fact that $b(x)\leq c(1+|x|)^{-(d+\de)}$
\beqa
\|\R_{i,i}\|_{\op}\leq O(\ti{M}^{-\de/2}) \sup_{x}\sum_{x'} b(x-x')=O(\ti{M}^{-\de/2})\sum_x b(x)=O(\ti{M}^{-\de/2}).
\eeqa
This concludes the discussion of the $i=j$ case.

 For $i\neq j$ we infer from (\ref{R-i-j}) that the expression can only be different from zero if
\beqa
x\in \supp(h_i)\cap (\mathbb{Z^d} \backslash \Box_j)\quad and \quad x'\in \supp\, h_j.
\eeqa
With this restriction we have $|x-x'|>\fr{1}{3} \ti{M} |i-j|$ by referring to (\ref{smooth-part}). 
We infer from $\R_{i,j}:=(\Box_j-1) h_i^2 A h_j$
that 
\beqa
|\R_{i,j}(x,x')|\leq |\A(x,x')|\leq a(x-x')=(1+|x-x'|)^{-(d+\de)} b(x-x'). \label{factors-R}
\eeqa
In the next step we apply Lemma~\ref{aux-lemma-one} 
\beqa
\|\R_{i,j}\|_{\op} \2\leq\2 \bigg( \sup_x\sum_{x'} |\R_{i,j}(x,x')|\bigg)^{1/2} \bigg( \sup_{x'} \sum_x |\R_{i,j}(x,x')|    \bigg)^{1/2}\non\\
\2\leq \2  O(\ti{M}^{-\de/2}) b(\ti{M}(i-j)/3),
\eeqa
where the factors $(1+|x-x'|)^{-(d+\de)}$ from (\ref{factors-R})  controlled the summation
which is  over $x\neq x'$  with the restriction $|x-x'|>\fr{1}{3} \ti{M} |i-j|$. More precisely, we used that Definition~\ref{short-range-localizing}~(v)
gives
\beqa
\sum_{x'} |\R_{i,j}(x,x')| \2\leq\2 \sum_{x'} (1+|x-x'|)^{-(d+\de)} b(x-x')\non\\
\2\leq\2 {\dgreen O(\ti{M}^{-\de/2})} b(\ti{M} |i-j|/3 ) \sum_{x'} (1+|x'|)^{-(d+\de/2)}. 
\eeqa
As the last sum is finite, this concludes the proof. \qed

\section{Some technical lemmas}\label{Lemma-SU(2)}
\setcounter{equation}{0}

\bel\label{SU(2)-lemma} If $U\in\Conf(\Om)$ then $\mcC(U)\in  \Conf(\Om_1)$.
\eel
\proof Let $c_1,c_2>0$, $U_1, U_2\in SU(2)$. We show that there exists $c_3\geq 0$ and $U_3\in SU(2)$ s.t.
\beqa
c_1U_1+c_2U_2=c_3U_3. \label{addition-of-unitaries}
\eeqa
Indeed, using  representation  (\ref{SU2}), we write $U_j=\del_jA_{j,0}{\cyan 1}+\i \vec{A}_j\cdot \vec{\si}$, 
$A_{j,0}:=\sqrt{1- |\vec{A}_j|^2}$, $\del_j\in \{ \pm 1\}$. Then
\beqa
c_1U_1+c_2U_2= (c_1\del_1A_{1,0}+c_2\del_2A_{2,0}){\cyan 1} + \i ( c_1\vec{A}_1+c_2\vec{A}_2)\cdot \vec{\si}.  
\eeqa
From this we read off conditions on $c_3$, $U_3$:
\beqa
\2 \2 c_1\vec{A}_1+c_2\vec{A}_2= c_3\vec{A}_3,  \label{sum-of-vectors} \\
\2 \2 c_1\del_1\sqrt{1- |\vec{A}_1|^2} +c_2\del_2\sqrt{1- |\vec{A_2}|^2}  = \del_3  \sqrt{c_3^2 - ( c_1\vec{A}_1+c_2\vec{A}_2    )^2 }. \label{sum-of-zero-components}
\eeqa
The latter equation gives 
\beqa
\2 \2c_3^2 - ( c_1\vec{A}_1+c_2\vec{A}_2    )^2= c_1^2 (1- |\vec{A}_{1}|^2) + c_2^2 (1- |\vec{A}_{2}|^2) + 2c_1c_2\del_1\del_2\sqrt{1- |\vec{A}_{1}|^2 } \sqrt{1- |\vec{A}_2 |^2}, \quad\quad\\
\2 \2c_3^2  = c_1^2 +c_2^2 +2c_1c_2\del_1\del_2\sqrt{1- |\vec{A}_{1}|^2} \sqrt{1- |\vec{A}_{2}|^2 } + 2 c_1c_2 \vec{A}_1\cdot \vec{A}_2.
\eeqa
Clearly the r.h.s. of the last equation is {\blue non-negative} as the mixed terms sum up to a  scalar product $A_1\cdot A_2$ of two unit Euclidean four-vectors.  Given $c_3$,
we determine $\vec{A}_3$ via (\ref{sum-of-vectors}) and $\del_3$ via (\ref{sum-of-zero-components}).  {\blue If $c_3=0$, by convention $\vec{A}=0$ and $U_3=1$.} This concludes the proof of (\ref{addition-of-unitaries}).
Now, by iterating this equality, we get
\beqa
 \mcC_0(U)(y)= \fr{1}{L^2}\sum_{x\in \B(y)} U(x)=  c U
\eeqa
for some $U\in SU(2)$ and $c\geq 0$. This completes the proof of the lemma. \qed

\bel\label{square-root} Let $M$ be a self-adjoint operator on a Hilbert space  with $\|M\|_{\op}<1/2$. Then
\beqa
\|(1+M)^{1/2}-1\|_{\op}\leq \ch\|M\|_{\op},
\eeqa
where $\ch\geq 1$ is a numerical constant and $\|\,\cdot\,\|_{\op}$ is the operator norm.
\eel
\proof We recall the standard representation
\beqa
(1+M)^{-1/2}=\fr{1}{\pi} \int_0^{\infty}  dy\fr{1}{\sqrt{y}} \fr{1}{y+1+M}.
\eeqa
This gives
\beqa
(1+M)^{1/2}-1\2=\2\fr{1}{\pi} \int_0^{\infty}  dy\fr{1}{\sqrt{y}} \bigg(\fr{1+M}{y+1+M}-\fr{1}{y+1}  \bigg)\non\\
\2=\2 \fr{1}{\pi} \int_0^{\infty}  dy\fr{1}{\sqrt{y}(y+1)} \bigg(\fr{(1+M)(y+1)- (y+1+M)  }{y+1+M}  \bigg)\non\\
\2=\2 \fr{1}{\pi} \int_0^{\infty}  dy\fr{1}{\sqrt{y}(y+1)} \bigg(\fr{My}{y+1+M}  \bigg).
\eeqa
Hence
\beqa
\|(1+M)^{1/2}-1\|_{\op}\2 \leq \2 \|M\|_{\op} \fr{1}{\pi} \int_0^{\infty}  dy\,\fr{y}{\sqrt{y}(y+1)} \|\bigg(\fr{1}{y+1+M}  \bigg) \|_{\op}\non\\
\2 \leq \2 \|M\|_{\op} \int_0^{\infty}  dy\,\fr{y}{\sqrt{y}(y+1)} \bigg(\fr{1}{y+1/2}  \bigg)=\ch \|M\|_{\op},
\eeqa
which concludes the proof. \qed

\bel\label{Pauli-estimates} The following estimates hold true for $|\vec{v}|, |\vec{w}_1|, |\vec{w}_2|\leq 1$
\beqa
\2 \2 \| v_0 +\i \vec{v}\cdot\vec{\si} \|^2\leq 2(v_0^2+\vec{v}^2), \\
\2 \2 \| (\sqrt{1-\vec{w}_1^2} -\sqrt{1-\vec{w}_2^2} )  +\i (\vec{w}_1-\vec{w}_2)\cdot \vec{\si} \|^2 \leq 6(|\vec{w}_1|^2+|\vec{w}_2|^2).
\eeqa
\eel
\proof  We note that, using $\|C\|\leq  (\mrm{Tr}(C^*C))^{1/2}$,
\beqa
\| v_0 +\i \vec{v}\cdot\vec{\si} \|^2 \2\leq\2 \mrm{Tr}( (v_0 -\i \vec{v}\cdot\vec{\si})( v_0 +\i \vec{v}\cdot\vec{\si}  )   )  \non\\
    \2=\2 \mrm{Tr}( v_0^2 + (\vec{v}\cdot\vec{\si})^2      )=2(v_0^2+\vec{v}^2).
\eeqa
Now choosing $\vec{v}=\vec{w}_1-\vec{w}_2$ and $v_0=\sqrt{1- \vec{w}_1^2}-\sqrt{1-\vec{w}_2^2}$. Thus we  have
\beqa
\|  (\sqrt{1-\vec{w}_1^2} -\sqrt{1-\vec{w}_2^2} )  +\i (\vec{w}_1-\vec{w}_2)\cdot \vec{\si}   \|^2\2\leq\2 2[ \big(\sqrt{1- \vec{w}_1^2}-\sqrt{1-\vec{w}_2^2} \big)^2+(\vec{w}_1-\vec{w}_2)^2]\non\\
\2=\2 2\fr{1-( 1-\vec{w}_1^2)(1-\vec{w}_2^2) }{1+\sqrt{1-\vec{w}_1^2}\sqrt{1-\vec{w}_2^2}}- 2 \vec{w}_1\cdot \vec{w}_2\non\\
\2\leq\2 2( \vec{w}_1^2+\vec{w}_2^2- \vec{w}_1^2\vec{w}_2^2)- 2 \vec{w}_1\cdot \vec{w}_2\leq 3(\vec{w}_1^2+\vec{w}_2^2)\quad
 \label{last-step-squares}
\eeqa 
which gives the required estimate.  \qed\\
\nin {\cyan Finally, we provide computations leading to formula~(\ref{terms}) for the remainder $\vec{r}$:
\bel\label{simple-computations} Setting $\ppU_{\pm}=\ppU(b_{\pm})$,  $\ppV=\pa V( y_b)$  and $\de(M)=1-M$ for any  $M\in \real$, we obtain  
\beqa
\vec{r}(b)\2=\2 - \de(\ppU^0_{-}\ppV^0) \vec{\ppU}_+  + \de(\ppU^0_{+} \ppV^0) \vec{\ppU}_-   + \ppU^0_{+} \ppU^0_{-} \vec{\ppV}\non\\        
                         \2 \2- \ppU^0_{+} (\vec{\ppU}_-\times \vec{\ppV}) + \ppU^0_{-} (\vec{\ppV} \times  \vec{\ppU}_+) +\ppV^0 (\vec{\ppU}_- \times  \vec{\ppU}_+)\non\\
                         \2 \2 + \vec{\ppU}_{+} (\vec{\ppU}_-\cdot \vec{\ppV}) - \vec{\ppU}_{-} (\vec{\ppV} \cdot  \vec{\ppU}_+) +\vec{\ppV} (\vec{\ppU}_- \cdot  \vec{\ppU}_+).                         \label{terms-app}
\eeqa
\eel
\newcommand{\vect}{\overrightarrow}
\proof In the notation introduced in Subsection~\ref{conf-W} we recall  the multiplication table for a product  $UV$ of two elements
of $SU(2)$:
\beqa
\2 \2(UV)^0=U^0V^0-\vec{U} \cdot \vec{V}, \label{multi-zero-app}\\
\2 \2\overrightarrow{(UV)}=U^0 \vec{V}+  V^0 \vec{U}  - (\vec{U}\times \vec{V}), \label{multi-one-app}
\eeqa
where we used (\ref{Pauli-multiplication}). 
Now we look at the product $UV\WZ$ of three elements from $SU(2)$
\beqa
(UV\WZ)^0 \2=\2(UV)^0 Z^0-\vect{(U V)}\cdot \vec{\WZ},\non\\
               \2=\2 U^0 V^0\WZ^0-\vec{U} \cdot \vec{V}\WZ^0- (U^0 \vec{V}+  V^0 \vec{U}  - (\vec{U}\times \vec{V}) )\cdot      \vec{\WZ} \non\\              
 \2=\2 U^0V^0\WZ^0-\WZ^0(\vec{U} \cdot \vec{V})- U^0 (\vec{V}\cdot \vec{\WZ}) -  V^0 (\vec{U}\cdot \vec{\WZ})  +(\vec{U}\times \vec{V}) \cdot  \vec{\WZ} \label{skip-first-app}
\eeqa
and similarly
\beqa
\vect{(U V \WZ)}\2=\2(UV)^0 \vec{\WZ}+  \WZ^0 \vect{UV}  - (\vect{UV} \times \vec{\WZ})\non\\
\2=\2 (U^0V^0-(\vec{U} \cdot \vec{V})) \vec{\WZ} +\WZ^0 (U^0 \vec{V}+  V^0 \vec{U}  - (\vec{U}\times \vec{V}) ) \non\\
\2 \2 -( U^0 \vec{V}+  V^0 \vec{U}  - (\vec{U}\times \vec{V})) \times  \vec{\WZ} \\
\2 \2 \phantom{33333}\non\\
\2=\2 U^0V^0 \vec{\WZ}+\WZ^0 U^0 \vec{V}+ \WZ^0 V^0 \vec{U}- \WZ^0 (\vec{U}\times \vec{V}) - U^0 (\vec{V} \times  \vec{\WZ}) -V^0 (\vec{U} \times  \vec{\WZ})\non\\
\2 \2 -(\vec{U} \cdot \vec{V}) \vec{\WZ} - (\vec{U}\times \vec{V}) \times  \vec{\WZ} \\
\2 \2 \phantom{33333}\non\\
\2=\2 U^0V^0 \vec{\WZ}+\WZ^0 U^0 \vec{V}+ \WZ^0 V^0 \vec{U}- \WZ^0 (\vec{U}\times \vec{V}) - U^0 (\vec{V} \times  \vec{\WZ}) -V^0 (\vec{U} \times  \vec{\WZ})\non\\
\2 \2 - \vec{\WZ}(\vec{U} \cdot \vec{V}) +  \vec{U}(\vec{\WZ}\cdot  \vec{V}) - \vec{V}(\vec{\WZ}\cdot \vec{U} ).   
\eeqa
Therefore, we have by substitution $\vec{\WZ} \to -\vec{\WZ}$.
\beqa
\vect{(U V \WZ^*)}\2=\2 -U^0V^0 \vec{\WZ}+\WZ^0 U^0 \vec{V}+ \WZ^0 V^0 \vec{U}- \WZ^0 (\vec{U}\times \vec{V}) + U^0 (\vec{V} \times  \vec{\WZ}) +V^0 (\vec{U} \times  \vec{\WZ})\non\\
\2 \2 + \vec{\WZ}(\vec{U} \cdot \vec{V}) -  \vec{U}(\vec{\WZ}\cdot  \vec{V}) + \vec{V}(\vec{\WZ}\cdot \vec{U} ). \label{long-formula}
\eeqa
Setting  $\vect{(U V \WZ^*)}=: \vec{U}-\vec{Z} +\vec{r}$, we obtain from  (\ref{long-formula})
\beqa
                     \vec{r}   \2=  \2   \de(U^0V^0) \vec{\WZ}   - \de(\WZ^0 V^0) \vec{U}   + \WZ^0 U^0 \vec{V}\non\\        
                        \2   \2 - \WZ^0 (\vec{U}\times \vec{V}) + U^0 (\vec{V} \times  \vec{\WZ}) +V^0 (\vec{U} \times  \vec{\WZ}) \non\\
\2 \2  +\vec{\WZ}(\vec{U} \cdot \vec{V}) -  \vec{U}(\vec{\WZ}\cdot  \vec{V}) + \vec{V}(\vec{\WZ}\cdot \vec{U} ).
\eeqa
We are interested in the case $U=\pU(b_-)$, $V=\pa V( y_b)$, $\WZ=\pU(b_+)$:
\beqa
 \vec{r}(b) 
 \2=  \2  - \de(\Uun^0 \Vvn^0 )\, \WZzv  + \de(\WZzn^0 \Vvn^0)\, \Uuv   + \WZzn^0 \Uun^0 \Vvv\non\\        
                        \2   \2 - \WZzn^0 (\Uuv\times \Vvv) + \Uun^0 (\Vvv \times  \WZzv) +\Vvn^0 (\Uuv \times \WZzv) \non \\
\2 \2  + {\WZzv}(\Uuv \cdot \Vvv) -   \Uuv  ( \WZzv    \cdot  \Vvv) +  \Vvv (  \WZzv \cdot   \Uuv). \label{r-def-app}
\eeqa
This gives the formula from the statement of the lemma. \qed 
}

\end{document}